\newif\ifOneCol

\OneColfalse

\ifOneCol

\documentclass[draftcls,onecolumn,12pt]{IEEEtran}
\else
\documentclass[twocolumn,10pt]{IEEEtran}
\fi

\usepackage{cite}
\usepackage{citesort}
\usepackage{graphicx}
\usepackage{amsmath}
\usepackage{amsfonts}
\usepackage{epstopdf}
\usepackage{xcolor}
\usepackage{amssymb,bm,upgreek}
\usepackage{algorithm}
\usepackage{algorithmic}
\usepackage{multirow}
\usepackage{transparent}
\usepackage{verbatim}
\usepackage{subfigure}

\usepackage{import}

\DeclareMathOperator\erf{erf}

\let\ss= \scriptscriptstyle

\newcommand{\RX}{\textnormal{RX}}
\newcommand{\R} {\textnormal{RX}}
\newcommand{\TX}{\textnormal{TX}}
\newcommand{\T}{\textnormal{TX}}
\newcommand{\FC} {\textnormal{FC}}
\newcommand{\trans}{\textnormal{trans}}
\newcommand{\report}{\textnormal{report}}
\newcommand{\metre}{\textnormal{m}}

\newcommand{\m}{\textnormal{m}}

\newcommand{\ob}{\textnormal{ob}}

\newcommand{\AF}{{\ss\textnormal{A}}}
\newcommand{\DF}{{\ss\textnormal{D}}}
\newcommand{\A}{{\ss\textnormal{A}}}
\newcommand{\D}{{\ss\textnormal{D}}}
\newcommand{\SD}{{\ss\textnormal{SD}}}

\newcommand{\SA}{{\ss\textnormal{SA}}}
\newcommand{\ca}{${\textnormal{Ca}^{2+}}$}

\newcommand{\s}{\textnormal{s}}

\newcommand{\N}{{\ss\textnormal{I}}}
\newcommand{\ad}{\textnormal{ad}}

\newcommand{\Tot}{{\textnormal{tot}}}

\newtheorem{theorem}{Theorem}

\newtheorem{proposition}{Proposition}
\newtheorem{lemma}{Lemma}

\begin{document}

\title{Symbol-by-Symbol Maximum Likelihood Detection for Cooperative Molecular Communication}

\ifOneCol
\author{Yuting Fang,
        Adam Noel,
        Nan Yang,
        Andrew W. Eckford,
        and Rodney A. Kennedy\vspace{-15mm}

\thanks{This work was presented in part at the IEEE ICC 2018~\cite{Fang2017C}.}
\thanks{Y. Fang, N. Yang, and R. A. Kennedy are with the Research School of Engineering, The Australian National University, Canberra, ACT 2601, Australia (e-mail: \{yuting.fang, nan.yang, rodney.kennedy\}@anu.edu.au).}
\thanks{A. Noel is with the School of Engineering, University of Warwick, Coventry, CV4 7AL, UK (e-mail: adam.noel@warwick.ac.uk).}
\thanks{A. W. Eckford is with the Department of Electrical Engineering and Computer Science, York University, Toronto, ON M3J 1P3, Canada (e-mail:aeckford@yorku.ca)}}
\else
\author{Yuting Fang, \IEEEmembership{Student Member, IEEE},
        Adam Noel, \IEEEmembership{Member, IEEE},
        Nan Yang, \IEEEmembership{Senior Member, IEEE},\\
        Andrew W. Eckford, \IEEEmembership{Senior Member, IEEE},
        and Rodney A. Kennedy,	\IEEEmembership{Fellow, IEEE}\vspace{-6mm}

\thanks{This work was presented in part at the IEEE ICC 2018~\cite{Fang2017C}.}
\thanks{Y. Fang, N. Yang, and R. A. Kennedy are with the Research School of Engineering, Australian National University, Canberra, ACT 2600, Australia (e-mail: \{yuting.fang, nan.yang, rodney.kennedy\}@anu.edu.au).}
\thanks{A. Noel is with the School of Engineering, University of Warwick, Coventry, CV4 7AL, UK (e-mail: adam.noel@warwick.ac.uk).}
\thanks{A. W. Eckford is with the Department of Electrical Engineering and Computer Science, York University, Toronto, ON M3J 1P3, Canada (e-mail:aeckford@yorku.ca)}}
\fi


\markboth{Submitted to IEEE Transactions on Communications}%
{Submitted paper}

\maketitle

\begin{abstract}
In this paper, symbol-by-symbol maximum likelihood (ML) detection is proposed for a cooperative diffusion-based molecular communication (MC) system. In this system, the transmitter (TX) sends a common information symbol to multiple receivers (RXs) and a fusion center (FC) chooses the TX symbol that is more likely, given the likelihood of its observations from all RXs. The transmission of a sequence of binary symbols and the resultant intersymbol interference are considered in the cooperative MC system. Three ML detection variants are proposed according to different RX behaviors and different knowledge at the FC. The system error probabilities for two ML detector variants are derived, one of which is in closed form. \textcolor{black}{The optimal molecule allocation among
RXs to minimize the system error probability of one variant is determined by solving a joint optimization problem. Also for this variant, the equal distribution of molecules among two symmetric RXs is analytically shown to achieve the local minimal error probability.} Numerical and simulation results show that the ML detection variants provide lower bounds on the error performance of simpler, non-ML cooperative variants and demonstrate that these simpler cooperative variants have error performance comparable to ML detectors.
\end{abstract}

\begin{IEEEkeywords}
Molecular communication, multi-receiver cooperation, symbol-by-symbol maximum likelihood detection, error performance, optimization
\end{IEEEkeywords}

\IEEEpeerreviewmaketitle
\ifOneCol
\vspace{-2mm}
\else
\fi

\section{Introduction}\label{sec:intro}

Molecular communication (MC) has been heralded as one of the most promising paradigms to implement communication in bio-inspired nanonetworks, due to the potential benefits of bio\text{-}compatibility and low energy consumption \cite{Nariman}. In MC, the information transmission between devices is realized through the exchange of molecules. Since no source of external energy is required for free diffusion, it is the simplest molecular propagation mechanism. One of the primary challenges posed by diffusion-based MC is that its reliability rapidly decreases when the transmitter (TX)-receiver (RX) distance increases. A naturally-inspired approach, which also makes use of the envisioned collaboration between nanomachines, is allowing multiple RXs to share information for cooperative detection. Often, cells or organisms share common information to achieve a specific task, e.g., calcium (\ca) signaling \cite{Tadashi2010}.

The majority of existing MC studies have focused on the modeling of a single-RX MC system \cite{Nariman}. Recent studies, e.g.,~\cite{Atakan2008,Nakano2013,Chun2013,Koo2016}, have expanded the single-RX MC system to a multi-RX MC system. Although these studies stand on their own merit, they did not establish the potential of \emph{active cooperation} among multiple RXs to determine a TX's intended symbol sequence in a multi-RX MC system, {i.e., the RXs do not actively share their available information to determine the transmitted information}. To address this gap, our work in~\cite{GC2016,TMBMC2016,simplified} analyzed the error performance of a cooperative diffusion-based MC system where a fusion center (FC) device combines the binary decisions of distributed RXs to improve the detection of a TX's symbols.

In other fields of communications, e.g., wireless communications, the maximum likelihood (ML) detector is commonly used to optimize detection performance \cite[Ch,\;5]{Digital}. In the MC domain, the ML sequence detector has been considered for optimality in several studies, e.g., \cite{Deniz2016,Adam2014}. However, the high complexity of sequence detection is a significant barrier to implementation in the MC domain, even when applying simplified algorithms.

The (suboptimal) symbol-by-symbol ML detector requires less computational complexity than the ML sequence detector. {Motivated by this, \cite{Mahfuz2015,Amit2016,Ghavami2017} considered symbol-by-symbol ML detection at a single RX for MC. Recently, \cite{Trang2017,Mosayebi2017,Rogers2016} considered cooperative ML detection for MC. However, in \cite{Trang2017,Mosayebi2017} the RXs communicate with an FC but do not detect information from a TX. Also, in \cite{Mosayebi2017,Rogers2016} the FC makes a single decision about the presence of an abnormality, such that there is only one information symbol and no symbol-by-symbol detection.}

{In this paper, we present symbol-by-symbol ML detection for a cooperative diffusion-based MC system, based on \cite{GC2016,TMBMC2016,simplified}, which consists of one TX, $K$ RXs, and an FC. The significance of this paper is that our results provide lower bounds on the error performance that can be achieved by the detectors considered in \cite{GC2016,TMBMC2016,simplified}. We consider relatively simple RXs with an energy detector or a signal amplifier. The computations required at the RXs can be implemented at the molecular level \cite{Silva}. We keep the relatively high complexity required for ML detection at the FC. This is because the FC could have a direct interface with the macroscopic world and easier access to computational resources. In our proposed system, the transmission of each information symbol from the TX to the FC via the RXs is completed in two phases, as shown in Fig.~\ref{system model}. In the first phase, the TX sends a symbol that is observed by all RXs. In the second phase, the RXs send their detected information to the FC and the FC chooses the TX symbol that is more likely, given the likelihood of its observations from all RXs.

\ifOneCol	
\begin{figure}[!t]
\centering
\includegraphics[height=1.5in]{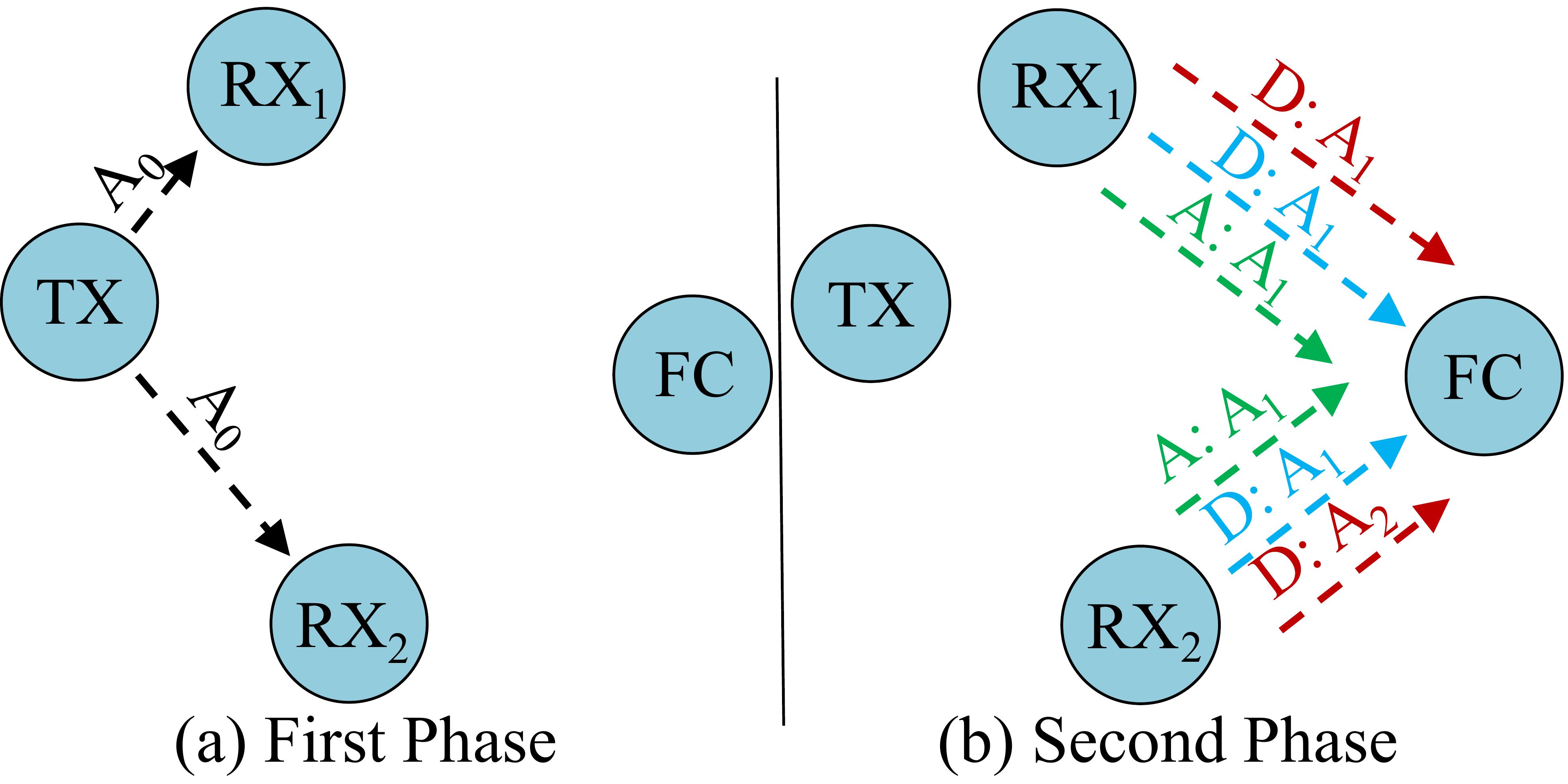}
\vspace{-6mm}
\caption{An example of a cooperative MC system with 2 RXs. The transmission from the TX to the RXs is represented by black dashed arrows. ``D'' and ``A'' denotes the RXs making decisions and amplifying observations, respectively, and $A_k$ denotes the type of released molecule. The transmission from the RXs to the FC in MD-ML, SD-ML, and SA-ML are represented by red, blue, and green arrows, respectively.}
\label{system model}
\vspace{-8mm}
\end{figure}
\else
\begin{figure}[!t]
\centering
\includegraphics[height=1.7in]{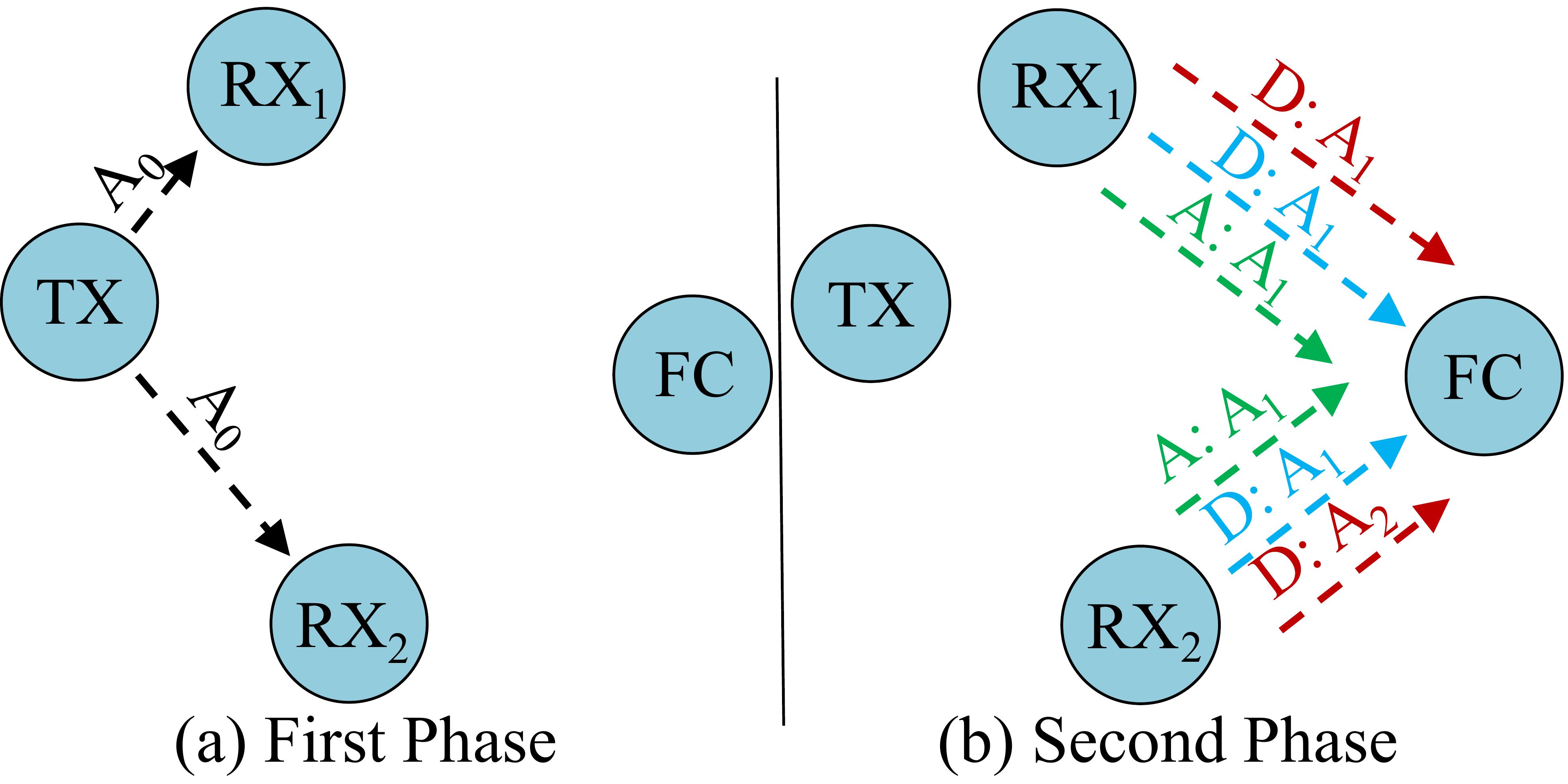}
\vspace{-2mm}
\caption{An example of a cooperative MC system with 2 RXs. The transmission from the TX to the RXs is represented by black dashed arrows. ``D'' and ``A'' denotes the RXs making decisions and amplifying observations, respectively, and $A_k$ denotes the type of released molecule. The transmission from the RXs to the FC in MD-ML, SD-ML, and SA-ML are represented by red, blue, and green arrows, respectively.}
\label{system model}
\vspace{-2mm}
\end{figure}
\fi

{Since binary symbols are the easiest to transmit and detect \cite{Kuran2011}, and we assume that the TX needs to send multiple bits of information in order to execute some complex task (such as disease localization)}, we consider the transmission of a \emph{sequence} of binary symbols and account for the resultant ISI due to previous symbols at the TX and the RXs in the design and analysis of the cooperative MC system. {The results of this paper could be applied to health and environmental monitoring and drug delivery scenarios. In these scenarios, the TX can be a nanomachine that transmits environmental sensor values, e.g., concentration, blood pressure, and temperature, or broadcasts the location of a target site. {To the best of the authors' knowledge, combined with our previous work in \cite{Fang2017C}, this work is the \emph{first} to apply \emph{symbol-by-symbol} ML detection to a cooperative MC system with \emph{multiple} communication phases.} {Although the system topology design and general communication processes can be adapted for traditional cooperative communications, our results cannot be directly applied to traditional cooperative communications due to unique ISI, the propagation channels, and the signal types in this work.

\ifOneCol	
\begin{table}[]
\renewcommand{\arraystretch}{0.8}
\centering
\caption{Variants of ML Detectors}
\label{tab:variants}\vspace{-3mm}
\begin{tabular}{c||c|c|c|c}
\hline
\textbf{Acronym} & \textbf{\begin{tabular}[c]{@{}c@{}}Relaying at RXs\end{tabular}} & \textbf{\begin{tabular}[c]{@{}c@{}}Molecule  Type used  in RXs\end{tabular}} & \textbf{\begin{tabular}[c]{@{}c@{}}Behavior  at FC\end{tabular}} & \textbf{\begin{tabular}[c]{@{}c@{}}Complexity  Comparison\end{tabular}}                                  \\ \hline \hline
MD-ML            & DF                                                                 & Multiple                                                                         & \begin{tabular}[c]{@{}c@{}}ML Detection\end{tabular}            & \multirow{3}{*}{\begin{tabular}[c]{@{}c@{}}MD-ML\\ \textgreater{}SD-ML\\ \textgreater{}SA-ML\end{tabular}} \\ \cline{1-4}
SD-ML            & DF                                                                 & Single                                                                           & \begin{tabular}[c]{@{}c@{}}ML  Detection\end{tabular}            &                                                                                                            \\ \cline{1-4}
SA-ML            & AF                                                                 & Single                                                                           & \begin{tabular}[c]{@{}c@{}}ML Detection\end{tabular}            &                                                                                                            \\ \hline
\end{tabular}
\vspace{-10mm}
\end{table}
\else
\begin{table}[]
\renewcommand{\arraystretch}{1}
\centering
\caption{Variants of ML Detectors}
\label{tab:variants}\vspace{-1mm}
\begin{tabular}{c||c|c|c|c}
\hline
\textbf{Acronym} & \textbf{\begin{tabular}[c]{@{}c@{}}Relaying\\ at RXs\end{tabular}} & \textbf{\begin{tabular}[c]{@{}c@{}}Molecule \\ Type used \\ in RXs\end{tabular}} & \textbf{\begin{tabular}[c]{@{}c@{}}Behavior \\ at FC\end{tabular}} & \textbf{\begin{tabular}[c]{@{}c@{}}Complexity \\ Comparison\end{tabular}}                                  \\ \hline \hline
MD-ML            & DF                                                                 & Multiple                                                                         & \begin{tabular}[c]{@{}c@{}}ML \\ Detection\end{tabular}            & \multirow{3}{*}{\begin{tabular}[c]{@{}c@{}}MD-ML\\ \textgreater{}SD-ML\\ \textgreater{}SA-ML\end{tabular}} \\ \cline{1-4}
SD-ML            & DF                                                                 & Single                                                                           & \begin{tabular}[c]{@{}c@{}}ML \\ Detection\end{tabular}            &                                                                                                            \\ \cline{1-4}
SA-ML            & AF                                                                 & Single                                                                           & \begin{tabular}[c]{@{}c@{}}ML \\ Detection\end{tabular}            &                                                                                                            \\ \hline
\end{tabular}
\vspace{-3mm}
\end{table}
\fi

In this paper, we present three symbol-by-symbol ML detectors: 1) decode-and-forward (DF) with multi-molecule-type and ML detection at the FC (MD-ML), 2) DF with single-molecule-type and ML detection at the FC (SD-ML), and 3) Amplify-and-forward (AF) with single-molecule-type and ML detection at the FC (SA-ML). We summarize these variants in Table \ref{tab:variants}. \textcolor{black}{We design the detectors according to different relaying modes and numbers of types of molecules available at RXs. These variants use either DF relaying or AF relaying and multi-type or single-type molecules. Generally, DF outperforms AF \cite{Ahmadzadeh2015C} and multi-type outperforms single-type molecules, but assumptions of AF and single-type molecules are more realistic in biological environments.} 
ML detection in the current symbol interval requires knowing the previously-transmitted symbols by the TX (and by all RXs for DF). For convenience, we refer to the FC-estimated previous symbols as \emph{local history} and the perfect knowledge of the previous symbols as \emph{genie-aided history}. {Furthermore, the memory required at the FC may be implemented by synthesizing a memory unit into the FC based on \cite{Siuti}.

Our major contributions are summarized as follows:
\begin{enumerate}
\item We present novel symbol-by-symbol ML detection designs for the cooperative MC system with all detector variants, i.e., SD-ML, MD-ML, and SA-ML. For practicality, we consider the FC chooses the current symbol using its \emph{local history} and design the methods for the FC to obtain the local history. We also derive the likelihood of observations for all detectors.
\item We derive analytical expressions for the system error probability for SD-ML and SA-ML using the \emph{genie-aided history}. The assumption of genie-aided history leads to tractable error performance analysis. The analytical error probabilities for SD-ML with $K=1$ and SA-ML are given in closed form. The error performance of MD-ML is mathematically intractable.
\item \textcolor{black}{We determine the optimal molecule allocation among RXs to minimize the system error probability of SD-ML. To achieve this, we formulate and solve a joint optimization problem in terms of molecule allocation and a constant threshold. In this problem, the objective function is the closed-form approximation of error probability of SD-ML since there is no closed-form expression for the error probability of SD-ML. We also analytically prove that the equal distribution of molecules among two symmetric RXs achieves the local minimal error probability of SD-ML.}
\item We validate the accuracy of our analytical expressions of error probability via a particle-based simulation method where we track the motions of molecules over time due to diffusion. Using simulation and numerical results, we also demonstrate the FC's effectiveness in estimating the previously-transmitted symbols \textcolor{black}{and confirm the effectiveness of our optimization method.} 
\end{enumerate}

In contrast to our preliminary work in \cite{Fang2017C}, which only presents ML detection design of SD-ML in a symmetric topology, and did not derive the system error probability, \textcolor{black}{this paper presents two additional detector variants, i.e., MD-ML and SA-ML, relaxes the constraint of symmetric topology, and derives and optimizes the system error probability.} 

\emph{Notations:} We use the following notations: $\textrm{Pr}(\cdot)$ denotes probability. $\lfloor x\rfloor$ denotes the greatest integer that is less than or equal to $x$, $\lceil x\rceil$ denotes the smallest integer that is greater than or equal to $x$, and $\lfloor\cdot\rceil$ denotes the nearest integer. $\log\left(\cdot\right)$ is the natural logarithm, $\erf\left(\cdot\right)$ is the error function, and $\exp\left(\cdot\right)$ is the exponential function. $|\cdot|$ is the cardinality of a set.

\ifOneCol
\vspace{-2mm}
\else
\fi
\section{System Model and Preliminaries}\label{sec:system model}

{In this section, we present the system model (i.e., physical environment and general behaviors of devices) for the cooperative MC system and some preliminary results that are needed in Section \ref{sec:ML detection design}. 
We will describe specific behaviors of the RXs and the FC for the ML detector variants in Section \ref{sec:ML detection design}.}

\vspace{-2mm}
\subsection{System Model}\label{subsec:System Model}

We consider a cooperative MC system in unbounded three-dimensional space. An example of the system is illustrated in Fig.~\ref{system model}. We assume that all RXs and the FC are passive spherical observers. Accordingly, we denote $V_{\ss\R_k}$ and $r_{\ss\R_k}$ as the volume and radius of the $k$th RX, $\RX_k$, respectively, where $k\in\{1,2,\ldots,K\}$. We also denote $V_{\ss\FC}$ and $r_{\ss\FC}$ as the volume and radius of the FC, respectively. We use the terms ``sample'' and ``observation'' interchangeably to refer to the number of molecules observed by a RX or the FC at some time $t$ and assume each observation is independent of each other\footnote{{Intuitively, we consider the time between samples sufficiently large and the distances between the RXs sufficiently large for all individual observations to be independent. The validity of assuming independence will be demonstrated by the excellent agreement between analytical and simulation results in Section \ref{sec:Numerical}.}}. The symbol interval time from the TX to the FC is given by $T = t_{\trans}+t_{\report}$, where $t_{\trans}$ is the transmission interval time from the TX to the RXs and $t_{\report}$ is the report interval time from the RXs to the FC.

\ifOneCol	
\begin{figure}[!t]
\centering
\includegraphics[height=2in]{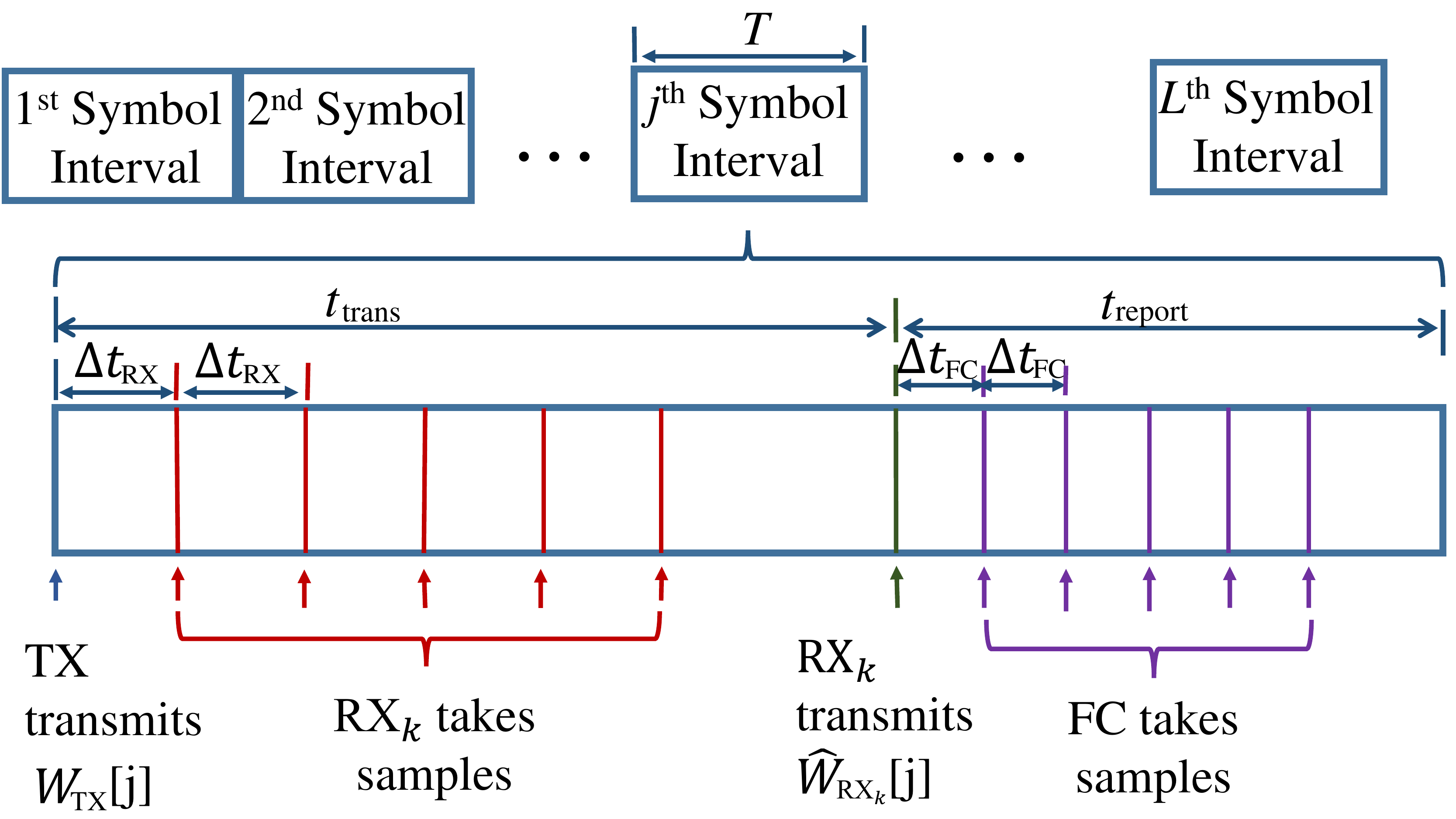}
\vspace{-6mm}
\caption{{An example of the timing schedule for the system with $M_{\ss\RX}=5$ and $M_{\ss\FC}=5$.}}
\label{time2}
\vspace{-8mm}
\end{figure}
\else
\begin{figure}[!t]
\centering
\includegraphics[height=1.8in]{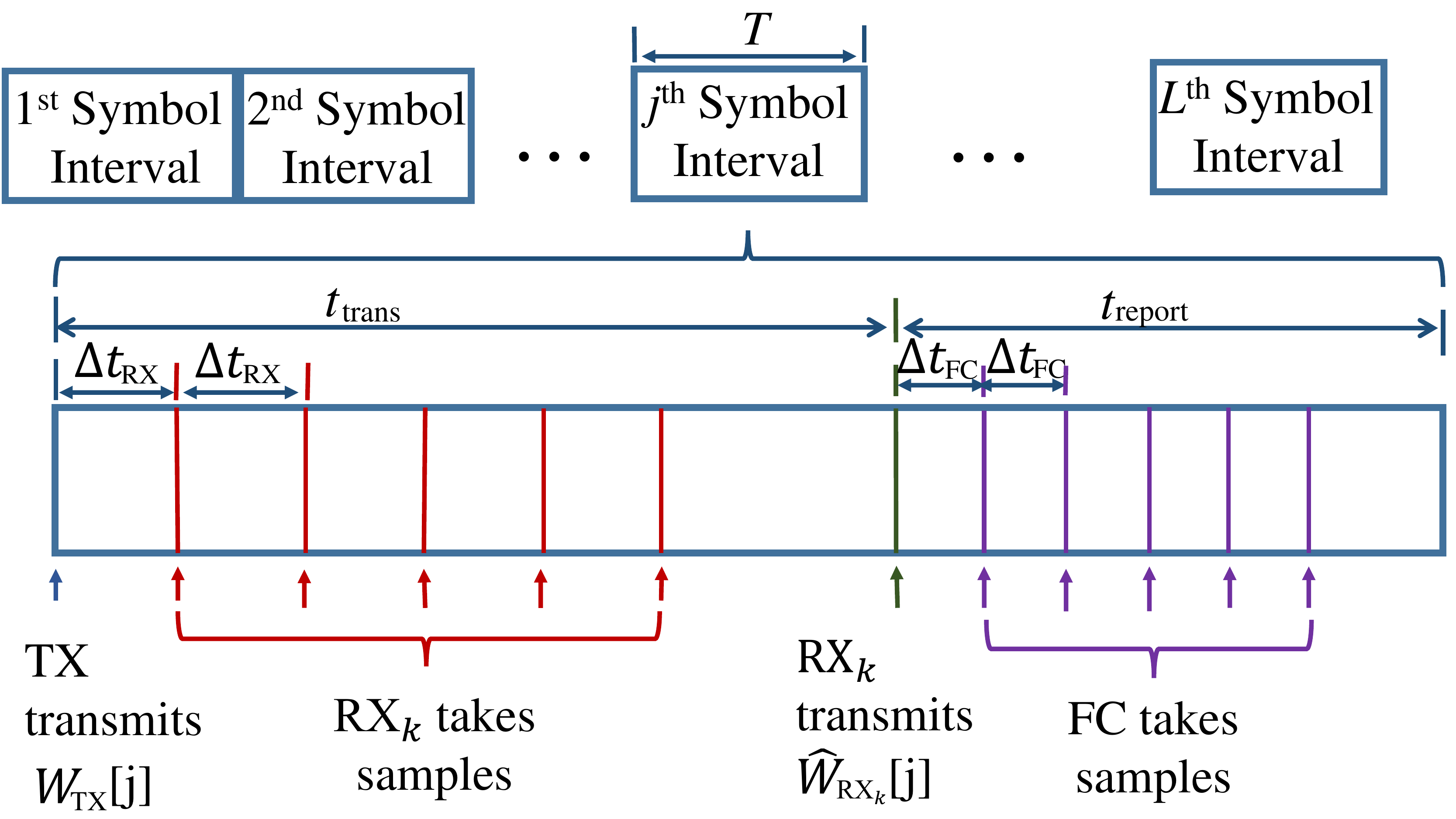}
\caption{{An example of the timing schedule for the system with $M_{\ss\RX}=5$ and $M_{\ss\FC}=5$.}}
\label{time2}
\vspace{-3mm}
\end{figure}
\fi

In the following, we describe the timing schedules and general behaviors of the TX, the RXs, and the FC. An example of the timing schedule for the system is shown in Fig.~\ref{time2}. {The timing schedules of the devices could be implemented by introducing oscillators to control the timing of releasing molecules \cite{Mondrag}. Also, various methods can be adopted to achieve time synchronization\footnote{{All RXs may not be perfectly synchronized. We make the assumption of identical sampling times at all RXs to get a bound on the best error performance achievable by a practical cooperative MC system.}} among nanomachines, e.g., \cite{ShahMohammadian2013,Abadal2011}.}

\underline{TX}: At the beginning of the $j$th symbol interval, i.e., $(j-1)T$, the TX transmits $W_{\ss\T}[j]$. The TX transmits $W_{\ss\T}[j]$ to the RXs over the diffusive channel via type $A_0$ molecules which diffuse independently. The TX uses ON/OFF keying~\cite{Kuran2011} to convey information, i.e., the TX releases $S_{0}$ molecules of type $A_0$ to convey information symbol ``1'' with probability $\textrm{Pr}(W_{\ss\T}[j]=1)=P_1$, but no molecules to convey information symbol ``0''.  The TX then keeps silent until the start of the $(j+1)$th symbol interval. We denote $L$ as the number of symbols transmitted by the TX. We define $\textbf{W}_{{\ss\T}}^{l}=\left\{W_{\ss\T}[1],\ldots,W_{\ss\T}[l]\right\}$ as an $l$-length subsequence of the symbols transmitted by the TX, where $l\leq{L}$. Throughout the paper, $W$ is a single symbol and $\textbf{W}$ is a vector of symbols. {We do not consider channel codes for this system since the required encoder and the decoder may not be practical for MC systems \cite{Lu2015,Nariman}.}

\underline{RX}: Each $\RX_{k}$ observes type $A_0$ molecules over the $\TX-\RX_{k}$ link and takes $M_{\ss\RX}$ samples\footnote{{We consider multiple samples at the RXs and the FC in each symbol interval to improve the detection performance.}} in each symbol interval at the same times. The time of the $m$th sample by each RX in the $j$th symbol interval is given by $t_{\ss\R}(j,m) = (j-1)T + m\Delta{t_{\ss\R}}$, where $\Delta{t_{\ss\R}}$ is the time step between two successive samples by each RX, $m\in\left\{1,2,\ldots,M_{\ss\RX}\right\}$. {The RXs operate in half-duplex mode, such that they do not receive the information and report their decisions at the same time. This is because half-duplex mode is more appropriate in a biological environment since it requires lower computational complexity than full-duplex mode.} At the time $(j-1)T + t_{\trans}$, each RX transmits molecules via a diffusion-based channel to the FC. For MD-ML and SD-ML, each RX detects with a relatively simple \emph{energy} detector \cite{Adam2014}. We denote $\hat{W}_{{\ss\R}_k}[j]$ as $\RX_k$'s binary decision on the $j$th transmitted symbol. Based on the energy detector, $\RX_k$ makes decision $\hat{W}_{\ss\R_k}[j]=1$ if $s_{k}[j]\geq\xi_{\ss\R_k}$, otherwise $\hat{W}_{\ss\R_k}[j]=0$, where $s_{k}[j]$ is the value of the realization of $S_{\ob}^{\ss{\R_k}}[j]$ and $\xi_{\ss\R_k}$ is the constant detection threshold at $\RX_k$, independent of $\textbf{W}_{\ss\T}^{j-1}$. We define $\hat{\textbf{W}}^{l}_{\ss\RX_k}=\left\{\hat{W}_{\ss\RX_k}[1],\ldots,\hat{W}_{\ss\RX_k}[l]\right\}$ as an $l$-length subsequence of $\RX_k$'s binary decisions.

\underline{FC}: The FC takes the $\tilde{m}$th sample in the $j$th symbol interval at $t_{\ss\FC}(j,\tilde{m})=(j-1)T+t_{\trans}+\tilde{m}\Delta{t_{\ss\FC}}$, where $\Delta{t_{\ss\FC}}$ is the time step between two successive samples by the FC and $\tilde{m}\in\left\{1,2,\ldots, M_{\ss\FC}\right\}$. We denote $\hat{W}_{\ss\FC}[j]$ as the FC's decision on the $j$th symbol transmitted by the TX. We define $\hat{\textbf{W}}^{l}_{\ss\FC}=\left\{\hat{W}_{\ss\FC}[1],\ldots,\hat{W}_{\ss\FC}[l]\right\}$ as an $l$-length subsequence of the FC's decisions on the symbols transmitted by the TX. We denote $\hat{W}_{\ss\FC_k}[j]$ as the FC's estimated binary decision of $\RX_k$ on the $j$th transmitted symbol. We define $\hat{\textbf{W}}^{l}_{\ss\FC_k}=\left\{\hat{W}_{\ss\FC_k}[1],\ldots,\hat{W}_{\ss\FC_k}[l]\right\}$ as the FC's estimate of the first $l$ binary decisions by $\RX_k$.

\ifOneCol
\vspace{-2mm}
\else
\fi

\subsection{Preliminaries}
In this subsection, we establish some preliminary results for a $\TX-\RX_k$ link and a $\RX_k-\FC$ link. We first evaluate the probability $P_{\ob}^{({\ss{\T},{\ss\R_k}})}\left(t\right)$ of observing a given type $A_0$ molecule, emitted from the TX at $t=0$, inside $V_{\ss\R_k}$ at time $t$. Based on \cite[Eq. (27)]{Noel2013}, we write $P_{\ob}^{({\ss{\T},{\ss\R_k}})}\left(t\right)$ as
\ifOneCol	
\begin{align}\label{general prob}
P_{\ob}^{({\ss{\T},{\ss\R_k}})}(t)=&\;\frac{1}{2}\left[\erf\left(\tau_{1}\right)+\erf\left(\tau_{2}\right)\right]
-\frac{\sqrt{D_{{0}}t}}{d_{\ss\T_k}\sqrt{\pi}}\left[\exp\left(-\tau_{1}^{2}\right)-\exp\left(-\tau_{2}^{2}\right)\right],
\end{align}
\else
\begin{align}\label{general prob}
P_{\ob}^{({\ss{\T},{\ss\R_k}})}(t)=&\;\frac{1}{2}\left[\erf\left(\tau_{1}\right)+\erf\left(\tau_{2}\right)\right]\nonumber\\
&-\frac{\sqrt{D_{{0}}t}}{d_{\ss\T_k}\sqrt{\pi}}\left[\exp\left(-\tau_{1}^{2}\right)-\exp\left(-\tau_{2}^{2}\right)\right],
\end{align}
\fi
where $\tau_{1}=\frac{r_{\ss\R_k}+d_{\ss\TX_k}}{2\sqrt{D_{{0}}t}}$,  $\tau_{2}=\frac{r_{\ss\R_k}-d_{\ss\TX_k}}{2\sqrt{D_{{0}}t}}$, $D_{0}$ is the diffusion coefficient of type $A_0$ molecules in $\m^{2}/{\s}$, $d_{\ss\T_k}$ is the distance between the TX and $\RX_k$ in $\m$. We denote the sum of $M_{\ss\RX}$ samples by $\RX_k$ in the $j$th symbol interval by $S_{\ob}^{\ss{\R_k}}[j]$. {As discussed in \cite{Pierobon,Mosayebi2014}, $S_{\ob}^{\ss{\R_k}}[j]$ can be accurately approximated by a Poisson random variable (RV). The mean of $S_{\ob}^{\ss{\R_k}}[j]$ is then given by
\ifOneCol
\begin{align}\label{observed molecular numbers R1}
\bar{S}_{\ob}^{\ss{\R_k}}[j]
=&\;\sum\limits^{j}_{i=1}S_{0}W_{\ss\T}[i]\sum\limits^{M_{\ss\RX}}_{m=1}P_{\ob}^{({\ss{\T},{\ss\R_k}})}\left(\left(j-i\right)T + m\Delta{t_{\ss\R}}\right).
\end{align}
\else
\begin{align}\label{observed molecular numbers R1}
\bar{S}_{\ob}^{\ss{\R_k}}[j]
\!=\!\sum\limits^{j}_{i=1}\!S_{0}W_{\ss\T}[i]\!\sum\limits^{M_{\ss\RX}}_{m=1}\!P_{\ob}^{({\ss{\T},{\ss\R_k}})}\!\left(\left(j-i\right)T \!+ \!m\Delta{t_{\ss\R}}\right).
\end{align}
\fi

We denote $P_{\ob,{k}}^{({\ss\R_k,\ss\FC})}(t)$ as the probability of observing a given $A_k$ molecule, emitted from the center of $\RX_k$ at $t=0$, inside $V_{\ss\FC}$ at time $t$.
We obtain $P_{\ob,{k}}^{(\ss\R_k,\ss\FC)}(t)$ by replacing $r_{\ss\R_k}$, $d_{\ss\TX_k}$, and $D_{{0}}$ with $r_{\ss\FC}$, $d_{\ss\FC_k}$, and $D_{{k}}$, respectively, where $D_{{k}}$ is the diffusion coefficient of type $A_{k}$ molecules in ${\m^{2}}/{\s}$ and $d_{\ss\FC_k}$ is the distance between $\RX_k$ and the $\FC$ in $\m$.

\ifOneCol
\vspace{-2mm}
\else
\fi
\section{ML Detection Design and Derivation}\label{sec:ML detection design}
In this section, we design and derive three symbol-by-symbol ML detectors. Throughout this section, the FC uses its \emph{local} history to choose the current symbol, i.e., the FC evaluates the likelihood of the observations $\hat{\textbf{W}}_{\ss\FC}^{j-1}$ and $\hat{\textbf{W}}^{j-1}_{\ss\FC_k}$ ($\hat{\textbf{W}}^{j-1}_{\ss\FC_k}$ is not needed for SA-ML) in the $j$th symbol interval, as shown in Table \ref{local2,ed}, where $k\in\{1,2,\ldots,K\}$. Using the local history at the FC, we formulate the general decision rule of ML detection in the $j$th interval as
\begin{align}\label{ML rule}
\hat{W}_{\ss\FC}[j]=\underset{W_{\ss\T}[j]\in\{0,1\}}{\text{argmax}}~ \mathcal{L}\left[j|W_{\ss\T}[j],\hat{\textbf{W}}_{\ss\FC}^{j-1}\right]
\end{align}
or
\begin{align}\label{ML rule,DF}
\hat{W}_{\ss\FC}[j]=\underset{W_{\ss\T}[j]\in\{0,1\}}{\text{argmax}}~ \mathcal{L}\left[j|W_{\ss\T}[j],\hat{\textbf{W}}_{\ss\FC}^{j-1},\hat{\textbf{W}}^{j-1}_{\ss\FC_k}\right],
\end{align}
where we define $\mathcal{L}\left[j|\cdot\right]\triangleq\text{Pr}\left(\text{FC's observations in $j$th interval}|\cdot\right)$. Eq. \eqref{ML rule} applies to SA-ML and \eqref{ML rule,DF} applies to SD-ML and MD-ML.
For simplicity, we also write the likelihoods in \eqref{ML rule} and \eqref{ML rule,DF} as $\mathcal{L}\left[j\right]$. In the following, we present the specific behaviors of the RXs and the FC of each ML detector, derive the corresponding $\mathcal{L}\left[j\right]$, and compare the complexities of the detectors. 
\ifOneCol	
\begin{table}[!t]
\renewcommand{\arraystretch}{0.8}
\centering\vspace{-3mm}
\caption{Illustration of the FC's local history}\vspace{-5mm}
\label{local2,ed}
\begin{tabular}{c||c|c}
\hline
\textbf{Interval} & \textbf{The FC's decisions}                      & \textbf{The FC's local history}                                                                      \\ \hline\hline
1                  &       $\hat{W}_{\ss\FC}[1]$ and $\hat{W}_{\ss\FC_k}[1]$            &          No History                                                                    \\ \hline
2                  & $\hat{W}_{\ss\FC}[2]$ and $\hat{W}_{\ss\FC_k}[2]$                  & $\hat{W}_{\ss\FC}[1]$ and $\hat{W}_{\ss\FC_k}[1]$                                                                      \\ \hline
$\vdots$             &   $\vdots$                 &  $\vdots$                                                                      \\ \hline
\multirow{2}{*}{L} & \multirow{2}{*}{$\hat{W}_{\ss\FC}[L]$ and $\hat{W}_{\ss\FC_k}[L]$} & \multirow{2}{*}{\begin{tabular}[c]{@{}l@{}}$\hat{W}_{\ss\FC}[L-1], \ldots, \hat{W}_{\ss\FC}[1]$ and $\hat{W}_{\ss\FC_k}[L-1], \ldots,\hat{W}_{\ss\FC_k}[1]$\end{tabular}} \\
                   &                          &                                                                              \\ \hline
\end{tabular}
\vspace{-8mm}
\end{table}
\else
\begin{table}[!t]
\renewcommand{\arraystretch}{1.3}
\centering
\caption{Illustration of the FC's local history}
\label{local2,ed}
\begin{tabular}{c||c|c}
\hline
\textbf{Interval} & \textbf{The FC's decisions}                      & \textbf{The FC's local history}                                                                      \\ \hline\hline
1                  &       $\hat{W}_{\ss\FC}[1]$ and $\hat{W}_{\ss\FC_k}[1]$            &          No History                                                                    \\ \hline
2                  & $\hat{W}_{\ss\FC}[2]$ and $\hat{W}_{\ss\FC_k}[2]$                  & $\hat{W}_{\ss\FC}[1]$ and $\hat{W}_{\ss\FC_k}[1]$                                                                      \\ \hline
$\vdots$             &   $\vdots$                 &  $\vdots$                                                                      \\ \hline
\multirow{2}{*}{L} & \multirow{2}{*}{$\hat{W}_{\ss\FC}[L]$ and $\hat{W}_{\ss\FC_k}[L]$} & \multirow{2}{*}{\begin{tabular}[c]{@{}l@{}}$\hat{W}_{\ss\FC}[L-1], \ldots, \hat{W}_{\ss\FC}[1]$\\ and $\hat{W}_{\ss\FC_k}[L-1], \ldots,\hat{W}_{\ss\FC_k}[1]$\end{tabular}} \\
                   &                          &                                                                              \\ \hline
\end{tabular}
\vspace{-4mm}
\end{table}
\fi


\ifOneCol
\vspace{-2mm}
\else
\fi
\subsection{MD-ML}\label{subsubsec:MD}
Each $\RX_k$ in MD-ML transmits type $A_{k}$ molecules, which can be independently detected by the FC, to report $\hat{W}_{{\ss\R}_k}[j]$ to the FC. Similar to the TX, each RX uses ON/OFF keying to report its decision to the FC and the RX releases $S_k$ molecules of type $A_k$ to convey information symbol ``1''. The FC receives type $A_k$ molecules over the $\RX_{k}-\FC$ link and takes $M_{\ss\FC}$ samples of each of the $K$ types of molecules transmitted by all RXs in every reporting interval. The FC adds $M_{\ss\FC}$ observations for each $\RX_k-\FC$ link in the $j$th symbol interval. We denote ${S}_{\ob,{k}}^{{\ss\FC},\D}[j]$ as the total number of $A_k$ molecules observed within $V_{\ss\FC}$ in the $j$th symbol interval, due to both current and previous emissions of molecules by $\RX_k$. The TX and $\RX_k$ use the same modulation method and the $\TX-\RX_k$ and $\RX_k-\FC$ links are both diffusion-based. Therefore, like $S_{\ob}^{\ss{\R_k}}[j]$, ${S}_{\ob,{k}}^{\ss\FC,\D}[j]$ can also be accurately approximated as a Poisson RV. We denote ${\bar{S}}_{\ob,{k}}^{\ss\FC,\D}[j]$ as the mean of ${S}_{\ob,{k}}^{\ss\FC,\D}[j]$. Values of realizations of ${S}_{\ob,{k}}^{{\ss\FC},\D}[j]$ are labeled $\tilde{s}_{k}[j]$. We assume that the $K$ $\RX_k-\FC$ links are independent, so the FC has $K$ independent sums $\tilde{s}_{k}[j]$ from the $K$ $\RX_k-\FC$ links. The FC chooses the symbol $\hat{W}_{\ss\FC}[j]$ that is more likely, given the joint likelihood of the $K$ sums $\tilde{s}_{k}[j]$ in the $j$th interval. We obtain $\mathcal{L}\left[j\right]$ by
\ifOneCol
\begin{align}\label{MD-ML}
\mathcal{L}\left[j\right]
=&\;\prod\limits^{K}_{k=1}\bigg[\textrm{Pr}\left(\hat{W}_{\ss\R_k}[j]=1|W_{\ss\T}[j],\hat{\textbf{W}}_{\ss\FC}^{j-1}\right)
\textrm{Pr}\left({S}_{\ob,{k}}^{{\ss\FC},\D}[j]=\tilde{s}_{k}[j]|\hat{W}_{\ss\R_k}[j]=1,\hat{\textbf{W}}_{\ss\FC_k}^{j-1}\right)\nonumber\\
&+\textrm{Pr}\left(\hat{W}_{\ss\R_k}[j]=0|W_{\ss\T}[j],\hat{\textbf{W}}_{\ss\FC}^{j-1}\right)
\textrm{Pr}\left({S}_{\ob,{k}}^{{\ss\FC},\D}[j]=\tilde{s}_{k}[j]|\hat{W}_{\ss\R_k}[j]=0,\hat{\textbf{W}}_{\ss\FC_k}^{j-1}\right)\bigg].
\end{align}
\else
\begin{align}\label{MD-ML}
\mathcal{L}\left[j\right]
=&\;\prod\limits^{K}_{k=1}\bigg[\textrm{Pr}\left(\hat{W}_{\ss\R_k}[j]=1|W_{\ss\T}[j],\hat{\textbf{W}}_{\ss\FC}^{j-1}\right)\nonumber\\
&\times\textrm{Pr}\left({S}_{\ob,{k}}^{{\ss\FC},\D}[j]=\tilde{s}_{k}[j]|\hat{W}_{\ss\R_k}[j]=1,\hat{\textbf{W}}_{\ss\FC_k}^{j-1}\right)\nonumber\\
&+\textrm{Pr}\left(\hat{W}_{\ss\R_k}[j]=0|W_{\ss\T}[j],\hat{\textbf{W}}_{\ss\FC}^{j-1}\right)\nonumber\\
&\times\textrm{Pr}\left({S}_{\ob,{k}}^{{\ss\FC},\D}[j]=\tilde{s}_{k}[j]|\hat{W}_{\ss\R_k}[j]=0,\hat{\textbf{W}}_{\ss\FC_k}^{j-1}\right)\bigg].
\end{align}
\fi

For the evaluation of the likelihood in all future intervals, i.e., $\mathcal{L}\left[j+1\right],\ldots,\mathcal{L}\left[L\right]$, the FC also chooses the symbol $\hat{W}_{\ss\FC_k}[j]$ in the $j$th interval given the likelihood of the sum $\tilde{s}_{k}[j]$ from the $\RX_k-\FC$ link in the $j$th interval. By doing so, $\hat{W}_{\ss\FC_k}[j]$ is obtained by
\ifOneCol
\begin{align}\label{ML rule_RX_MD}
\hat{W}_{\ss\FC_k}[j]=\underset{\hat{W}_{\ss\R_k}[j]\in\{0,1\}}{\text{argmax}}~\textrm{Pr}\left({S}_{\ob,{k}}^{{\ss\FC},\D}[j]=\tilde{s}_{k}[j]|\hat{W}_{\ss\R_k}[j],\hat{\textbf{W}}_{\ss\FC_k}^{j-1}\right).
\end{align}
\else
\begin{align}\label{ML rule_RX_MD}
\hat{W}_{\ss\FC_k}[j]\!=\hspace{-2mm}\underset{\hat{W}_{\ss\R_k}[j]\in\{0,1\}}{\text{argmax}}\hspace{-2mm}\textrm{Pr}\left(\!{S}_{\ob,{k}}^{{\ss\FC},\D}[j]\!=\!\tilde{s}_{k}[j]|\hat{W}_{\ss\R_k}[j],\hat{\textbf{W}}_{\ss\FC_k}^{j-1}\!\right).
\end{align}
\fi

Eqs. \eqref{MD-ML} and \eqref{ML rule_RX_MD} can be evaluated by applying the conditional cumulative distribution function (CDF) of the Poisson RV $S_{\ob}^{\ss{\R_k}}[j]$ and the conditional PMF of the Poisson RV ${S}_{\ob,{k}}^{{\ss\FC},\D}[j]$.
The conditional means ${\bar{S}}_{\ob,{k}}^{\ss\FC,\D}[j]$ given $\hat{\textbf{W}}_{\ss\FC_k}^{j-1}$ are obtained by replacing $S_{0}$, $W_{\ss\T}[i]$, $P_{\ob}^{({\ss{\T},{\ss\R_k}})}$, $M_{\ss\RX}$, $m$, and $\Delta{t_{\ss\R}}$ in \eqref{observed molecular numbers R1} with $S_{k}$, $\hat{W}_{\ss\FC_k}[i]$, $P_{\ob,{k}}^{({\ss\R_k,\ss\FC})}$, $M_{\ss\FC}$, $\tilde{m}$, and $\Delta{t_{\ss\FC}}$, respectively.
\ifOneCol
\vspace{-2mm}
\else
\fi
\subsection{SD-ML}\label{subsubsec:SD-ML}
The behavior of each $\RX_k$ in SD-ML is the same as that in MD-ML, except we assume that each $\RX_k$ transmits type $A_{1}$ molecules to report $\hat{W}_{{\ss\R}_k}[j]$ to the FC. This is because it may not be realistic for each RX to release a unique type of molecule. \textcolor{black}{For simplicity, the number of released type $A_1$ molecules for each $\RX_k$ in SD-ML is also denoted by $S_k$.} The FC receives type $A_1$ molecules over all $K$ $\RX_{k}-\FC$ links and takes $M_{\ss\FC}$ samples of type $A_1$ molecules in each symbol interval.
The FC adds $M_{\ss\FC}$ observations for all $\RX_k-\FC$ links in the $j$th symbol interval. We denote ${S}_{\ob}^{{\ss\FC},\D}[j]$ as the total number of $A_1$ molecules observed within $V_{\ss\FC}$ in the $j$th symbol interval, due to both current and previous emissions of molecules by all RXs. We note that ${S}_{\ob}^{\ss\FC,\D}[j]=\sum^{K}_{k=1}S_{\ob,k}^{\ss{\FC},\D}[j]$ is also a Poisson RV whose mean is given by ${\bar{S}}_{\ob}^{\ss\FC,\D}[j]=\sum^{K}_{k=1}\bar{S}_{\ob,k}^{\ss{\FC},\D}[j]$. Values of realizations of ${S}_{\ob}^{\ss\FC,\D}[j]$ are labeled $\tilde{s}[j]$. The FC chooses the symbol $\hat{W}_{\ss\FC}[j]$ that is more likely, given the likelihood of $\tilde{s}[j]$ in the $j$th interval. \textcolor{black}{To facilitate the evaluation of $\mathcal{L}\left[j\right]$ for SD-ML, we define $\hat{\mathcal{W}}^{{\ss\FC}}_l=\{\hat{W}_{\ss\FC_1}[l],\ldots,\hat{W}_{\ss\FC_K}[l]\}$. Using the notation $\hat{\mathcal{W}}^{{\ss\FC}}_l$, we derive $\mathcal{L}\left[j\right]$ as
\ifOneCol
\begin{align}\label{SD-ML}
\mathcal{L}\left[j\right]=
&\;\sum_{h=1}^{2^K}\left[\textrm{Pr}\left(\hat{\mathcal{W}}^{{\ss\RX}}_{j,h} |W_{\ss\T}[j],\hat{\textbf{W}}_{\ss\FC}^{j-1}\right)\right.
\left.\textrm{Pr}\left({S}_{\ob}^{\ss\FC,\D}[j]=\tilde{s}[j]|\hat{\mathcal{W}}^{{\ss\RX}}_{j,h} ,\hat{\mathcal{W}}^{{\ss\FC}}_{j-1},\ldots,\hat{\mathcal{W}}^{{\ss\FC}}_1\right)\right],
\end{align}
\else
\begin{align}\label{SD-ML}
\mathcal{L}\left[j\right]=
&\;\sum_{h=1}^{2^K}\left[\textrm{Pr}\left(\hat{\mathcal{W}}^{{\ss\RX}}_{j,h} |W_{\ss\T}[j],\hat{\textbf{W}}_{\ss\FC}^{j-1}\right)\right.\nonumber\\
&\left.\times\textrm{Pr}\left({S}_{\ob}^{\ss\FC,\D}[j]=\tilde{s}[j]|\hat{\mathcal{W}}^{{\ss\RX}}_{j,h} ,\hat{\mathcal{W}}^{{\ss\FC}}_{j-1},\ldots,\hat{\mathcal{W}}^{{\ss\FC}}_1\right)\right],
\end{align}
\fi
where $\hat{\mathcal{W}}^{{\ss\RX}}_{j,h}$ is the $h$th realization of the vector $\{\hat{W}_{\ss\R_1}[j],\ldots,\hat{W}_{\ss\R_K}[j]\}$, $h\in\{1,2,\ldots,2^K\}$.} For \eqref{SD-ML}, we need to consider each $\hat{\mathcal{W}}^{{\ss\RX}}_{j,h} $ and the corresponding probability leading to $\tilde{s}[j]$. For the evaluation of the likelihood in all future intervals, \textcolor{black}{the FC chooses $\hat{\mathcal{W}}^{{\ss\FC}}_j$ that gives the maximum likelihood of $\tilde{s}[j]$. By doing so, $\hat{\mathcal{W}}^{{\ss\FC}}_j$ is obtained by
\begin{align}\label{ML rule_RX_SD}
\hat{\mathcal{W}}^{{\ss\FC}}_j\!=\!\underset{\hat{\mathcal{W}}^{{\ss\RX}}_{j,h}}{\text{argmax}}\;\textrm{Pr}\!\left({S}_{\ob}^{\ss\FC,\D}[j]\!=\!\tilde{s}[j]|\hat{\mathcal{W}}^{{\ss\RX}}_{j,h} ,\hat{\mathcal{W}}^{{\ss\FC}}_{j-1},\ldots,\hat{\mathcal{W}}^{{\ss\FC}}_1\right)\!.
\end{align}}
We now derive the conditional mean of ${S}_{\ob}^{\ss\FC,\D}[j]$ given $\hat{\mathcal{W}}^{{\ss\RX}}_{j,h} $ and $\hat{\mathcal{W}}^{{\ss\FC}}_{j-1},\ldots,\hat{\mathcal{W}}^{{\ss\FC}}_1$. To this end, we evaluate ${\bar{S}}_{\ob}^{\ss\FC,\D}[j]$ as
\ifOneCol	
\begin{align}\label{observed molecular numbers R1_SD1}
{\bar{S}}_{\ob}^{\ss\FC,\D}[j]
=\!&\sum\limits^{K}_{k=1}\sum\limits^{M_{\ss\FC}}_{\tilde{m}=1}S_{k}\left(\!\sum\limits^{j-1}_{i=1}\hat{W}_{\ss\FC_k}[i]
P_{\ob,{k}}^{({\ss\R_k,\ss\FC})}\left(\left(j-i\right)T \!+\! \tilde{m}\Delta{t_{\ss\FC}}\right)\!+\!\hat{W}_{\ss\RX_k}[j]P_{\ob,{k}}^{({\ss\R_k,\ss\FC})}\left(\tilde{m}\Delta{t_{\ss\FC}}\right)\!\right).
\end{align}
\else
\begin{align}\label{observed molecular numbers R1_SD1}
{\bar{S}}_{\ob}^{\ss\FC,\D}[j]
=&\;\sum\limits^{K}_{k=1}\sum\limits^{M_{\ss\FC}}_{\tilde{m}=1}S_{k}\Big(\hat{W}_{\ss\RX_k}[j]P_{\ob,{k}}^{({\ss\R_k,\ss\FC})}\left(\tilde{m}\Delta{t_{\ss\FC}}\right)\nonumber\\
&+\sum\limits^{j-1}_{i=1}\hat{W}_{\ss\FC_k}[i]P_{\ob,{k}}^{({\ss\R_k,\ss\FC})}\left(\left(j-i\right)T + \tilde{m}\Delta{t_{\ss\FC}}\right)\Big).
\end{align}
\fi

\ifOneCol
\vspace{-2mm}
\else
\fi
\subsection{SA-ML}\label{subsubsec:SA}
For SA-ML, each RX amplifies the number of molecules observed in the $j$th symbol interval, i.e., $S_{k}^\AF[j]= \alpha_k S_{\ob}^{\ss{\R_k}}[j]$, where $S_{k}^\AF[j]$ denotes the number of molecules released by $\RX_k$ in the $j$th symbol interval and $\alpha_k$ is the \emph{constant} amplification factor at $\RX_k$. The RXs retransmit $S_{k}^\AF[j]$ molecules of type $A_1$ to the FC at the same time. Since all RXs in both SA-ML and SD-ML release molecules of the same type $A_1$, the description of the behavior of the FC in SA-ML is analogous to that in SD-ML. We denote ${S}_{\ob,{k}}^{{\ss\FC},\A}[j]$ as the number of molecules observed within $V_{\ss\FC}$ in the $j$th symbol interval, due to the emissions of molecules from the current and the previous intervals by $\RX_k$. The $\TX-\RX_k$ and $\RX_k-\FC$ links are both diffusion-based. Therefore, ${S}_{\ob,{k}}^{{\ss\FC},\A}[j]$ can be accurately approximated as a Poisson RV. We denote ${\bar{S}}_{\ob,{k}}^{\ss\FC,\A}[j]$ as the mean of ${S}_{\ob,{k}}^{{\ss\FC},\A}[j]$. The FC adds $M_{\ss\FC}$ observations for all $\RX_k-\FC$ links in the $j$th symbol interval and this sum is denoted by the RV ${S}_{\ob}^{\ss\FC,\A}[j]$. We note that ${S}_{\ob}^{{\ss\FC},\A}[j] = \sum^{K}_{k=1}{S}_{\ob,{k}}^{{\ss\FC},\A}[j]$ is also a Poisson RV whose mean is given by ${\bar{S}}_{\ob}^{{\ss\FC},\A}[j] = \sum^{K}_{k=1}{\bar{S}}_{\ob,{k}}^{{\ss\FC},\A}[j]$. Values of realizations of ${S}_{\ob}^{\ss\FC,\A}[j]$ are labeled $\tilde{s}[j]$. The FC chooses the symbol $\hat{W}_{\ss\FC}[j]$ that is more likely given the likelihood of $\tilde{s}[j]$ in the $j$th interval and $\mathcal{L}\left[j\right]$ is given
\ifOneCol	
by
{\begin{align}\label{SA-ML}
\mathcal{L}\left[j\right]
= &\;\sum_{s_{1}[1]= 0}^{S_0}\ldots\sum_{s_{1}[j] = 0}^{S_0}\ldots\sum_{s_{K}[1]= 0}^{S_0}\ldots\sum_{s_{K}[j] = 0}^{S_0} \nonumber\\
&\hspace{-15mm}\!\times\!\textrm{Pr}\Big(\!S_{\ob}^{\ss{\R_1}}[1]=s_{1}[1]\!,
\ldots,\!S_{\ob}^{\ss{\R_1}}[j]=s_{1}[j],\!\ldots,\!S_{\ob}^{\ss{\R_K}}[1]=s_{K}[1],\!\ldots,\!
S_{\ob}^{\ss{\R_K}}[j]=s_{K}[j]|W_{\ss\T}[j],\!\hat{\textbf{W}}_{\ss\FC}^{j-1}\!\Big)\nonumber\\
&\hspace{-15mm}\!\times\!\textrm{Pr}\Big(\!{S}_{\ob}^{\ss\FC,\A}[j]=\tilde{s}[j]|S_{\ob}^{\ss{\R_1}}[1]=s_{1}[1]\!,\ldots,\!S_{\ob}^{\ss{\R_1}}[j]=s_{1}[j],\!\ldots,\!S_{\ob}^{\ss{\R_K}}[1]=s_{K}[1],\!\ldots,\!S_{\ob}^{\ss{\R_K}}[j]=s_{K}[j] \!\Big),%
\end{align}}
\else
in \eqref{SA-ML} at the top of the following page.
\begin{figure*}
{\begin{align}\label{SA-ML}
\mathcal{L}\left[j\right]
= &\;\sum_{s_{1}[1]= 0}^{S_0}\ldots\sum_{s_{1}[j] = 0}^{S_0}\ldots\sum_{s_{K}[1]= 0}^{S_0}\ldots\sum_{s_{K}[j] = 0}^{S_0} \nonumber\\
&\times\textrm{Pr}\Big(S_{\ob}^{\ss{\R_1}}[1]=s_{1}[1],
\ldots,S_{\ob}^{\ss{\R_1}}[j]=s_{1}[j],\ldots,S_{\ob}^{\ss{\R_K}}[1]=s_{K}[1], \ldots,
S_{\ob}^{\ss{\R_K}}[j]=s_{K}[j]|W_{\ss\T}[j],\hat{\textbf{W}}_{\ss\FC}^{j-1}\Big)\nonumber\\
&\times\textrm{Pr}\Big({S}_{\ob}^{\ss\FC,\A}[j]=\tilde{s}[j]|S_{\ob}^{\ss{\R_1}}[1]=s_{1}[1],\ldots,S_{\ob}^{\ss{\R_1}}[j]=s_{1}[j],\ldots,S_{\ob}^{\ss{\R_K}}[1]=s_{K}[1], \ldots,S_{\ob}^{\ss{\R_K}}[j]=s_{K}[j] \Big),%
\end{align}}\hrulefill
\end{figure*}
\fi
where $S_{\ob}^{\ss{\R_k}}[i]$ and $s_{k}[i]$, $i\in\{1,\ldots,j\}$ and $k\in\{1,\ldots,K\}$, are defined in Section~\ref{subsec:System Model}.

{Theoretically, any number of molecules between 0 and $S_0$ can be observed at each RX. Thus, there is a large number of realizations for each Poisson RV $S_{\ob}^{\ss{\R_k}}[i]$ in \eqref{SA-ML}, which makes the complete evaluation of \eqref{SA-ML} cumbersome. To simplify the evaluation of \eqref{SA-ML}, we consider finitely many random realizations of each Poisson RV $S_{\ob}^{\ss{\R_k}}[i]$\footnote{We assume that the FC may have sufficiently high computational capabilities such that it can generate random realizations. This assumption is because the FC could have a direct interface to additional computational resources.}.} For example, we generate $5000$ random realizations of each $S_{\ob}^{\ss{\R_k}}[i]$ for a given $\hat{\textbf{W}}_{\ss\FC}^{j-1}$, which is sufficient to ensure the accuracy of \eqref{SA-ML}. It is shown that \eqref{SA-ML} can be evaluated by applying the conditional PMF of the Poisson RV ${S}_{\ob}^{\ss\FC,\A}[j]$. We obtain the conditional mean of ${S}_{\ob,{k}}^{\ss\FC,\A}[j]$ by replacing $S_{0}W_{\ss\T}[i]$, $P_{\ob}^{({\ss{\T},{\ss\R_k}})}$, $M_{\ss\RX}$, $m$, and $\Delta{t_{\ss\R}}$ in \eqref{observed molecular numbers R1} with $S_{k}^\AF[j]$, $P_{\ob,{k}}^{({\ss\R_k,\ss\FC})}$, $M_{\ss\FC}$, $\tilde{m}$, and $\Delta{t_{\ss\FC}}$, respectively. Based on ${\bar{S}}_{\ob}^{{\ss\FC},\A}[j] = \sum^{K}_{k=1}{\bar{S}}_{\ob,{k}}^{{\ss\FC},\A}[j]$, we can then obtain the conditional mean of ${S}_{\ob}^{\ss\FC,\A}[j]$.

\ifOneCol
\vspace{-2mm}
\else
\fi
\subsection{Comparison of Complexity}
We summarize the complexity comparison in Table \ref{tab:variants}. MD-ML requires higher complexity than SD-ML. This is because each RX releases a unique type of molecule in MD-ML, whereas in SD-ML the RXs release a single type of molecule. SD-ML requires higher complexity than SA-ML. This is because the RXs need to decode the TX's symbols and the FC needs to estimate the RXs' decisions in SD-ML, but in SA-ML the RXs only need to amplify the received signal and the FC does not need to estimate the RXs' decisions. 

\ifOneCol
\vspace{-2mm}
\else
\fi
\section{Error Performance Analysis}\label{sec:error performance}
In this section, we derive the error probability of SD-ML and SA-ML using the \emph{genie-aided history}, which leads to tractable expressions. {Also, the error probability with genie-aided history provides a lower bound on that with local history.} We denote $Q_{\ss\FC}[j]$ as the error probability of the system in the $j$th symbol interval for a TX sequence $\textbf{W}_{\ss\T}^{j-1}$. The closed-form expressions of $Q_{\ss\FC}[j]$ for SD-ML with $K=1$ and SA-ML are mathematically tractable. 

To derive $Q_{\ss\FC}[j]$, we first derive equivalent decision rules with lower-complexity than \eqref{ML rule} and \eqref{ML rule,DF} for SD-ML and SA-ML in Theorems \ref{SD-ML,dec} and \ref{SA-ML,dec}, respectively. 
The decision rules when not all previously-transmitted symbols are ``0'' cannot be directly applied to the case where all previously-transmitted symbols are ``0''. Based on these theorems, the general forms of these lower-complexity decision rules are that the FC compares the observation with \emph{adaptive} thresholds when not all previously-transmitted symbols are ``0'' and the FC compares the observation with 0 when all previously-transmitted symbols are ``0''. Notably, these adaptive thresholds adapt to different ISI in different symbol intervals. 

\ifOneCol
\vspace{-2mm}
\else
\fi
\subsection{SD-ML}
We now derive $Q_{\ss\FC}[j]$ for the SD-ML variant.
To this end, we first define $\hat{\lambda}_\N^\D[j]$ as the expected ISI at the FC in the $j$th symbol interval due to the previous symbols transmitted by all RXs, $\hat{\textbf{W}}_{\ss\R_1}^{j-1},\hat{\textbf{W}}_{\ss\R_2}^{j-1},\ldots, \hat{\textbf{W}}_{\ss\R_K}^{j-1}$, i.e.,
\begin{align}\label{noise}
\hat{\lambda}_\N^\D[j] =\sum\limits^{K}_{k=1}S_{k}\sum\limits^{j-1}_{i=1}\hat{W}_{\ss\RX_k}[i]\sum\limits^{M_{\ss\FC}}_{\tilde{m}=1}P_{\ob,{k}}^{({\ss\R_k,\ss\FC})}\left(\left(j-i\right)T +\tilde{m}\Delta{t_{\ss\FC}}\right).
\end{align}
If not all previous symbols transmitted by all RXs are ``0''. i.e., $\sum\limits^{j-1}_{i=1}\sum\limits^{K}_{k=1}\hat{W}_{\ss\RX_k}[i] \neq 0$, we have $\hat{\lambda}_\N^\D[j]>0$; otherwise, we have $\hat{\lambda}_\N^\D[j]=0$. \textcolor{black}{We then define $\hat{\lambda}_{\s,h}^{\D,\Tot}[j]$ as the total number of signal molecules at the $\FC$ in the $j$th symbol interval due to the $h$th realization of currently-transmitted RX symbols $\hat{\mathcal{W}}^{{\ss\RX}}_{j,h}$, i.e.,
\begin{align}\label{signal}
\hat{\lambda}_{\s,h}^{\D,\Tot}[j] = \sum\limits^{K}_{k=1}S_{k}\hat{W}_{\ss\RX_k}[j]\sum\limits^{M_{\ss\FC}}_{\tilde{m}=1}P_{\ob,{k}}^{({\ss\R_k,\ss\FC})}\left(\tilde{m}\Delta{t_{\ss\FC}}\right).
\end{align}}
For the sake of brevity, for SD-ML, we define $\mathcal{L}\left[j|{W}_{\ss\TX}[j]=1,\textbf{W}_{\ss\TX}^{j-1},\hat{\textbf{W}}_{\ss\R_k}^{j-1}\right]\triangleq\mathcal{L}_1^\SD\left[j\right]$ and $\mathcal{L}\left[j|W_{\ss\TX}[j]=0,\textbf{W}_{\ss\TX}^{j-1},\hat{\textbf{W}}_{\ss\R_k}^{j-1}\right]\triangleq\mathcal{L}_0^\SD\left[j\right]$.
Applying the conditional PMF of ${S}_{\ob}^{\ss\FC,\D}[j]$ to \eqref{SD-ML}, \textcolor{black}{we write $\mathcal{L}_b^\SD\left[j\right]$ as
\ifOneCol	
\begin{align}\label{SD-ML,L,asy}
\mathcal{L}_b^\SD\left[j\right]=\!\sum_{h=1}^{2^K}\!\bigg[\textrm{Pr}\left(\hat{\mathcal{W}}^{{\ss\RX}}_{j,h} |W_{\ss\T}[j]=b,\textbf{W}_{\ss\TX}^{j-1}\right)\!\exp\left(-\hat{\lambda}_\N^\D[j]-\hat{\lambda}_{\s,h}^{\D,\Tot}[j]\right)\frac{\left(\hat{\lambda}_\N^\D[j]+\hat{\lambda}_{\s,h}^{\D,\Tot}[j]\right)^{\tilde{s}[j]}}{\left(\tilde{s}[j]!\right)}\bigg],
\end{align}
\else
\begin{align}\label{SD-ML,L,asy}
\mathcal{L}_b^\SD\left[j\right]=&\;\sum_{h=1}^{2^K}\bigg[\textrm{Pr}\left(\hat{\mathcal{W}}^{{\ss\RX}}_{j,h} |W_{\ss\T}[j]=b,\textbf{W}_{\ss\TX}^{j-1}\right){\left(\tilde{s}[j]!\right)^{-1}}\nonumber\\
&\hspace{-5mm}\times\!\exp\!\left(-\hat{\lambda}_\N^\D[j]-\hat{\lambda}_{\s,h}^{\D,\Tot}[j]\!\right)\left(\!\hat{\lambda}_\N^\D[j]+\hat{\lambda}_{\s,h}^{\D,\Tot}[j]\!\right)^{\tilde{s}[j]}\bigg],
\end{align}
\fi
where $b\in\{0,1\}$.} Based on \eqref{SD-ML,L,asy}, we rederive the decision rule of SD-ML in \eqref{ML rule,DF} as a lower-complexity decision rule in the following theorem.
\begin{theorem}\label{SD-ML,dec}
When $\hat{\lambda}_\N^\D[j]>0$, the decision rule of SD-ML is 
\begin{align}\label{dec 1}
\hat{W}_{\ss\FC}[j]=
\begin{cases}
1,&\mbox{if $\tilde{s}[j]\geq\xi_{\ss\FC}^{\ad,\SD}[j]$,}\\
0,&\mbox{otherwise},
\end{cases}
\end{align}
where $\xi_{\ss\FC}^{\ad,\SD}[j]$ is the solution to $\mathcal{L}_1^\SD\left[j\right]=\mathcal{L}_0^\SD\left[j\right]$ in terms of $\tilde{s}[j]$. We note that $\mathcal{L}_1^\SD\left[j\right]=\mathcal{L}_0^\SD\left[j\right]$ has a solution only when $\hat{\lambda}_\N^\D[j]>0$. When $\hat{\lambda}_\N^\D[j]=0$, the decision rule for SD-ML is 
\begin{align}\label{dec 0}
\hat{W}_{\ss\FC}[j]=
\begin{cases}
1,&\mbox{if $\tilde{s}[j]>0$,}\\
0,&\mbox{$\tilde{s}[j]=0$}.
\end{cases}
\end{align}
\end{theorem}
\begin{IEEEproof}
\textcolor{black}{Please see Appendix \ref{app}.}
\end{IEEEproof}

Based on Theorem \ref{SD-ML,dec}, when $\hat{\lambda}_\N^\D[j]>0$, we evaluate the conditional $Q_{\ss\FC}$ as
\ifOneCol	
\begin{align}\label{Pe,noisy,SD,>0}
Q_{\ss\FC}\left[j|\hat{\lambda}_\N^\D[j]>0\right]=&\;
P_1\textrm{Pr}\left({S}_{\ob}^{\ss\FC,\D}[j] < \xi_{\ss\FC}^{\ad,\SD}[j]|W_{\ss\T}[j]=1, \hat{\lambda}_\N^\D[j]> 0\right)\nonumber\\
&+(1-P_1)\textrm{Pr}\left({S}_{\ob}^{\ss\FC,\D}[j] \geq \xi_{\ss\FC}^{\ad,\SD}[j]|W_{\ss\T}[j]=0, \hat{\lambda}_\N^\D[j]> 0\right),
\end{align}
\else
\begin{align}\label{Pe,noisy,SD,>0}
&Q_{\ss\FC}\left[j|\hat{\lambda}_\N^\D[j]>0\right]\nonumber\\
&=P_1\textrm{Pr}\left({S}_{\ob}^{\ss\FC,\D}[j] < \xi_{\ss\FC}^{\ad,\SD}[j]|W_{\ss\T}[j]=1, \hat{\lambda}_\N^\D[j]> 0\right)\nonumber\\
&\!+\!(1-P_1)\textrm{Pr}\!\left(\!{S}_{\ob}^{\ss\FC,\D}[j] \!\geq\! \xi_{\ss\FC}^{\ad,\SD}[j]|W_{\ss\T}[j]\!=0, \hat{\lambda}_\N^\D[j]\!>\! 0\right),
\end{align}
\fi
where the conditional CDF of the Poisson RV ${S}_{\ob}^{\ss\FC,\D}[j]$ can be evaluated by \textcolor{black}{
\ifOneCol	
\begin{align}\label{SD-ML  CDF}
\textrm{Pr}\left({S}_{\ob}^{\ss\FC,\D}[j] < \xi_{\ss\FC}^{\ad,\SD}[j]|W_{\ss\T}[j], \hat{\lambda}_\N^\D[j]\right)
=&\;\sum_{h=1}^{2^K}\Bigg[\textrm{Pr}\left(\hat{\mathcal{W}}^{{\ss\RX}}_{j,h} |W_{\ss\T}[j],\textbf{W}_{\ss\TX}^{j-1}\right)\exp\left(-\hat{\lambda}_\N^\D[j]-\hat{\lambda}_{\s,h}^{\D,\Tot}[j]\right)\nonumber\\
&\times\sum\limits^{\xi_{\ss\FC}^{\ad,\SD}[j]-1}_{\eta=0}\left(\hat{\lambda}_\N^\D[j]+ \hat{\lambda}_{\s,h}^{\D,\Tot}[j]\right)^{\eta}/{\left(\eta!\right)}\Bigg],
\end{align}
\else
\begin{align}\label{SD-ML  CDF}
&\textrm{Pr}\left({S}_{\ob}^{\ss\FC,\D}[j] < \xi_{\ss\FC}^{\ad,\SD}[j]|W_{\ss\T}[j], \hat{\lambda}_\N^\D[j]\right)\nonumber\\
&=\sum_{h=1}^{2^K}\Bigg[\textrm{Pr}\left(\hat{\mathcal{W}}^{{\ss\RX}}_{j,h} |W_{\ss\T}[j],\textbf{W}_{\ss\TX}^{j-1}\right)\exp\left(-\hat{\lambda}_\N^\D[j]-\hat{\lambda}_{\s,h}^{\D,\Tot}[j]\right)\nonumber\\
&\times\sum\limits^{\xi_{\ss\FC}^{\ad,\SD}[j]-1}_{\eta=0}\left(\hat{\lambda}_\N^\D[j]+ \hat{\lambda}_{\s,h}^{\D,\Tot}[j]\right)^{\eta}/{\left(\eta!\right)}\Bigg],
\end{align}
\fi}
where $\hat{\lambda}_\N^\D[j]$ can be evaluated by \eqref{noise} via the approximated $\hat{\textbf{W}}_{\ss\R_k}^{j-1}$, $k\in\{1,2,\ldots,K\}$. The approximated $\hat{\textbf{W}}_{\ss\R_k}^{j-1}$ can be obtained using the biased coin toss method introduced in \cite{Ahmadzadeh2015}. Specifically, we model the $i$th decision at $\RX_k$, $\hat{W}_{\ss\R_k}[i]$, as $\hat{W}_{\ss\R_k}[i] = |\lambda-W_{\ss\T}[i]|$, where $i\in\{1,\ldots,j-1\}$ and $\lambda\in\{0,1\}$ is the outcome of the coin toss with $\textrm{Pr}(\lambda=1)=\textrm{Pr}\left(\hat{W}_{\ss\RX_k}[i]=0|W_{\ss\T}[i]=1\right)$ if $W_{\ss\T}[i]=1$ and $\textrm{Pr}(\lambda=1)=\textrm{Pr}\left(\hat{W}_{\ss\RX_k}[i]=1|W_{\ss\T}[i]=0\right)$ if $W_{\ss\T}[i]=0$. When $\hat{\lambda}_\N^\D[j] = 0$, we evaluate the conditional $Q_{\ss\FC}$ as
\ifOneCol	
\begin{align}\label{Pe,noisy,SD,=0}
Q_{\ss\FC}\left[j|\hat{\lambda}_\N^\D[j]=0\right]=&P_1\textrm{Pr}\left({S}_{\ob}^{\ss\FC,\D}[j] = 0|W_{\ss\T}[j]=1, \hat{\lambda}_\N^\D[j]= 0\right)\nonumber\\
&+(1-P_1)\textrm{Pr}\left({S}_{\ob}^{\ss\FC,\D}[j] > 0|W_{\ss\T}[j]=0, \hat{\lambda}_\N^\D[j]= 0\right),
\end{align}
\else
\begin{align}\label{Pe,noisy,SD,=0}
&Q_{\ss\FC}\left[j|\hat{\lambda}_\N^\D[j]=0\right]\nonumber\\
&=P_1\textrm{Pr}\left({S}_{\ob}^{\ss\FC,\D}[j] = 0|W_{\ss\T}[j]=1, \hat{\lambda}_\N^\D[j]= 0\right)\nonumber\\
&+(1-P_1)\textrm{Pr}\left({S}_{\ob}^{\ss\FC,\D}[j] > 0|W_{\ss\T}[j]=0, \hat{\lambda}_\N^\D[j]= 0\right),
\end{align}
\fi
where the conditional CDF of the Poisson RV ${S}_{\ob}^{\ss\FC,\D}[j]$ can be evaluated analogously to \eqref{SD-ML  CDF}. Combining \eqref{Pe,noisy,SD,=0} and \eqref{Pe,noisy,SD,>0}, we obtain $Q_{\ss\FC}[j]$ for SD-ML as
\ifOneCol	
\begin{align}\label{Pe,noisy,DF}
Q_{\ss\FC}[j] = &\textrm{Pr}\left(\hat{\lambda}_\N^\D[j]>0|\textbf{W}_{\ss\T}^{j-1}\right)Q_{\ss\FC}\!\left[j|\hat{\lambda}_\N^\D[j]>0\right]
\!+\!\textrm{Pr}\left(\hat{\lambda}_\N^\D[j]= 0|\textbf{W}_{\ss\T}^{j-1}\right)Q_{\ss\FC}\!\left[j|\hat{\lambda}_\N^\D[j]=0\right].
\end{align}
\else
\begin{align}\label{Pe,noisy,DF}
Q_{\ss\FC}[j] = &\;\textrm{Pr}\left(\hat{\lambda}_\N^\D[j]>0|\textbf{W}_{\ss\T}^{j-1}\right)Q_{\ss\FC}\left[j|\hat{\lambda}_\N^\D[j]>0\right]\nonumber\\
&+\textrm{Pr}\left(\hat{\lambda}_\N^\D[j]= 0|\textbf{W}_{\ss\T}^{j-1}\right)Q_{\ss\FC}\left[j|\hat{\lambda}_\N^\D[j]=0\right].
\end{align}
\fi

Finally, we derive the closed-form expression for $Q_{\ss\FC}[j]$ for SD-ML with $K=1$. To this end, we first rewrite $\mathcal{L}_1^\SD\left[j\right]$ and $\mathcal{L}_0^\SD\left[j\right]$ using \eqref{SD-ML,L,asy} with $K=1$.
We then solve $\mathcal{L}_1^\SD\left[j\right]=\mathcal{L}_0^\SD\left[j\right]$ in terms of $\xi_{\ss\FC}^{\ad,\SD}[j]$ and obtain the closed-form expression for $\xi_{\ss\FC}^{\ad,\SD}[j]$ when $K=1$ as
${\xi}_{\ss\FC}^{\ad,\SD}[j] =
\left\lfloor{\hat{\lambda}_{\s}^\D[j]}/{\log\left({\hat{\lambda}_\N^\D[j]+\hat{\lambda}_{\s}^\D[j]}/{\hat{\lambda}_\N^\D[j]}\right)}\right\rceil$.
We note that $Q_{\ss\FC}[j]$ for SD-ML with $K=1$ can be obtained using \eqref{Pe,noisy,DF} via this expression.

\ifOneCol
\vspace{-2mm}
\else
\fi
\subsection{SA-ML}\label{subsubsec:SA-error}
We now derive $Q_{\ss\FC}[j]$ for SA-ML. {In \eqref{SA-ML}, multiple possible realizations of each Poisson RV $S_{\ob}^{\ss{\R_k}}[i]$ make the analytical error performance analysis cumbersome. To facilitate the error performance analysis, we consider only one random realization of $S_{\ob}^{\ss{\R_k}}[i]$ with the mean $\bar{S}_{\ob}^{\ss{\R_k}}[i]$ for the \emph{given} previous symbols transmitted by the TX, $\textbf{W}_{\ss\T}^{j-1}$.} We define $\hat{\lambda}_{\N}^\A[j]$ as the expected ISI at the FC in the $j$th symbol interval due to $\textbf{W}_{\ss\T}^{j-1}$. We define $\hat{\lambda}_{\s}^\A[j]$ as the number of the signal molecules at the FC in the $j$th symbol interval due to $W_{\ss\T}[j]=1$. By modeling the realization of $S_{\ob}^{\ss{\R_k}}[i]$ as its mean $\bar{S}_{\ob}^{\ss{\R_k}}[i]$, we write $\hat{\lambda}_{\N}^\A[j]$ and $\hat{\lambda}_{\s}^\A[j]$ as
\ifOneCol	
\begin{align}\label{noise, SA}
\hat{\lambda}_{\N}^\A[j] = &\;\sum\limits^{K}_{k=1}\Bigg(\sum\limits^{j-1}_{i=1}\alpha_k \bar{S}_{\ob}^{\ss{\R_k}}[i]\sum\limits^{M_{\ss\FC}}_{\tilde{m}=1}P_{\ob,{k}}^{({\ss\R_k,\ss\FC})}\left(\left(j-i\right)T + \tilde{m}\Delta{t_{\ss\FC}}\right)\nonumber\\
&+\alpha_k S_0 \sum\limits^{j-1}_{i=1}W_{\ss\T}[i]\sum\limits^{M_{\ss\RX}}_{m=1}P_{\ob}^{({\ss{\TX},{\RX}})}((j-i)T + m\Delta{t_{\ss\RX}})
\sum\limits^{M_{\ss\FC}}_{\tilde{m}=1}P_{\ob,{k}}^{({\ss\R_k,\ss\FC})}\left(\tilde{m}\Delta{t_{\ss\FC}}\right)\Bigg)
\end{align}
\else
\begin{align}\label{noise, SA}
\hat{\lambda}_{\N}^\A[j] = &\;\sum\limits^{K}_{k=1}\Bigg(\sum\limits^{j-1}_{i=1}\alpha_k \bar{S}_{\ob}^{\ss{\R_k}}[i]\sum\limits^{M_{\ss\FC}}_{\tilde{m}=1}P_{\ob,{k}}^{({\ss\R_k,\ss\FC})}\left(\left(j-i\right)T + \tilde{m}\Delta{t_{\ss\FC}}\right)\nonumber\\
&+\alpha_k S_0 \sum\limits^{j-1}_{i=1}W_{\ss\T}[i]\sum\limits^{M_{\ss\RX}}_{m=1}P_{\ob}^{({\ss{\TX},{\RX}})}((j-i)T + m\Delta{t_{\ss\RX}})\nonumber\\
&\times\sum\limits^{M_{\ss\FC}}_{\tilde{m}=1}P_{\ob,{k}}^{({\ss\R_k,\ss\FC})}\left(\tilde{m}\Delta{t_{\ss\FC}}\right)\Bigg)
\end{align}
\fi
and
\begin{align}\label{signal, SA}
\hat{\lambda}_{\s}^\A[j] = &\;\sum\limits^{K}_{k=1}\alpha_k S_0 \sum\limits^{M_{\ss\RX}}_{m=1}P_{\ob}^{({\ss{\TX},{\RX}})}(m\Delta{t_{\ss\RX}})
\sum\limits^{M_{\ss\FC}}_{\tilde{m}=1}P_{\ob,{k}}^{({\ss\R_k,\ss\FC})}\left(\tilde{m}\Delta{t_{\ss\FC}}\right),
\end{align}
respectively. 
$\hat{\lambda}_{\N}^\A[j]$ in \eqref{noise, SA} consists of two components. The first summation over $i$ is the expected ISI at the FC in the $j$th symbol interval due to the molecules released by the RXs but without the amplification of the RXs' ISI from the TX. The second summation over $i$ accounts for the amplification of the ISI in the $j$th symbol interval at all RXs due to $\textbf{W}_{\ss\T}^{j-1}$. We note that the conditional mean of ${S}_{\ob}^{\ss\FC,\A}[j]$ is $\hat{\lambda}_{\s}^\A[j]+\hat{\lambda}_{\N}^\A[j]$ when $W_{\ss\TX}[j]=1$, and the conditional mean of  ${S}_{\ob}^{\ss\FC,\A}[j]$ is $\hat{\lambda}_{\N}^\A[j]$ when $W_{\ss\TX}[j]=0$. If not all previous symbols transmitted by the TX are ``0''. i.e., $\textbf{W}_{\ss\T}^{j-1} \neq \textbf{0}$, then we have $\hat{\lambda}_{\N}^\A[j]>0$. If all previous symbols transmitted by the TX are ``0'', i.e., $\textbf{W}_{\ss\T}^{j-1} = \textbf{0}$, then we have $\hat{\lambda}_{\N}^\A[j]=0$. For the sake of brevity, for SA-ML, we define $\mathcal{L}\left[j|W_{\ss\T}[j]=1,\textbf{W}_{\ss\T}^{j-1}\right]\triangleq\mathcal{L}_1^{\SA}$ and $\mathcal{L}\left[j|W_{\ss\T}[j]=0,\textbf{W}_{\ss\T}^{j-1}\right]\triangleq\mathcal{L}_0^{\SA}$. Applying the conditional PMF of the Poisson RV ${S}_{\ob}^{\ss\FC,\A}[j]$ to \eqref{SA-ML}, we derive $\mathcal{L}_1^{\SA}$ and $\mathcal{L}_0^{\SA}$ as
\ifOneCol
\begin{align}\label{V2,L1,SA}
\mathcal{L}_1^{\SA}=&\;\frac{\left(\hat{\lambda}_{\s}^\A[j]+\hat{\lambda}_{\N}^\A[j]\right)^{\tilde{s}[j]}
\exp\left(-\left(\hat{\lambda}_{\s}^\A[j]+\hat{\lambda}_{\N}^\A[j]\right)\right)}{\tilde{s}[j]!}
\end{align}
\else
\begin{align}\label{V2,L1,SA}
\mathcal{L}_1^{\SA}=&\;\frac{\left(\hat{\lambda}_{\s}^\A[j]+\hat{\lambda}_{\N}^\A[j]\right)^{\tilde{s}[j]}
\exp\left(-\left(\hat{\lambda}_{\s}^\A[j]+\hat{\lambda}_{\N}^\A[j]\right)\right)}{\tilde{s}[j]!}
\end{align}
\fi
and
\ifOneCol
\begin{align}\label{V2,L0,SA}
\mathcal{L}_0^{\SA}=&\;\frac{{\left(\hat{\lambda}_{\N}^\A[j]\right)^{\tilde{s}[j]}}\exp\left(-\hat{\lambda}_{\N}^\A[j]\right)}{\left({\tilde{s}[j]!}\right)},
\end{align}
\else
\begin{align}\label{V2,L0,SA}
\mathcal{L}_0^{\SA}=&\;\frac{{\left(\hat{\lambda}_{\N}^\A[j]\right)^{\tilde{s}[j]}}\exp\left(-\hat{\lambda}_{\N}^\A[j]\right)}{\left({\tilde{s}[j]!}\right)},
\end{align}
\fi
respectively. Based on \eqref{V2,L1,SA} and \eqref{V2,L0,SA}, we rewrite the general decision rule of SA-ML in \eqref{ML rule} as a lower-complexity decision rule in the following theorem.
\begin{theorem}\label{SA-ML,dec}
When $\hat{\lambda}_\N^\A[j]>0$, the decision rule of SA-ML is 
\begin{align}\label{dec 1, SA}
\hat{W}_{\ss\FC}[j]=
\begin{cases}
1,&\mbox{if $\tilde{s}[j]\geq\xi_{\ss\FC}^{\ad,\SA}[j]$,}\\
0,&\mbox{otherwise},
\end{cases}
\end{align}
where $\xi_{\ss\FC}^{\ad,\SA}[j]=
\left\lfloor{\hat{\lambda}_\s^\A[j]}/{\log\left({\hat{\lambda}_{\N}^\A[j]+\hat{\lambda}_\s^\A[j]}/{\hat{\lambda}_{\N}^\A[j]}\right)}\right\rceil$. 
When $\hat{\lambda}_{\N}^\A[j]=0$, the decision rule is 
\begin{align}\label{dec 0, SA}
\hat{W}_{\ss\FC}[j]=
\begin{cases}
1,&\mbox{if $\tilde{s}[j]>0$,}\\
0,&\mbox{$\tilde{s}[j]=0$}.
\end{cases}
\end{align}
\end{theorem}
\begin{IEEEproof}
Applying \eqref{V2,L1,SA} and \eqref{V2,L0,SA} to \eqref{ML rule}, we rewrite the decision rule for SA-ML as
\ifOneCol
\begin{align}\label{variant 2 ML1}
{\left(\hat{\lambda}_{\s}^\A[j]+\hat{\lambda}_{\N}^\A[j]\right)^{\tilde{s}[j]}}
\exp\left(-\left(\hat{\lambda}_{\s}^\A[j]+\hat{\lambda}_{\N}^\A[j]\right)\right)
\overset{\hat{W}_{\ss\FC}[j]=1}{\underset{\hat{W}_{\ss\FC}[j]=0} \gtreqless\;}
{\left(\hat{\lambda}_{\N}^\A[j]\right)^{\tilde{s}[j]}}\exp\left(-\hat{\lambda}_{\N}^\A[j])\right).
\end{align}
\else
\begin{align}\label{variant 2 ML1}
&{\left(\hat{\lambda}_{\s}^\A[j]+\hat{\lambda}_{\N}^\A[j]\right)^{\tilde{s}[j]}}
\exp\left(-\left(\hat{\lambda}_{\s}^\A[j]+\hat{\lambda}_{\N}^\A[j]\right)\right)\nonumber\\
&\overset{\hat{W}_{\ss\FC}[j]=1}{\underset{\hat{W}_{\ss\FC}[j]=0} \gtreqless\;}
{\left(\hat{\lambda}_{\N}^\A[j]\right)^{\tilde{s}[j]}}\exp\left(-\hat{\lambda}_{\N}^\A[j])\right).
\end{align}
\fi
We then discuss the cases when $\textbf{W}_{\ss\T}^{j-1}=\mathbf{0}$ and $\textbf{W}_{\ss\T}^{j-1}\neq\mathbf{0}$. When $\textbf{W}_{\ss\T}^{j-1} \neq \textbf{0}$, then we have $\hat{\lambda}_{\N}^\A[j]>0$ and we rewrite \eqref{variant 2 ML1}  as
\ifOneCol
\begin{align}\label{variant 2 ML3}
\left(\frac{\hat{\lambda}_{\s}^\A[j]}{\hat{\lambda}_{\N}^\A[j]}+1\right)^{\tilde{s}[j]}\overset{\hat{W}_{\ss\FC}[j]=1}{\underset{\hat{W}_{\ss\FC}[j]=0} \gtreqless\;}\exp\left(\hat{\lambda}_{\s}^\A[j]\right).
\end{align}
\else
\begin{align}\label{variant 2 ML3}
\left({\hat{\lambda}_{\s}^\A[j]}/{\hat{\lambda}_{\N}^\A[j]}+1\right)^{\tilde{s}[j]}\overset{\hat{W}_{\ss\FC}[j]=1}{\underset{\hat{W}_{\ss\FC}[j]=0} \gtreqless\;}\exp\left(\hat{\lambda}_{\s}^\A[j]\right).
\end{align}
\fi
We rearrange \eqref{variant 2 ML3} and obtain \eqref{dec 1, SA}. We next discuss the case $\textbf{W}_{\ss\T}^{j-1}=\mathbf{0}$, which leads to $\hat{\lambda}_{\N}^\A[j]=0$. If $\hat{\lambda}_{\N}^\A[j]=0$ and $\tilde{s}[j]=0$, we write \eqref{variant 2 ML1} as
\begin{align}\label{variant 2 ML5}
\exp\left(-\hat{\lambda}_{\s}^\A[j]\right)\overset{\hat{W}_{\ss\FC}[j]=1}{\underset{\hat{W}_{\ss\FC}[j]=0} \gtreqless\;}1,
\end{align}
where the decision at the FC is always $\hat{W}_{\ss\FC}[j]=0$ since $<$ always holds. If $\hat{\lambda}_{\N}^\A[j]=0$ and any $\tilde{s}[j]>0$, we write \eqref{variant 2 ML1} as
\begin{align}\label{variant 2 ML6}
{\left(\hat{\lambda}_{\s}^\A[j]\right)^{\tilde{s}[j]}}\exp\left(-(\hat{\lambda}_{\s}^\A[j])\right)/{\tilde{s}[j]!}\overset{\hat{W}_{\ss\FC}[j]=1}{\underset{\hat{W}_{\ss\FC}[j]=0} \gtreqless\;}0,
\end{align}
where the decision at the FC is always $\hat{W}_{\ss\FC}[j]=1$ since $>$ always holds. Thus, we obtain the decision rule in \eqref{dec 0, SA}.
\end{IEEEproof}
Based on Theorem \ref{SA-ML,dec}, when $\textbf{W}_{\ss\T}^{j-1} \neq \textbf{0}$, we evaluate $Q_{\ss\FC}[j]$ for SA-ML as
\begin{align}\label{perfect probability1,SA}
Q_{\ss\FC}[j] = &\;\left(1-P_{1}\right)\textrm{Pr}\left({S}_{\ob}^{\ss\FC,\A}[j] \geq\xi_{\ss\FC}^{\ad,\SA}[j]|W_{\ss\T}[j]=0,\textbf{W}_{\ss\T}^{j-1}\right)\nonumber\\
&+P_1\textrm{Pr}\left({S}_{\ob}^{\ss\FC,\A}[j] <\xi_{\ss\FC}^{\ad,\SA}[j]|W_{\ss\T}[j]=1,\textbf{W}_{\ss\T}^{j-1}\right),
\end{align}
where $\textrm{Pr}\left({S}_{\ob}^{\ss\FC,\A}[j] <\xi_{\ss\FC}^{\ad,\SA}[j]|\textbf{W}_{\ss\T}^{j}\right)$ can be evaluated by replacing $\hat{\textbf{W}}_{\ss\FC}^{j-1}$ and ${S}_{\ob}^{\ss\FC,\A}[j]=\tilde{s}[j]$ in \eqref{SA-ML} with $\textbf{W}_{\ss\T}^{j-1}$ and ${S}_{\ob}^{\ss\FC,\A}[j]<\xi_{\ss\FC}^{\ad,\SA}[j]$, respectively. Similar to the evaluation of \eqref{SA-ML}, we consider finitely many random realizations of $S_{\ob}^{\ss{\R_k}}[i]$ in \eqref{perfect probability1,SA}. When $\textbf{W}_{\ss\T}^{j-1} = \textbf{0}$, $Q_{\ss\FC}[j]$ for SA-ML can be obtained by replacing $\geq$, $<$, and $\xi_{\ss\FC}^{\ad,\SA}[j]$ with $>$, $=$, and $0$ in \eqref{perfect probability1,SA}, respectively.

\ifOneCol
\vspace{-2mm}
\else
\fi
\section{\textcolor{black}{Error Performance Optimization}}\label{subsub:SD-MLopti}
\textcolor{black}{In this section, we determine the optimal molecule distribution among RXs that minimizes the error probability of SD-ML using the genie-aided history, inspired by the fact
the quantity of any type of molecule is usually constrained in practical biological environments. We also analytically prove that the equal allocation of molecules among two symmetric RXs achieves the local minimal error probability of SD-ML.}

\textcolor{black}{To this end, we first formulate the optimization problem as follows:
\begin{equation}\label{opti}
\begin{aligned}
& \underset{\mathbf{S}}{\text{min}}
& & Q_{\ss\FC}[j]~\text{in \eqref{Pe,noisy,DF}} \\
& \text{s.t.}
& & S_1+S_2+\cdots+S_K-N=0,\\
& & & S_k\geq0,\\
\end{aligned}
\end{equation}
where $\mathbf{S}=\{S_1,S_2,\ldots,S_K\}$, $k\in\{1,2,\ldots,K\}$, and $N$ is the total number of molecules released by $K$ RXs for symbol ``1''. Combining \eqref{Pe,noisy,SD,>0} and \eqref{Pe,noisy,DF}, we note that $\xi_{\ss\FC}^{\ad,\SD}[j]$ is required to evaluate $Q_{\ss\FC}[j]$. Based on Theorem \ref{SD-ML,dec},
the adaptive threshold $\xi_{\ss\FC}^{\ad,\SD}[j]$ is obtained by numerically solving $\mathcal{L}_1^\SD\left[j\right]=\mathcal{L}_0^\SD\left[j\right]$ in terms of $\tilde{s}[j]$, while the closed-form expression for $\xi_{\ss\FC}^{\ad,\SD}[j]$ is mathematically intractable. Therefore, there is no closed-form expression for $Q_{\ss\FC}[j]$, which makes it very hard to optimize $Q_{\ss\FC}[j]$ in \eqref{Pe,noisy,DF}. To tackle this challenge, we find a closed-form approximation for $Q_{\ss\FC}[j]$ in \eqref{Pe,noisy,DF} by considering a constant threshold $\xi$ in \eqref{Pe,noisy,SD,>0}. By doing so, we find the approximation of
$Q_{\ss\FC}[j]$ as
\ifOneCol
\begin{align}\label{Pe,noisy,SD,app}
Q^\sharp_{\ss\FC}[j]=&\;
P_1\sum_{h=1}^{2^K}\Big[\textrm{Pr}\left(\hat{\mathcal{W}}^{{\ss\RX}}_{j,h} |W_{\ss\T}[j]=1,\textbf{W}_{\ss\TX}^{j-1}\right)\Lambda\Big]\nonumber\\
&+(1-P_1)\sum_{h=1}^{2^K}\Big[\textrm{Pr}\left(\hat{\mathcal{W}}^{{\ss\RX}}_{j,h} |W_{\ss\T}[j]=0,\textbf{W}_{\ss\TX}^{j-1}\right)\left(1-\Lambda\right)\Big],
\end{align}	
\else
\begin{align}\label{Pe,noisy,SD,app}
Q^\sharp_{\ss\FC}[j]=&\;
P_1\sum_{h=1}^{2^K}\Big[\textrm{Pr}\left(\hat{\mathcal{W}}^{{\ss\RX}}_{j,h} |W_{\ss\T}[j]=1,\textbf{W}_{\ss\TX}^{j-1}\right)\Lambda\Big]\nonumber\\
&+(1-P_1)\sum_{h=1}^{2^K}\Big[\textrm{Pr}\left(\hat{\mathcal{W}}^{{\ss\RX}}_{j,h} |W_{\ss\T}[j]=0,\textbf{W}_{\ss\TX}^{j-1}\right)\nonumber\\
&\times\left(1-\Lambda\right)\Big],
\end{align}
\fi
where $Q^\sharp_{\ss\FC}[j]$ is the approximation of $Q_{\ss\FC}[j]$, $\Lambda$ is given by
\begin{align}\label{CDF,possion}
\Lambda=\sum\limits^{\xi-1}_{\eta=0}\exp\left(-\hat{\lambda}_\N^\D[j]-\hat{\lambda}_{\s,h}^{\D,\Tot}[j]\right)\frac{\left(\hat{\lambda}_\N^\D[j]+ \hat{\lambda}_{\s,h}^{\D,\Tot}[j]\right)^{\eta}}{{\left(\eta!\right)}},
\end{align}
and $\xi$ is a constant. In \eqref{CDF,possion}, $\hat{\lambda}_\N^\D[j]$ and $\hat{\lambda}_{\s,h}^{\D,\Tot}[j]$ are the functions of $\mathbf{S}$ based on \eqref{noise} and \eqref{signal}.
\begin{lemma}\label{tight}
The approximation of $Q_{\ss\FC}[j]$ by $Q^\sharp_{\ss\FC}[j]$ is tight when $\xi=\xi_{\ss\FC}^{\ad,\SD}[j]$.
\end{lemma}
\begin{IEEEproof}
We note that the likelihood of the occurrence that all previous symbols transmitted by all RXs are ``0'' is very small. Thus, we approximate $\textrm{Pr}\left(\hat{\lambda}_\N^\D[j]= 0|\textbf{W}_{\ss\T}^{j-1}\right)\approx0$ and $\textrm{Pr}\left(\hat{\lambda}_\N^\D[j]>0|\textbf{W}_{\ss\T}^{j-1}\right)\approx1$. Using these approximations in \eqref{Pe,noisy,DF}, we obtain $Q_{\ss\FC}[j]\approx Q_{\ss\FC}\left[j|\hat{\lambda}_\N^\D[j]>0\right]$. We then note that $Q^\sharp_{\ss\FC}[j]|_{\xi=\xi_{\ss\FC}^{\ad,\SD}[j]}=Q_{\ss\FC}\left[j|\hat{\lambda}_\N^\D[j]>0\right]$. Thus, $Q_{\ss\FC}[j]$ is accurately approximated by $Q^\sharp_{\ss\FC}[j]$ when $\xi=\xi_{\ss\FC}^{\ad,\SD}[j]$.
\end{IEEEproof}
\begin{lemma}\label{opti-thres}
Since the adaptive threshold $\xi_{\ss\FC}^{\ad,\SD}[j]$ adapts to different ISI for different symbol intervals, $\xi_{\ss\FC}^{\ad,\SD}[j]$ is the optimal $\xi$ that minimizes $Q^\sharp_{\ss\FC}[j]$ if $P_1=\frac{1}{2}$, i.e., $\xi_{\ss\FC}^{\ad,\SD}[j]=\underset{\xi}{\text{argmin}}~ Q^\sharp_{\ss\FC}[j]$.
\end{lemma}
\begin{IEEEproof}
Please see Appendix \ref{app2}.
\end{IEEEproof}
Based on Lemma \ref{tight} and Lemma \ref{opti-thres}, the approximation of $Q_{\ss\FC}[j]$ by $Q^\sharp_{\ss\FC}[j]$ is tight when $\xi=\xi_{\ss\FC}^{\ad,\SD}[j]$ and $\xi_{\ss\FC}^{\ad,\SD}[j]$ is the optimal $\xi$ which minimizes $Q^\sharp_{\ss\FC}[j]$. Therefore, the optimal $\mathbf{S}$ that minimizes $Q_{\ss\FC}[j]$ in \eqref{Pe,noisy,DF} can be obtained by finding the jointly optimal $\mathbf{S}$ and $\xi$ to minimize $Q^\sharp_{\ss\FC}[j]$ in \eqref{Pe,noisy,SD,app}, i.e., the approximate solution to the problem \eqref{opti} can be obtained by solving the optimization problem given by:
\begin{equation}\label{opti,app}
\begin{aligned}
& \underset{\mathbf{S},\;\xi}{\text{min}}
& & Q^\sharp_{\ss\FC}[j] \\
& \text{s.t.}
& & S_1+S_2+\cdots+S_K-N=0,\\
& & & S_k\geq0.\\
\end{aligned}
\end{equation}}

\textcolor{black}{To solve \eqref{opti,app}, we examine its convexity. The convexity of an optimization problem can be proven by showing that its objective function and constraints are convex with respect to the optimization variables. Since the constraints in \eqref{opti,app} are affine, they are convex. The convexity of the objective function, i.e., $Q^\sharp_{\ss\FC}[j]$, can be proven by showing that its Hessian is positive semidefinite with respect to its optimization variables. For the convexity of $Q^\sharp_{\ss\FC}[j]$, we have the following proposition:
\begin{proposition}\label{convexity}
The Hessian of $Q^\sharp_{\ss\FC}[j]$ is not positive semidefinite with respect to $\mathbf{S}$ and $\xi$.
\end{proposition}
\begin{IEEEproof}
Please see Appendix \ref{app3}.
\end{IEEEproof}
Based on Proposition \ref{convexity}, the multi-dimensional optimization problem \eqref{opti,app} is not a convex optimization problem. To overcome this challenge, we use GlobalSearch in MATLAB to repeatedly run a local solver with the sequential quadratic programming (SQP) algorithm until convergence is achieved (i.e., the global minimum is found) to solve the problem  \eqref{opti,app}. Our numerical results in Section \ref{sec:Numerical} confirm the effectiveness of this optimization method.}

\textcolor{black}{To obtain additional analytical insights in molecule distribution, we discuss the optimal distribution of the number of molecules in a symmetric topology. Intuitively, we expect that an equal distribution of molecules among symmetric RXs is the optimal allocation to minimize the error probability. To confirm this conjecture, we first find that the equal distribution locally minimizes $Q^\sharp_{\ss\FC}[j]$ under certain conditions. We derive such conditions in the following Lemma:
\begin{lemma}\label{molDistri}
In the symmetric topology with $K=2$, if $\Upsilon(\xi)>0$, $Q^\sharp_{\ss\FC}[j]$ achieves a local minimum when $S_1=\frac{N}{2}$; otherwise, it achieves a local maximum, where $\Upsilon(\xi)$ is given by
\begin{align}\label{cond}
\Upsilon (\xi)= \left(\alpha(P_1-1)+\beta P_1\right)\left(2+N(\nu+2\sigma)-2\lceil\xi\rceil\right),
\end{align}
where
\ifOneCol	
\begin{align}\label{cond2}
&\sigma_1=\sigma_2=\sigma,~\nu_1=\nu_2=\nu,~\alpha(1,0)=\alpha(0,1)=\alpha,~\text{and}~\beta(1,0)=\beta(0,1)=\beta,
\end{align}
\else
\begin{align}\label{cond2}
&\sigma_1=\sigma_2=\sigma,~\nu_1=\nu_2=\nu,\nonumber\\
&\alpha(1,0)=\alpha(0,1)=\alpha,~\text{and}~\beta(1,0)=\beta(0,1)=\beta,
\end{align}
\fi
\begin{align}\label{nu_n}
\sigma_k = \sum\limits^{j-1}_{i=1}\hat{W}_{\ss\RX_k}[i]\sum\limits^{M_{\ss\FC}}_{\tilde{m}=1}P_{\ob,{k}}^{({\ss\R_k,\ss\FC})}\left(\left(j-i\right)T +\tilde{m}\Delta{t_{\ss\FC}}\right),
\end{align}
\begin{align}\label{nu}
\nu_{k} = \sum\limits^{M_{\ss\FC}}_{\tilde{m}=1}P_{\ob,{k}}^{({\ss\R_k,\ss\FC})}\left(\tilde{m}\Delta{t_{\ss\FC}}\right),
\end{align}
\ifOneCol
\begin{align}\label{alpha}
\alpha(a_1,a_2)=&\;\textrm{Pr}\left(\hat{W}_{\ss\R_1}[j]=a_1|W_{\ss\T}[j]=0,\textbf{W}_{\ss\TX}^{j-1}\right)\textrm{Pr}\left(\hat{W}_{\ss\R_2}[j]=a_2|W_{\ss\T}[j]=0,\textbf{W}_{\ss\TX}^{j-1}\right),
\end{align}	
\else
\begin{align}\label{alpha}
\alpha(a_1,a_2)=&\;\textrm{Pr}\left(\hat{W}_{\ss\R_1}[j]=a_1|W_{\ss\T}[j]=0,\textbf{W}_{\ss\TX}^{j-1}\right)\nonumber\\
&\times\textrm{Pr}\left(\hat{W}_{\ss\R_2}[j]=a_2|W_{\ss\T}[j]=0,\textbf{W}_{\ss\TX}^{j-1}\right),
\end{align}
\fi
and\footnote{\textcolor{black}{In the symmetric topology, $\sigma_1=\sigma_2$ is valid because the observations at symmetric RXs are independently and identically distributed (even though symmetric RXs may not necessarily make the same decisions). We need to consider all possible realizations of $\hat{\textbf{W}}_{\ss\RX_k}^{j-1}$ at each $\RX_k$ to evaluate $Q^\sharp_{\ss\FC}[j]$, but this requires high complexity. To facilitate the calculation, we only consider one realization of $\hat{\textbf{W}}_{\ss\RX_k}^{j-1}$ at each RX and it is sufficiently accurate for the evaluation of $Q^\sharp_{\ss\FC}[j]$ to assume that this realization is the same for all RXs.}}
\ifOneCol
\begin{align}\label{beta}
\beta(a_1,a_2)=&\;\textrm{Pr}\left(\hat{W}_{\ss\R_1}[j]=a_1|W_{\ss\T}[j]=1,\textbf{W}_{\ss\TX}^{j-1}\right)\textrm{Pr}\left(\hat{W}_{\ss\R_2}[j]=a_2|W_{\ss\T}[j]=1,\textbf{W}_{\ss\TX}^{j-1}\right).
\end{align}	
\else
\begin{align}\label{beta}
\beta(a_1,a_2)=&\;\textrm{Pr}\left(\hat{W}_{\ss\R_1}[j]=a_1|W_{\ss\T}[j]=1,\textbf{W}_{\ss\TX}^{j-1}\right)\nonumber\\
&\times\textrm{Pr}\left(\hat{W}_{\ss\R_2}[j]=a_2|W_{\ss\T}[j]=1,\textbf{W}_{\ss\TX}^{j-1}\right).
\end{align}
\fi
\end{lemma}
\begin{IEEEproof}
Please see Appendix \ref{app4}.
\end{IEEEproof}
Using Lemma \ref{tight}, Lemma \ref{opti-thres}, and Lemma \ref{molDistri}, we find that the equal distribution of molecules always achieves the local minimal error probability for SD-ML in a two-RX system, as stated in the following theorem:
\begin{theorem}\label{molDistri,asy}
In the symmetric topology with two RXs, ${Q}_{\ss\FC}[j]$ achieves a local minimal value when $S_1=\frac{N}{2}$ if $P_1=\frac{1}{2}$.
\end{theorem}
\begin{IEEEproof}
Please see Appendix \ref{app5}.
\end{IEEEproof}}
\ifOneCol
\vspace{-4mm}
\else
\fi
\section{Numerical Results and Simulations}\label{sec:Numerical}
In this section, we present numerical and simulation results to examine the error performance of the ML detectors. We simulate using a particle-based method considered in \cite{Andrew2004}, where we track the precise locations of all individual molecules. 
Unless otherwise noted, we consider the environmental parameters in Table~\ref{tab:table1}.

\ifOneCol	
\begin{table}[!t]
\renewcommand{\arraystretch}{0.8}
\centering
\caption{Environmental Parameters}\label{tab:table1}\vspace{-3mm}
\begin{tabular}{c|c|c||c|c|c}
\hline
\bfseries Parameter &  \bfseries Symbol&  \bfseries Value & \bfseries Parameter &  \bfseries Symbol&  \bfseries Value \\
\hline\hline
Radius of each RX & $r_{\ss\R_k}$ & $0.2\,{\mu}\metre$ & Report time interval & $t_{\report}$ & $0.3\,{\m}\s$\\\hline
Radius of FC & $r_{\ss\FC}$ & $0.2\,{\mu}\metre$ & Bit interval time& $T$ & $1.3\,{\m}\s$ \\\hline
Time step at RXs & $\Delta{t_{\ss\R}}$ & $100\,{\mu}\s$ & Diffusion coefficient & $D_0=D_{k}$ & $5\times10^{-9}{\m^{2}}/{\s}$\\\hline
Time step at FC & $\Delta{t_{\ss\FC}}$ & $30\,{\mu}\s$ & Length of symbol sequence & $L$ & $20$\\\hline
Number of samples by RXs& $M_{\ss\RX}$ & 5 & Probability of binary 1 & $P_1$ & $0.5$\\\hline
Number of samples by FC& $M_{\ss\FC}$ & 10 & Transmission time interval & $t_{\trans}$ & $1\,{\m}\s$\\\hline
\end{tabular}
\vspace{-8mm}
\end{table}
\else
\begin{table}[!t]
\renewcommand{\arraystretch}{1}
\centering
\caption{Environmental Parameters}\label{tab:table1}\vspace{-1mm}
\begin{tabular}{c||c|c}
\hline
\bfseries Parameter &  \bfseries Symbol&  \bfseries Value \\
\hline\hline
Volume of each RX & $V_{\ss\R_k}$ & $\frac{4}{3}\times\pi\times0.2^3\,{\mu}\metre^3$\\\hline
Radius of FC & $r_{\ss\FC}$ & $0.2\,{\mu}\metre$ \\\hline
Time step at RXs & $\Delta{t_{\ss\R}}$ & $100\,{\mu}\s$\\\hline
Time step at FC & $\Delta{t_{\ss\FC}}$ & $30\,{\mu}\s$ \\\hline
Number of samples by RXs& $M_{\ss\RX}$ & 5 \\\hline
Number of samples by FC& $M_{\ss\FC}$ & 10 \\\hline
Transmission time interval & $t_{\trans}$ & $1\,{\m}\s$\\\hline
Report time interval & $t_{\report}$ & $0.3\,{\m}\s$\\\hline
Bit interval time& $T$ & $1.3\,{\m}\s$\\\hline
Diffusion coefficient & $D_0=D_{k}$ & $5\times10^{-9}{\m^{2}}/{\s}$\\\hline
Length of symbol sequence & $L$ & $20$ \\\hline
Probability of binary 1 & $P_1$ & $0.5$ \\\hline
\end{tabular}
\vspace{-3mm}
\end{table}
\fi

Throughout this section, we keep the TX and the FC fixed at $(0{\mu}\metre, 0{\mu}\metre, 0{\mu}\metre)$ and
\ifOneCol
\\$(2{\mu}\metre, 0{\mu}\metre, 0{\mu}\metre)$,	
\else
$(2{\mu}\metre, 0{\mu}\metre, 0{\mu}\metre)$,
\fi
respectively. \textcolor{black}{To clearly demonstrate the impact of the number of samples and the number of RXs on the error probability of the system, we consider a symmetric topology in Section \ref{Symmetric}. To clearly show the impact of asymmetric RX location on the error probability of the system and the corresponding optimal molecule distribution, we consider an asymmetric topology in Section \ref{Asymmetric}.}

We assume that the TX releases $10^4$ molecules for symbol ``1''. We also assume the total number of molecules released by all RXs for symbol ``1'' is \emph{fixed} at $2000$ throughout this section {to ensure the fairness of error performance comparison for different $K$}. For MD-ML and SD-ML, in Figs. \ref{fig:Pe_M_FC}-\ref{MD-SA-SD-dist}, each RX releases $S_\DF=\lfloor2000/K\rceil$ molecules to report a decision of ``1''. For SA-ML, in Figs. \ref{fig:Pe_M_FC}--\ref{MD-SA-SD-dist}, each RX uses an amplification factor to ensure that the average number of molecules released by all RXs for transmission of one symbol is $1000$ for the fair comparison among SA-ML, SD-ML, and MD-ML. $\overline{Q}_{\ss\FC}$ is obtained by averaging $Q_{\ss\FC}[j]$ over all symbol intervals and 50000 random-generated realizations of $\textbf{W}_{\ss\T}^{j-1}$, and then the value of $\overline{Q}_{\ss\FC}^{\ast}$ is the minimum $\overline{Q}_{\ss\FC}$ found by numerically optimizing the corresponding constant decision thresholds via exhaustive search. To decrease the complexity of exhaustive search, we consider the same decision threshold at all RXs such that $\xi_{\ss\RX_k} = \xi_{\ss\RX}, \forall k$. 

In Figs. \ref{fig:Pe_M_FC}-\ref{MD-SA-SD-dist}, for each ML detection variant, we plot the error probability with the local history and genie-aided history. We observe that the error performance using the local history has a very small degradation from that using the genie-aided history. This demonstrates the effectiveness of our proposed method to estimate the previous symbols. We also observe that the simulations have very strong agreement with the analytical results, thereby validating our analytical results. In Figs. \ref{fig:Pe_M_FC}-\ref{MD-SA-SD-dist}, we observe that the error performance degradation with the local history compared to the genie-aided history for SA-ML is more noticeable than that for SD-ML and MD-ML. This is because in SD-ML and MD-ML, the FC directly estimates previous RX symbols from the RX-FC links. However, for SA-ML, the FC does not directly estimate the previous RX emissions from the RX-FC links and the error in the estimation of previous TX symbols propagates to the estimated previous RX emissions.
\ifOneCol
\vspace{-2mm}
\else
\fi
\subsection{Symmetric Topology}\label{Symmetric}


\ifOneCol	
\begin{table}[]
\renewcommand{\arraystretch}{0.8}
\centering
\caption{Summary of Considered Variants}
\label{tab:variant2}\vspace{-3mm}
\begin{tabular}{c||c|c|c}
\hline
\textbf{Variants}     & \begin{tabular}[c]{@{}c@{}}\textbf{Relaying} \textbf{at RXs}\end{tabular} & \begin{tabular}[c]{@{}c@{}}\textbf{Molecule} \textbf{Type Used}\textbf{in RXs}\end{tabular} & \begin{tabular}[c]{@{}c@{}}\textbf{Behavior}  \textbf{at FC}\end{tabular}     \\ \hline\hline
Majority Rule \cite{TMBMC2016,GC2016}      & DF                                                        & Multiple                                                                & \begin{tabular}[c]{@{}c@{}}Constant Threshold \end{tabular} \\ \hline
MD-ML       & DF                                                        & Multiple                                                                & \begin{tabular}[c]{@{}c@{}}ML Detection\end{tabular}       \\ \hline
SD-Constant\cite{simplified} & DF                                                        & Single                                                                  & \begin{tabular}[c]{@{}c@{}}Constant Threshold\end{tabular} \\ \hline
SD-ML       & DF                                                        & Single                                                                  & \begin{tabular}[c]{@{}c@{}}ML Detection\end{tabular}       \\ \hline
SA-Constant & AF                                                        & Single                                                                  & \begin{tabular}[c]{@{}c@{}}Constant Threshold\end{tabular} \\ \hline
SA-ML       & AF                                                        & Single                                                                  & \begin{tabular}[c]{@{}c@{}}ML Detection\end{tabular}       \\ \hline
\end{tabular}
\vspace{-8mm}
\end{table}
\else
\begin{table}[]
\renewcommand{\arraystretch}{1}
\centering
\caption{Summary of Considered Variants}
\label{tab:variant2}\vspace{-1mm}
\begin{tabular}{c||c|c|c}
\hline
\textbf{Variants}     & \begin{tabular}[c]{@{}c@{}}\textbf{Relaying}\\ \textbf{at RXs}\end{tabular} & \begin{tabular}[c]{@{}c@{}}\textbf{Molecule}\\ \textbf{Type Used}\\ \textbf{in RXs}\end{tabular} & \begin{tabular}[c]{@{}c@{}}\textbf{Behavior} \\ \textbf{at FC}\end{tabular}     \\ \hline\hline
Majority Rule \cite{TMBMC2016,GC2016}      & DF                                                        & Multiple                                                                & \begin{tabular}[c]{@{}c@{}}Constant Threshold \end{tabular} \\ \hline
MD-ML       & DF                                                        & Multiple                                                                & \begin{tabular}[c]{@{}c@{}}ML Detection\end{tabular}       \\ \hline
SD-Constant\cite{simplified} & DF                                                        & Single                                                                  & \begin{tabular}[c]{@{}c@{}}Constant Threshold\end{tabular} \\ \hline
SD-ML       & DF                                                        & Single                                                                  & \begin{tabular}[c]{@{}c@{}}ML Detection\end{tabular}       \\ \hline
SA-Constant & AF                                                        & Single                                                                  & \begin{tabular}[c]{@{}c@{}}Constant Threshold\end{tabular} \\ \hline
SA-ML       & AF                                                        & Single                                                                  & \begin{tabular}[c]{@{}c@{}}ML Detection\end{tabular}       \\ \hline
\end{tabular}
\vspace{-3mm}
\end{table}
\fi

We consider at most 6 RXs in this subsection and the specific locations of RXs are:
\ifOneCol	
\\$(2{\mu}\metre, \pm0.6{\mu}\metre,0)$ and $(2{\mu}\metre,\pm0.3{\mu}\metre,\pm0.5196{\mu}\metre)$,
\else
$(2{\mu}\metre, \pm0.6{\mu}\metre,0)$ and $(2{\mu}\metre,\pm0.3{\mu}\metre,\pm0.5196{\mu}\metre)$,
\fi
\textcolor{black}{where the RXs are placed on a circle perpendicular to the line passing from the TX, the FC, and the center of the circle.}

In order to provide trade-offs between the performance versus the information available, we compare the error performance of the ML detectors with the majority rule \cite{TMBMC2016,GC2016} and SD-Constant \cite{simplified}. Notably, we also propose a \emph{new} variant for comparison, namely, SA-Constant. In SA-Constant, the behavior of each RX is the same as that in SA-ML, but the FC makes a decision $\hat{W}_{\ss\FC}[j]$ by comparing $\tilde{s}[j]$ with a constant threshold $\xi_{\ss\FC}$, independent of $\textbf{W}_{\ss\T}^{j-1}$. It can be shown that $Q_{\ss\FC}[j]$ for SA-Constant with \emph{any} realization of $\textbf{W}_{\ss\T}^{j-1}$ can be obtained by replacing $\xi_{\ss\FC}^{\ad,\SA}[j]$ with the threshold $\xi_{\ss\FC}$ in \eqref{perfect probability1,SA}. We summarize all variants considered in this subsection in Table \ref{tab:variant2}. For these variants, we consider the same parameters as the ML detectors for the fairness of our comparisons.

\ifOneCol	
\begin{figure}[!t]
\centering
\includegraphics[height=3in]{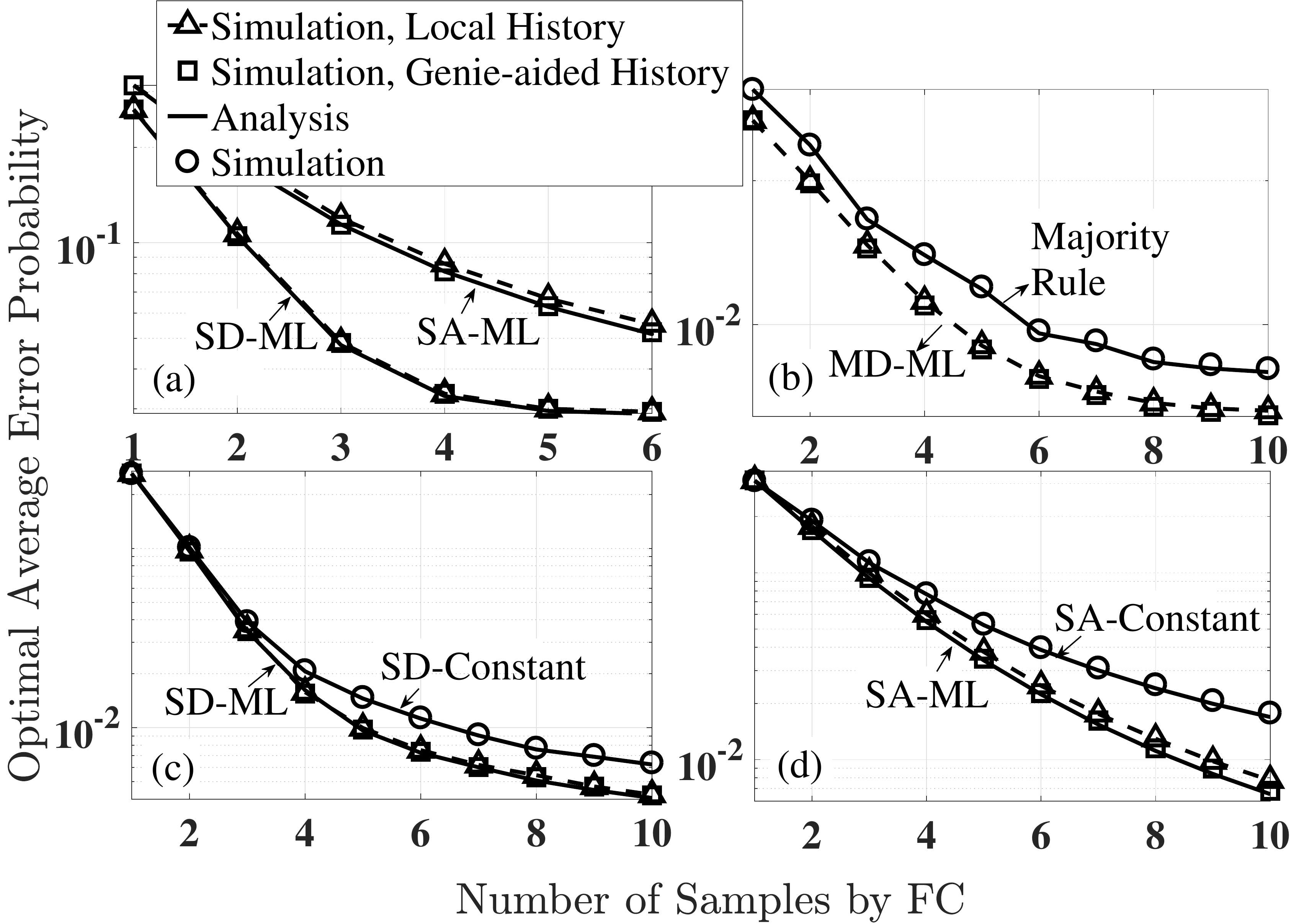}
\vspace{-7mm}
\caption{Optimal average error probability $\overline{Q}_{\ss\FC}^{\ast}$ versus the number of samples by FC $M_{\ss\FC}$ for (a) SD-ML and SA-ML, (b) MD-ML and the majority rule, (c) SD-ML and SD-Constant, and (d) SA-ML and SA-Constant. The analytical error performance of the majority rule and SD-Constant is presented in \cite{TMBMC2016} and \cite{simplified}, respectively.}
\label{fig:Pe_M_FC}
\vspace{-8mm}
\end{figure}
\else
\begin{figure}[!t]
\centering
\includegraphics[height=2.4in]{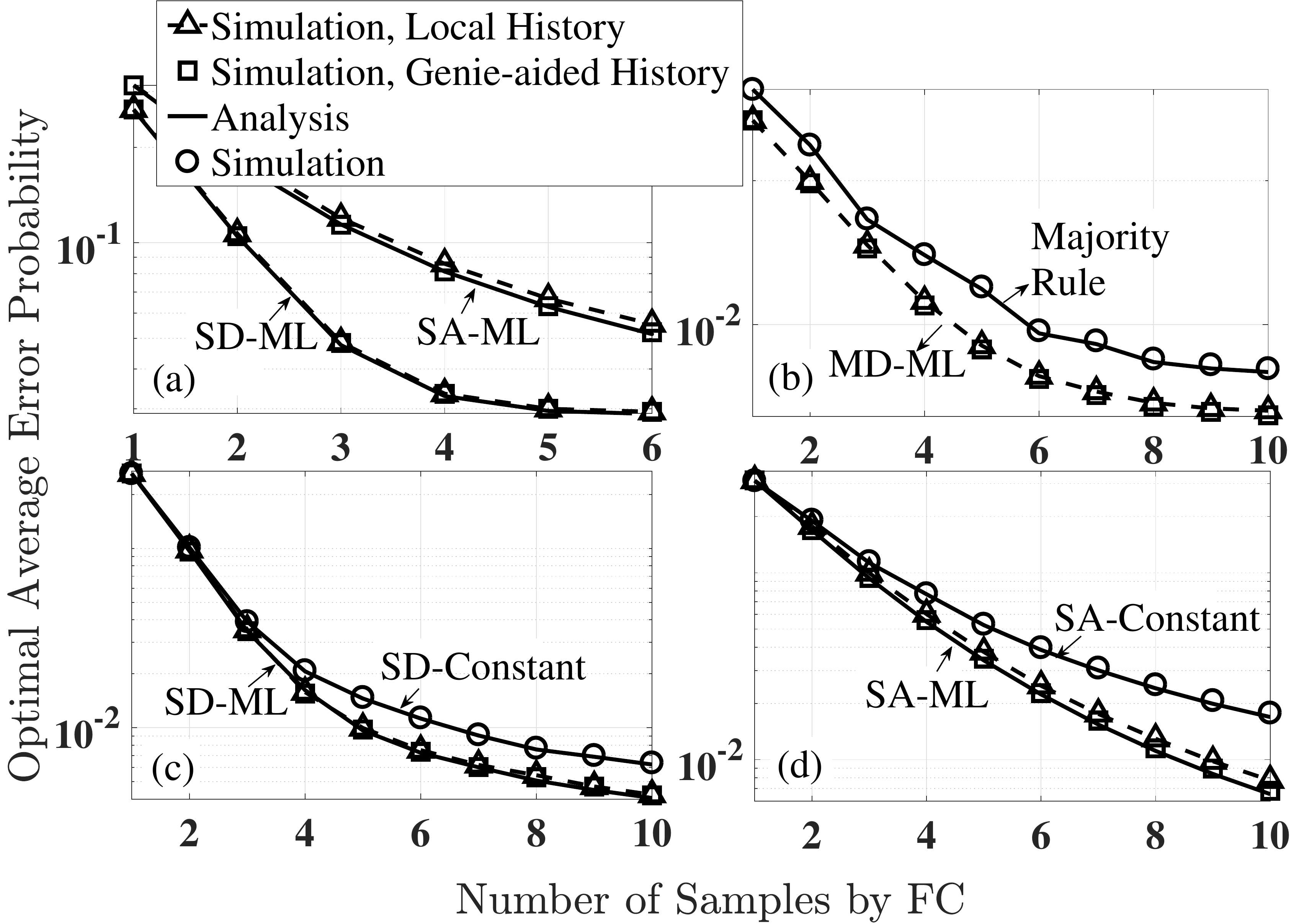}
\vspace{-2mm}
\caption{Optimal average error probability $\overline{Q}_{\ss\FC}^{\ast}$ versus the number of samples by FC $M_{\ss\FC}$ for (a) SD-ML and SA-ML, (b) MD-ML and the majority rule, (c) SD-ML and SD-Constant, and (d) SA-ML and SA-Constant. The analytical error performance of the majority rule and SD-Constant is presented in \cite{TMBMC2016} and \cite{simplified}, respectively.}
\label{fig:Pe_M_FC}
\vspace{-4mm}
\end{figure}
\fi

In Fig. \ref{fig:Pe_M_FC}, we plot the optimal average global error probability $\overline{Q}_{\ss\FC}^{\ast}$ of different variants versus the number $M_{\ss\FC}$ of samples by the FC. In Fig.~\ref{fig:Pe_M_FC}, the report time interval is fixed at $t_{\report}=0.3\,{\m}\s$ as in Table~\ref{tab:table1} and the time step at the FC for each $M_{\ss\FC}$ is $\Delta{t_{\ss\FC}}=0.3\,{\m}\s/M_{\ss\FC}$. We observe that the system error performance improves as $M_{\ss\FC}$ increases. This is because when $M_{\ss\FC}$ increases, the number of molecules expected to be observed at each RX increases.

In Fig. \ref{fig:Pe_M_FC}(a), we consider a single-RX system (which is analogous to the two-hop environment considered in \cite{Ahmadzadeh2015}). 
We observe that SD-ML outperforms SA-ML. 
In Fig. \ref{fig:Pe_M_FC}(b)-(d), we consider a three-RX system. 
We observe that MD-ML, SD-ML, and SA-ML outperform the majority rule, SD-Constant, and SA-Constant, respectively. 
However, the error performance degradation with these simpler cooperative variants are all within an order of magnitude for the range of $M_{\ss\FC}$ considered. This demonstrates the relatively good performance of the simpler variants.

\ifOneCol	
\begin{figure}[!t]
\centering
\includegraphics[height=3in]{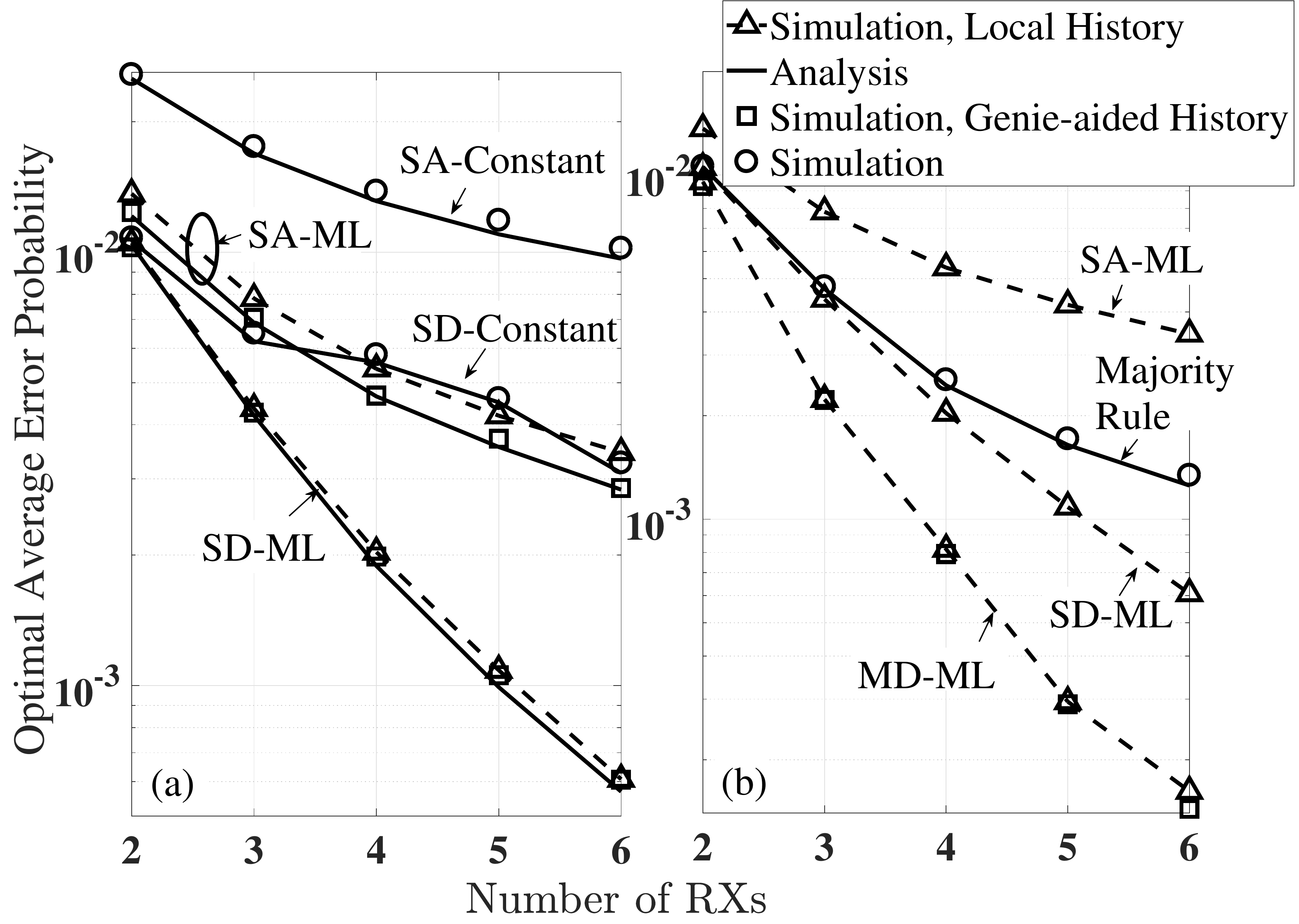}
\vspace{-7mm}
\caption{Optimal average error probability $\overline{Q}_{\ss\FC}^{\ast}$ of different variants versus the number of RXs $K$. The analytical error performance of SD-Constant and the majority rule is presented in \cite{simplified} and \cite{TMBMC2016}, respectively.}
\label{MD-SA-SD-K}
\vspace{-8mm}
\end{figure}
\else
\begin{figure}[!t]
\centering
\includegraphics[height=2.4in]{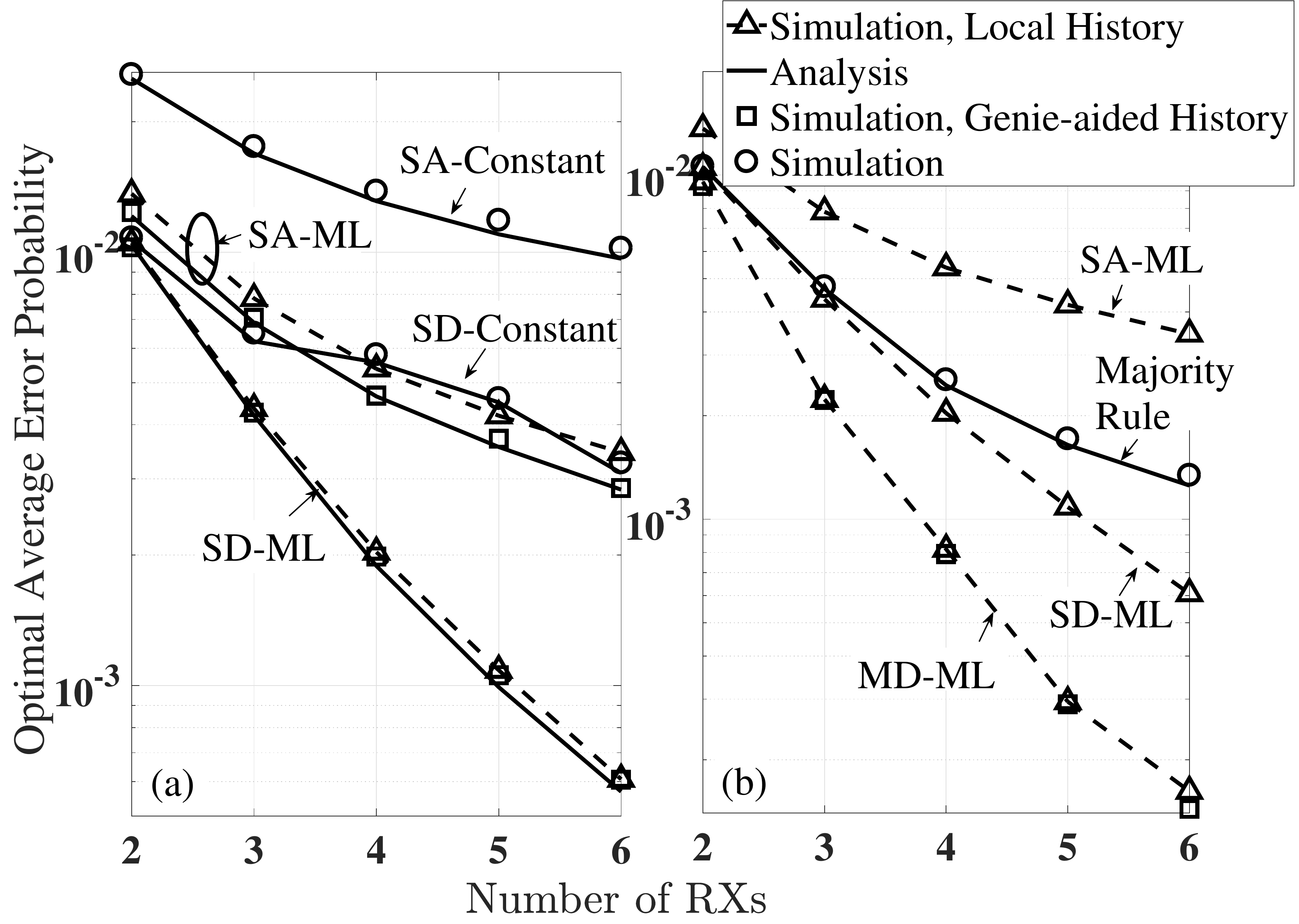}
\vspace{-2mm}
\caption{Optimal average error probability $\overline{Q}_{\ss\FC}^{\ast}$ of different variants versus the number of RXs $K$. The analytical error performance of SD-Constant and the majority rule is presented in \cite{simplified} and \cite{TMBMC2016}, respectively.}
\label{MD-SA-SD-K}
\vspace{-4mm}
\end{figure}
\fi

In Fig.~\ref{MD-SA-SD-K}, we plot the optimal average global error probability versus the number $K$ of cooperative RXs for different variants. We see that the system error performance improves as $K$ increases, even though the total number of molecules is constrained. The same observation of error performance improvement may be observed in a channel with additive signal dependent noise if our results can be well approximated by the Gaussian signal dependent noise model\cite{Aminian}. The system error performance does not always improve as $K$ increases. This is because if we keep increasing $K$, the number of released molecules for each $\RX_k$ decreases, which leads to the $\RX_k-\FC$ link becoming unreliable. The system error performance would improve as the volume of the FC increases for the fixed $K$, since the FC can observe more molecules, but the volume of microorganisms cannot be easily altered.


In Fig.~\ref{MD-SA-SD-K}(a), we observe that SD-Constant and SA-ML using the local history achieve similar error performance. In Fig.~\ref{MD-SA-SD-K}(b), we observe that the majority rule has similar error performance with SD-ML and the majority rule outperforms SA-ML using the local history. These observations demonstrate the good performance of the majority rule, relative to SD-ML and SA-ML. Importantly, we observe that MD-ML outperforms SD-ML and SD-ML outperforms SA-ML. This is because the knowledge of individual $\tilde{s}_{k}[j]$ for each $\RX_k-\FC$ link in MD-ML improves detection performance over only knowing the sum $\tilde{s}[j]$ in SD-ML. Comparing to $\RX_k$ making a binary decision in the current symbol interval in SD-ML, $\RX_k$ in SA-ML amplifies the ISI at $\RX_k$ in the current symbol interval due to the previous TX symbols.

The system error performance in the subsection would degrade relative to the independent case if any of the links become dependent. This can be explained by a special case where all RXs overlap each other and thus have the same observations. Then, the error performance of this case would be the same as that of a cooperative system with $K=1$.
\ifOneCol
\vspace{-4mm}
\else
\fi
\subsection{\textcolor{black}{Asymmetric Topology}}\label{Asymmetric}

\ifOneCol	
\begin{figure}[!t]
\centering
\includegraphics[height=2.8in]{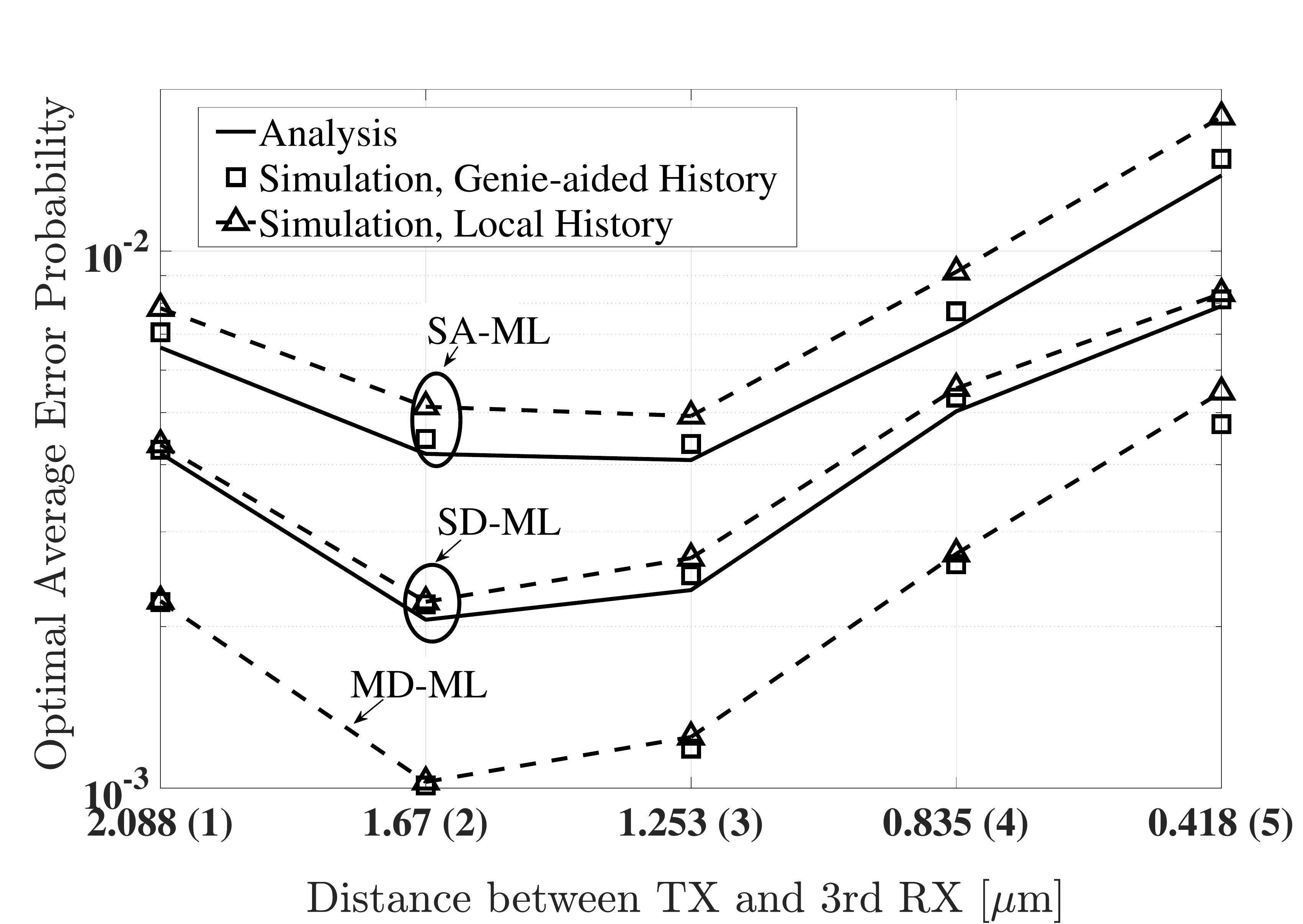}\vspace{-7mm}
\caption{\textcolor{black}{Optimal average error probability $\overline{Q}_{\ss\FC}^{\ast}$ of different variants versus the distance $d_{\ss\TX_3}$ between the TX and $\RX_3$. $\RX_1$ and $\RX_2$ are fixed at $(2{\mu}\metre, 0, 0.6{\mu}\metre)$ and $(2{\mu}\metre, 0, -0.6{\mu}\metre)$, respectively. The locations of $\RX_3$ are (1) $(2{\mu}\metre, 0.6{\mu}\metre, 0)$, (2) $(1.6{\mu}\metre, 0.48{\mu}\metre, 0)$, (3) $(1.2{\mu}\metre, 0.36{\mu}\metre, 0)$, (4) $(0.8{\mu}\metre, 0.24{\mu}\metre, 0)$, (5) $(0.4{\mu}\metre, 0.12{\mu}\metre, 0)$. }}
\label{MD-SA-SD-dist}\vspace{-8mm}
\end{figure}
\else
\begin{figure}[!t]
\centering
\includegraphics[height=2.4in]{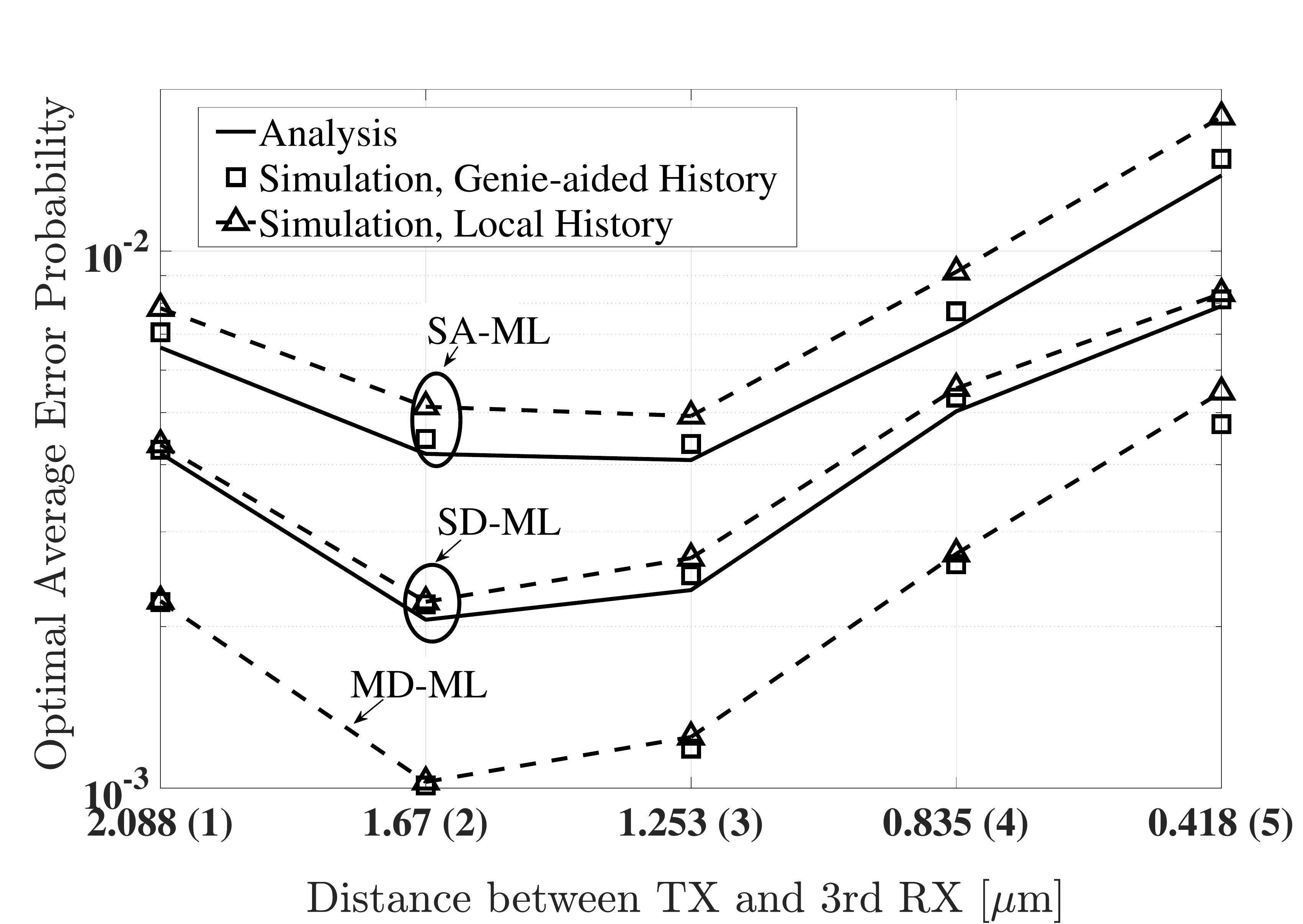}\vspace{-2mm}
\caption{\textcolor{black}{Optimal average error probability $\overline{Q}_{\ss\FC}^{\ast}$ of different variants versus the distance $d_{\ss\TX_3}$ between the TX and $\RX_3$. $\RX_1$ and $\RX_2$ are fixed at $(2{\mu}\metre, 0, 0.6{\mu}\metre)$ and $(2{\mu}\metre, 0, -0.6{\mu}\metre)$, respectively. The locations of $\RX_3$ are (1) $(2{\mu}\metre, 0.6{\mu}\metre, 0)$, (2) $(1.6{\mu}\metre, 0.48{\mu}\metre, 0)$, (3) $(1.2{\mu}\metre, 0.36{\mu}\metre, 0)$, (4) $(0.8{\mu}\metre, 0.24{\mu}\metre, 0)$, (5) $(0.4{\mu}\metre, 0.12{\mu}\metre, 0)$.}}
\label{MD-SA-SD-dist}
\vspace{-4mm}
\end{figure}
\fi

\textcolor{black}{In Fig. \ref{MD-SA-SD-dist}, we consider a three-RX system and plot the optimal average error probability of different variants versus the distance between the TX and $\RX_3$. We keep the positions of $\RX_1$ and $\RX_2$ fixed and move $\RX_3$ along the line segment between the symmetric position and the TX, as indicated in the caption. We observe for our three variants that the error performance first improves and then decreases as $\R_3$ moves toward the TX. This is because both the $\TX$-$\RX_3$ link and the $\RX_3$-$\FC$ link contribute to the error performance of the system. When $d_{\ss\TX_3}$ is relatively large, the system error performance is dominated by the $\TX$-$\RX_3$ link and this link becomes more reliable as $d_{\ss\TX_3}$ decreases. For $d_{\ss\TX_3}$ is relatively small, the system error performance is dominated by the $\RX_3$-$\FC$ link, which becomes weaker when $d_{\ss\TX_3}$ decreases. We also observe that MD-ML outperforms SD-ML and SD-ML outperforms SA-ML, which is consistent with our observations in Fig.~\ref{MD-SA-SD-K}(b).}

\textcolor{black}{In the following figures, we present results to assess the accuracy of our proposed optimization method in Section \ref{subsub:SD-MLopti}. We denote the solution to problem \eqref{opti,app} by $\mathbf{S}^\dagger=\{S_1^\dagger,S_2^\dagger,\ldots,S_K^\dagger\}$. We denote the optimal solution via exhaustive search by $\mathbf{S}^\star=\{S_1^\star,S_2^\star,\ldots,S_K^\star\}$.}

\ifOneCol	
\begin{figure}[!t]
\centering
\includegraphics[height=3in]{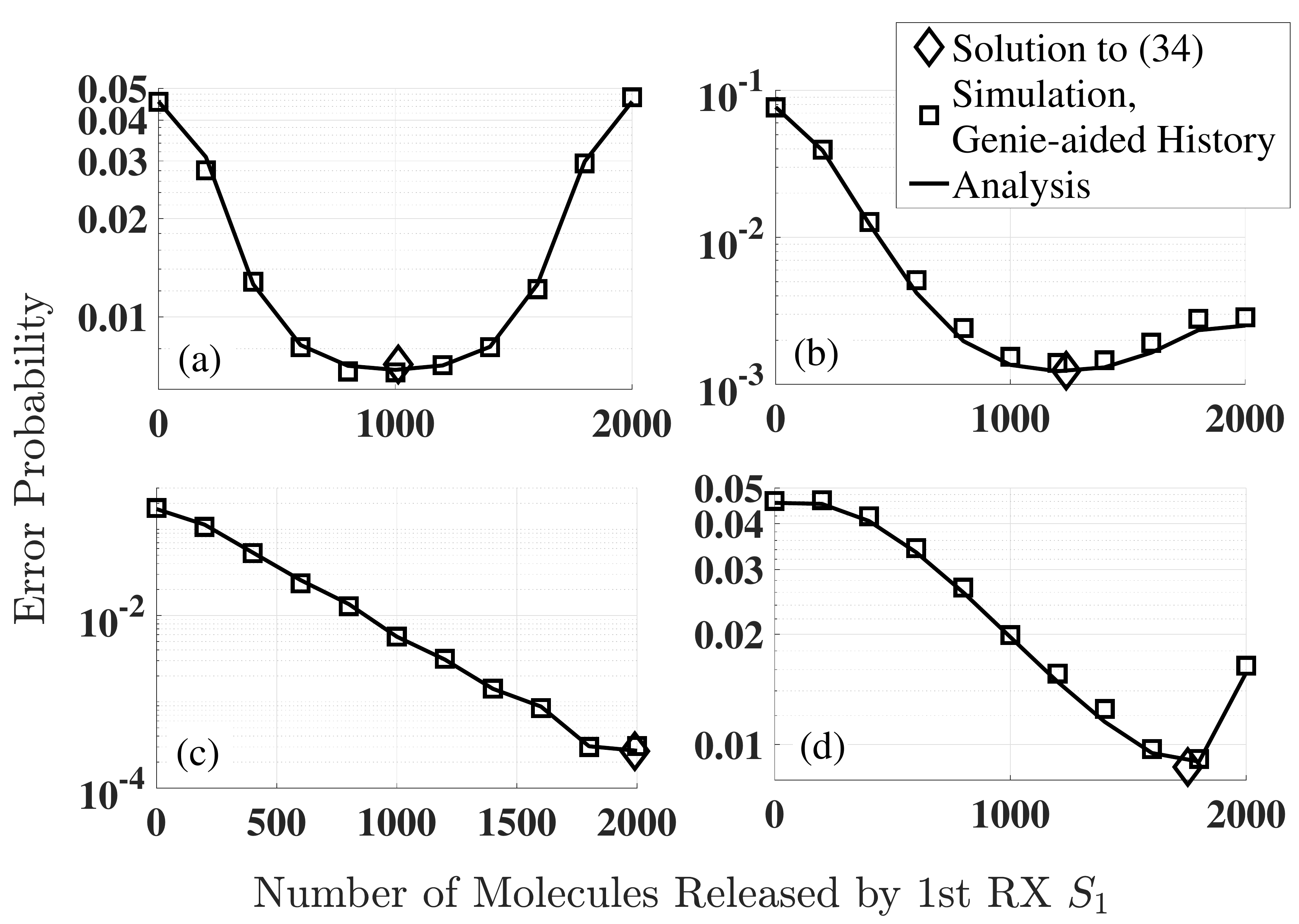}\vspace{-7mm}
\caption{\textcolor{black}{Error probability $Q_{\ss\FC}[j]$ of SD-ML versus the number of molecules released by $\RX_1$, $S_1$, for different locations of $\RX_1$: (a) $(2{\mu}\metre, 0.6{\mu}\metre, 0{\mu}\metre)$, (b) $(1.5{\mu}\metre, 0.45{\mu}\metre, 0{\mu}\metre)$, (c) $(1{\mu}\metre, 0.3{\mu}\metre, 0{\mu}\metre)$, (d) $(0.5{\mu}\metre, 0.15{\mu}\metre, 0{\mu}\metre)$. The location of $\RX_2$ is fixed at $(2{\mu}\metre, -0.6{\mu}\metre, 0{\mu}\metre)$.}}
\label{SD-opti}\vspace{-8mm}
\end{figure}
\else
\begin{figure}[!t]
\centering
\includegraphics[height=2.4in]{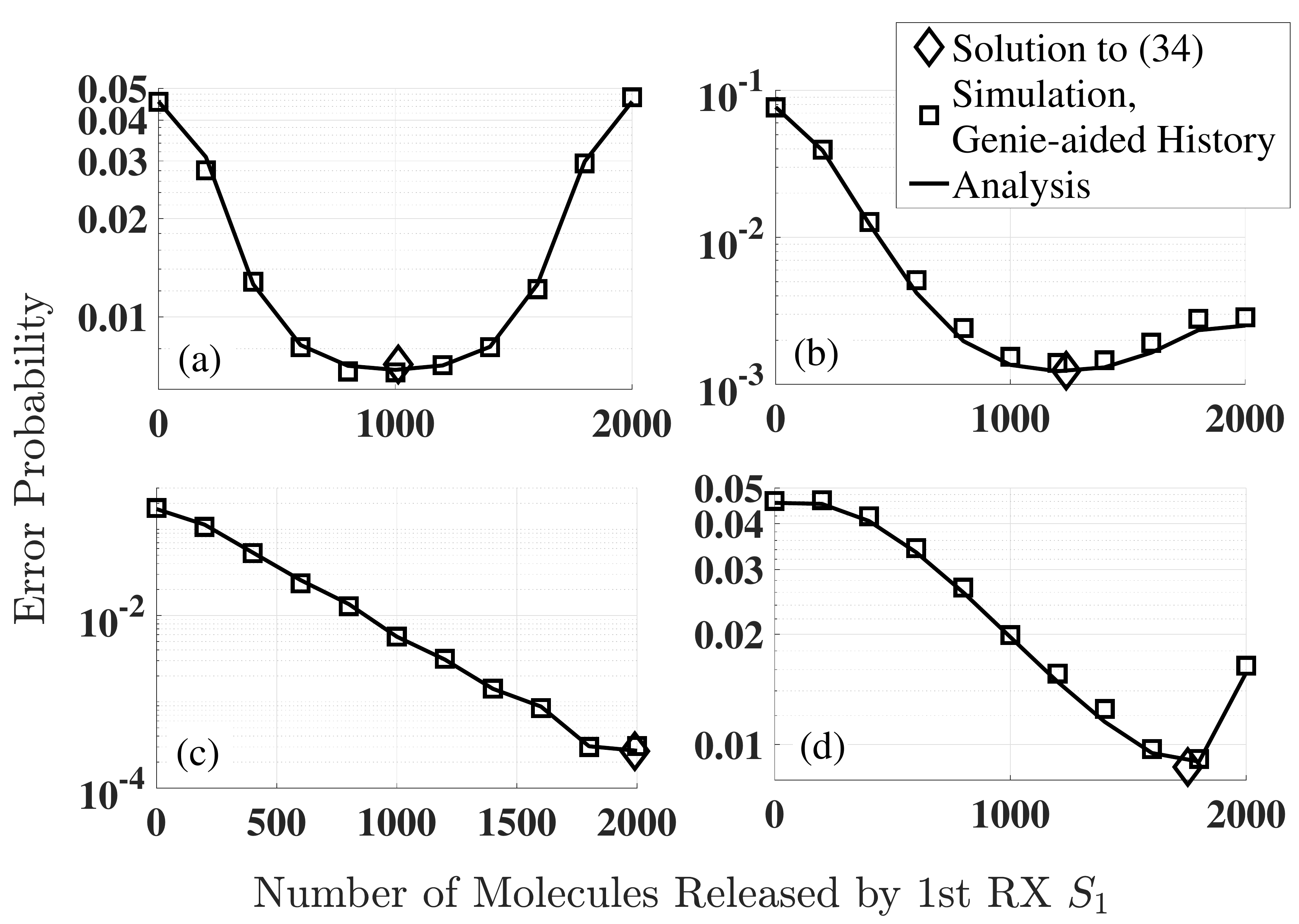}\vspace{-2mm}
\caption{\textcolor{black}{Error probability $Q_{\ss\FC}[j]$ of SD-ML versus the number of molecules released by $\RX_1$, $S_1$, for different locations of $\RX_1$: (a) $(2{\mu}\metre, 0.6{\mu}\metre, 0{\mu}\metre)$, (b) $(1.5{\mu}\metre, 0.45{\mu}\metre, 0{\mu}\metre)$, (c) $(1{\mu}\metre, 0.3{\mu}\metre, 0{\mu}\metre)$, (d) $(0.5{\mu}\metre, 0.15{\mu}\metre, 0{\mu}\metre)$. The location of $\RX_2$ is fixed at $(2{\mu}\metre, -0.6{\mu}\metre, 0{\mu}\metre)$.}}
\label{SD-opti}\vspace{-4mm}
\end{figure}
\fi

\textcolor{black}{In Fig. \ref{SD-opti}, we consider a two-RX system and plot the error probability of SD-ML versus the number of molecules released by $\RX_1$ for different location of $\RX_1$, where we keep $\RX_2$ fixed at $(2{\mu}\metre, 0.6{\mu}\metre, 0{\mu}\metre)$ and move $\RX_1$ along the line segment between the symmetric position and the TX, as indicated in the caption. The x-axis coordinate of $\diamond$ is the solution $S_1^\dagger$ to problem \eqref{opti,app} and the corresponding y-axis coordinate is the $Q^\sharp_{\ss\FC}[j]$ achieved at $\mathbf{S}^\dagger$. We observe that $S_1^\dagger$ and $Q^\sharp_{\ss\FC}[j]|_{\mathbf{S}=\mathbf{S}^\dagger}$ are almost identical to $S_1^\star$ and $Q_{\ss\FC}[j]|_{\mathbf{S}=\mathbf{S}^\star}$, respectively, which confirms the validity of Lemma \ref{tight} and Lemma \ref{opti-thres}, the effectiveness of problem \eqref{opti,app}, and the accuracy of our method to solve problem \eqref{opti,app}. In Fig. \ref{SD-opti}(a), we observe that $S_1 = 1000$ achieves the minimal $Q_{\ss\FC}[j]$, which verifies Theorem \ref{molDistri,asy}. Interestingly, we observe that from Figs. \ref{SD-opti}(a)--(d), when we move $\RX_1$ towards the TX, the optimal molecule allocation for $\RX_1$ first increases and then decreases. This is because, when $\RX_1$ approaches to the TX, the $\TX-\RX_1-\FC$ link becomes more reliable, so increasing the number of molecules for $\RX_1$ optimizes the whole system; and when $\RX_1$ is very close to the TX, the $\TX-\RX_1-\FC$ link becomes less reliable due to a weak $\RX_1-\FC$ link. In particular, in Fig.~\ref{SD-opti}(c), the optimal solution is to allocate all molecules to $\RX_1$. This is because when $\RX_1$ is at $(1{\mu}\metre, 0.3{\mu}\metre, 0{\mu}\metre)$, $\RX_1$ is very close to the optimal relay location, i.e., the midpoint between the TX and the FC, thus the $\TX-\RX_1-\FC$ link is much more reliable than the $\TX-\RX_2-\FC$ link and allocating all molecules to $\RX_1$ optimizes the whole system.}

\ifOneCol	
\begin{figure}[!t]
\centering
\includegraphics[height=3in]{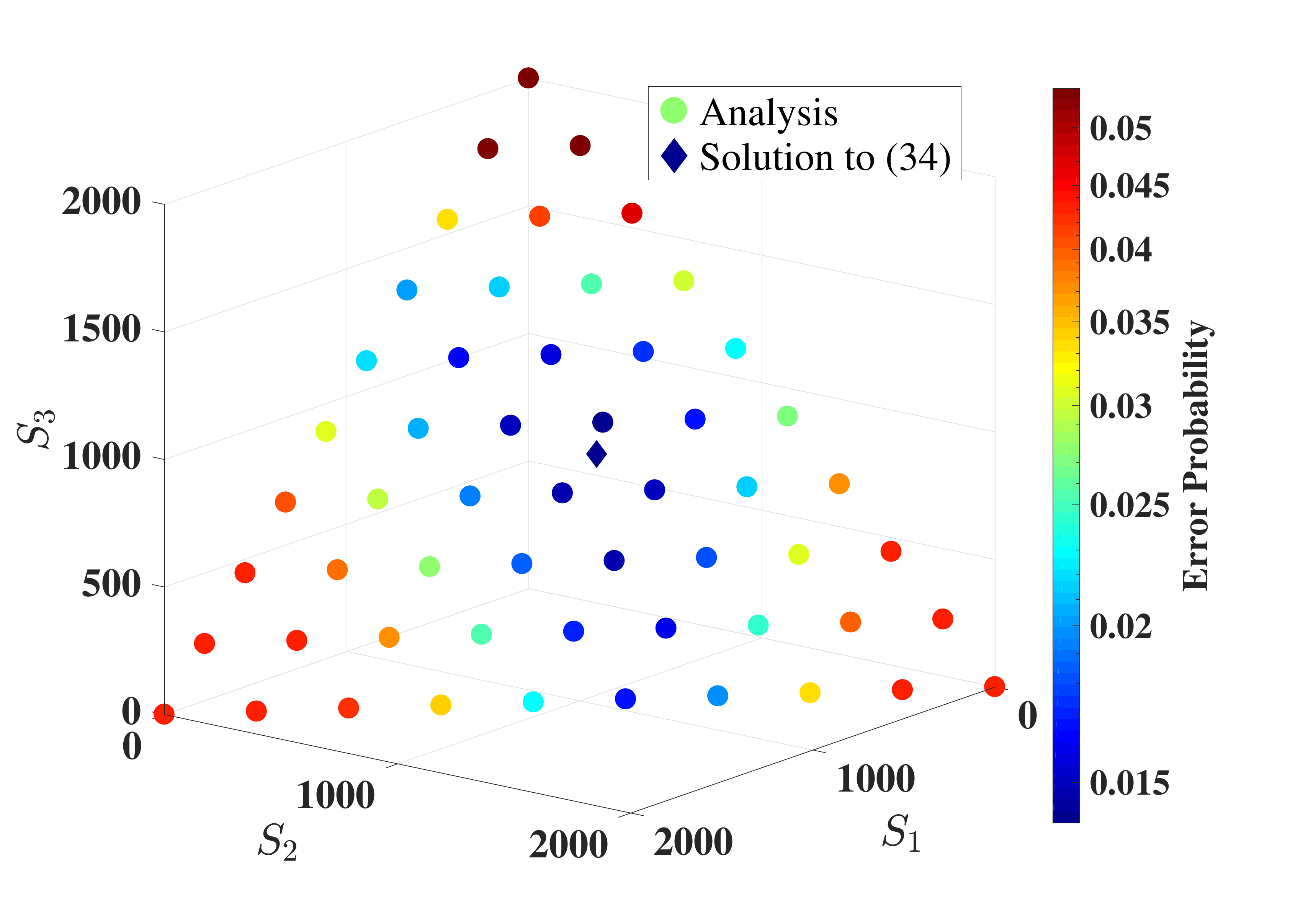}\vspace{-7mm}
\caption{\textcolor{black}{Error probability $Q_{\ss\FC}[j]$ of SD-ML versus the number of molecules released by $\RX_1$, $S_1$, the number of molecules released by $\RX_2$, $S_2$, and the number of molecules released by $\RX_3$, $S_3$. The X-axis, Y-axis, and Z-axis coordinates of `\textcolor{blue}{$\blacklozenge$}' are the solutions to problem \eqref{opti,app}. $\RX_1$, $\RX_2$, and $\RX_3$ are at $(1.915{\mu}\metre, 0.58{\mu}\metre, 0)$, $(1.827{\mu}\metre, 0.579{\mu}\metre, 0)$, and $(1.265{\mu}\metre, 0.328{\mu}\metre, 0)$, respectively. The x-axis and y-axis coordinates of the locations of the RXs are randomly generated.}}
\label{SD-opti1}\vspace{-8mm}
\end{figure}
\else
\begin{figure}[!t]
\centering
\includegraphics[height=2.4in]{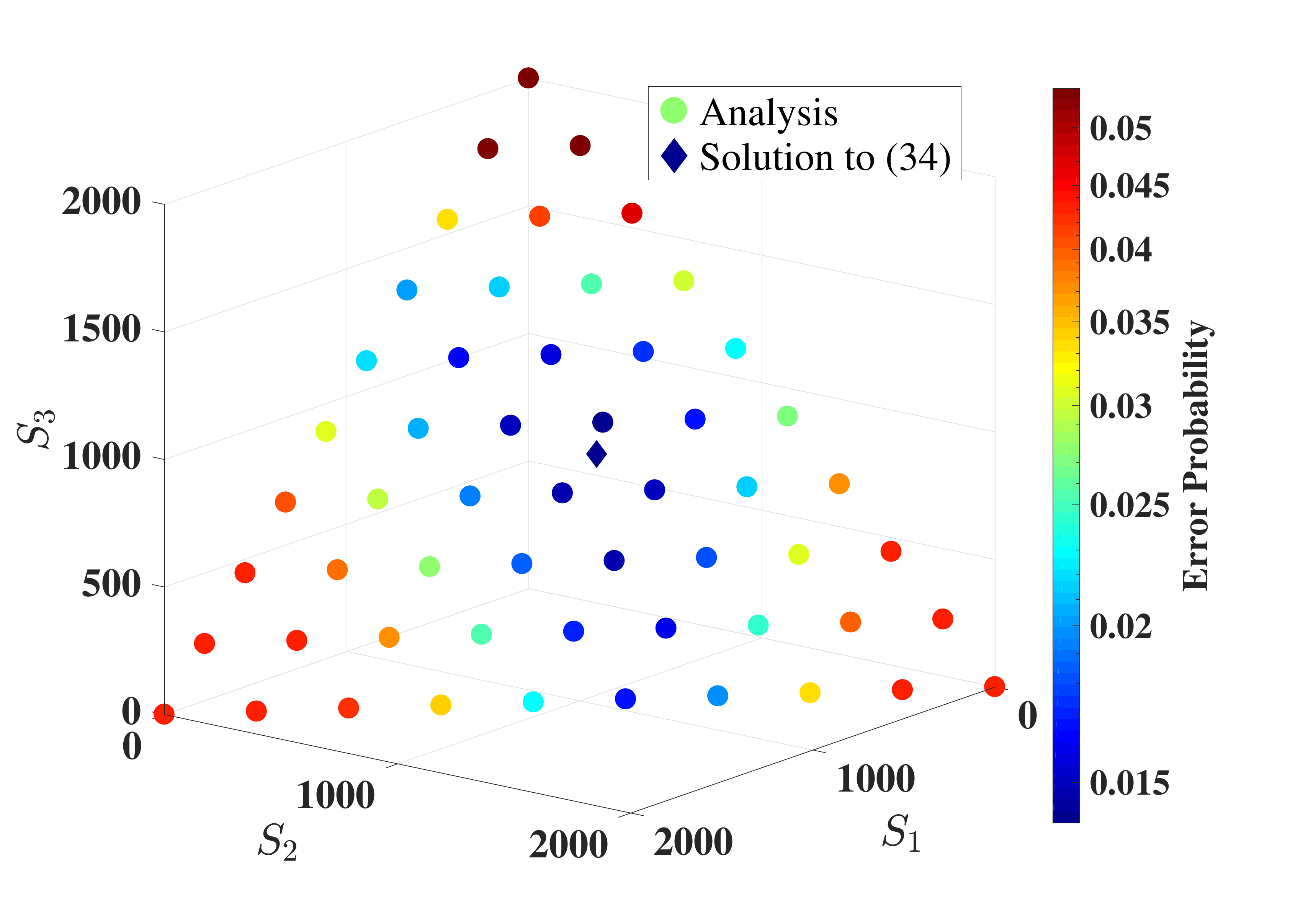}\vspace{-2mm}
\caption{\textcolor{black}{Error probability $Q_{\ss\FC}[j]$ of SD-ML versus the number of molecules released by $\RX_1$, $S_1$, the number of molecules released by $\RX_2$, $S_2$, and the number of molecules released by $\RX_3$, $S_3$. The X-axis, Y-axis, and Z-axis coordinates of `\textcolor{blue}{$\blacklozenge$}' are the solutions to problem \eqref{opti,app}. $\RX_1$, $\RX_2$, and $\RX_3$ are at $(1.915{\mu}\metre, 0.58{\mu}\metre, 0)$, $(1.827{\mu}\metre, 0.579{\mu}\metre, 0)$, and $(1.265{\mu}\metre, 0.328{\mu}\metre, 0)$, respectively. The x-axis and y-axis coordinates of the locations of the RXs are randomly generated.}}
\label{SD-opti1}\vspace{-4mm}
\end{figure}
\fi

\textcolor{black}{In Fig. \ref{SD-opti1}, we consider a three-RX system and plot the error probability of SD-ML versus the number of molecules released by $\RX_1$, $\RX_2$, and $\RX_3$. The locations of the three RXs are generated randomly, as indicated in the caption. The x-axis, y-axis, and z-axis coordinates of `\textcolor{blue}{$\blacklozenge$}' are the solutions $S_1^\dagger$, $S_2^\dagger$, and $S_3^\dagger$ to problem \eqref{opti,app}, respectively. The corresponding 4th coordinate (i.e., color bar) is the $Q^\sharp_{\ss\FC}[j]$ achieved at $\mathbf{S}^\dagger$. We observe that $\mathbf{S}^\dagger$ and $Q^\sharp_{\ss\FC}[j]|_{\mathbf{S}=\mathbf{S}^\dagger}$ are almost identical to $\mathbf{S}^\star$ and $Q_{\ss\FC}[j]|_{\mathbf{S}=\mathbf{S}^\star}$, respectively, which again verifies Lemma \ref{tight}, Lemma \ref{opti-thres}, and the effectiveness of problem \eqref{opti,app}. }
\ifOneCol
\vspace{-6mm}
\else
\vspace{-2mm}
\fi
\section{Conclusions}\label{sec:con}
Combined with our initial work in \cite{Fang2017C}, we presented for the first time symbol-by-symbol ML detection for the cooperative diffusion-based MC system with multiple communication phases.
We considered the transmission of a sequence of binary symbols and accounted for the resultant ISI in the design and analysis of the cooperative MC system.
We presented three ML detectors, i.e., MD-ML, SD-ML, and SA-ML. For practicality, the FC chooses the current symbol using its own local history. For tractability, we derived the system error probabilities for SD-ML and SA-ML using the genie-aided history. \textcolor{black}{We formulated and solved a multi-dimensional optimization problem to find the optimal molecule allocation among RXs that minimizes the system error probability of SD-ML. We analytically proved that the equal distribution of molecules
among two symmetric RXs obtains the local minimal error probability of SD-ML.} Using numerical and simulation results, we corroborated the accuracy of these analytical expressions and the effectiveness of the formulated optimization problem. 
Our results revealed trade-offs between the performance, knowledge of previous symbols, the types of molecule available, relaying modes, and computational complexity.


\appendices

\ifOneCol
\vspace{-5mm}
\else
\vspace{-3mm}
\fi
\section{\textcolor{black}{Proof of Theorem \ref{SD-ML,dec}}}\label{app}
\textcolor{black}{We first prove the decision rule for SD-ML when $\hat{\lambda}_\N^\D[j]>0$. To this end, based on \eqref{ML rule,DF}, we first rewrite the general decision rule for SD-ML as $\hat{W}_{\ss\FC}[j]=1$ if $\frac{\mathcal{L}_1^\SD\left[j\right]}{\mathcal{L}_0^\SD\left[j\right]}\geq1$, otherwise $\hat{W}_{\ss\FC}[j]=0$. Thus, if $\frac{\mathcal{L}_1^\SD\left[j\right]}{\mathcal{L}_0^\SD\left[j\right]}$ is a monotonically increasing function with respect to $\tilde{s}[j]$, then we can obtain the decision rule \eqref{dec 1}. We can prove that $\frac{\mathcal{L}_1^\SD\left[j\right]}{\mathcal{L}_0^\SD\left[j\right]}$ is a monotonically increasing function with respect to $\tilde{s}[j]$ by proving that $\left(\frac{\mathcal{L}_1^\SD\left[j\right]}{\mathcal{L}_0^\SD\left[j\right]}\right)^{'}>0$. Based on \eqref{SD-ML,L,asy}, we first rewrite $\mathcal{L}_1^\SD\left[j\right]$ and $\mathcal{L}_0^\SD\left[j\right]$ as
\ifOneCol
\begin{align}\label{SD-ML,L,asy,1}
{\mathcal{L}_1^\SD\left[j\right]}=&\;\sum_{h_1=1}^{2^K}\bigg[\textrm{Pr}\left(h_1|1\right){\left(\tilde{s}[j]!\right)^{-1}}
\exp\left(-\hat{\lambda}_\N^\D[j]-\hat{\lambda}_{\s,h_1}^{\D,\Tot}[j]\right)\left(\hat{\lambda}_\N^\D[j]+\hat{\lambda}_{\s,h_1}^{\D,\Tot}[j]\right)^{\tilde{s}[j]}\bigg]
\end{align}	
\else
\begin{align}\label{SD-ML,L,asy,1}
{\mathcal{L}_1^\SD\left[j\right]}=&\;\sum_{h_1=1}^{2^K}\bigg[\textrm{Pr}\left(h_1|1\right)
\exp\left(-\hat{\lambda}_\N^\D[j]-\hat{\lambda}_{\s,h_1}^{\D,\Tot}[j]\right)\nonumber\\
&\times\left(\hat{\lambda}_\N^\D[j]+\hat{\lambda}_{\s,h_1}^{\D,\Tot}[j]\right)^{\tilde{s}[j]}{\left(\tilde{s}[j]!\right)^{-1}}\bigg]
\end{align}
\fi
and
\ifOneCol	
\begin{align}\label{SD-ML,L,asy,0}
{\mathcal{L}_0^\SD\left[j\right]}=&\;\sum_{h_0=1}^{2^K}\bigg[\textrm{Pr}\left(h_0|0\right){\left(\tilde{s}[j]!\right)^{-1}}\exp\left(-\hat{\lambda}_\N^\D[j]-\hat{\lambda}_{\s,h_0}^{\D,\Tot}[j]\right)\left(\hat{\lambda}_\N^\D[j]+\hat{\lambda}_{\s,h_0}^{\D,\Tot}[j]\right)^{\tilde{s}[j]}\bigg],
\end{align}
\else
\begin{align}\label{SD-ML,L,asy,0}
{\mathcal{L}_0^\SD\left[j\right]}=&\;\sum_{h_0=1}^{2^K}\bigg[\textrm{Pr}\left(h_0|0\right)\exp\left(-\hat{\lambda}_\N^\D[j]-\hat{\lambda}_{\s,h_0}^{\D,\Tot}[j]\right)\nonumber\\
&\times\left(\hat{\lambda}_\N^\D[j]+\hat{\lambda}_{\s,h_0}^{\D,\Tot}[j]\right)^{\tilde{s}[j]}{\left(\tilde{s}[j]!\right)^{-1}}\bigg],
\end{align}
\fi
respectively, where $\textrm{Pr}\left(h|b\right)=\textrm{Pr}\left(\hat{\mathcal{W}}^{{\ss\RX}}_{j,h} |W_{\ss\T}[j]=b,\textbf{W}_{\ss\TX}^{j-1}\right)$. Based on \eqref{SD-ML,L,asy,1} and \eqref{SD-ML,L,asy,0}, we find the first derivative of $\frac{\mathcal{L}_1^\SD\left[j\right]}{\mathcal{L}_0^\SD\left[j\right]}$ with respect to $\tilde{s}[j]$ as
\ifOneCol	
\begin{align}\label{incre1}
\left(\frac{\mathcal{L}_1^\SD\left[j\right]}{\mathcal{L}_0^\SD\left[j\right]}\right)^{'}=&\;\sum_{h_1=1}^{2^K}\sum_{h_0=1}^{2^K}\left[
\textrm{Pr}\left(h_1|1\right)\right.
\left.\textrm{Pr}\left(h_0|0\right)\Pi(h_1,h_0)\right]\left(\mathcal{L}_0^\SD\left[j\right]\left(\tilde{s}[j]!\right)\right)^{-2},
\end{align}
\else
\begin{align}\label{incre1}
\left(\frac{\mathcal{L}_1^\SD\left[j\right]}{\mathcal{L}_0^\SD\left[j\right]}\right)^{'}\!=\!\sum_{h_1=1}^{2^K}\sum_{h_0=1}^{2^K}\frac{\left[
\textrm{Pr}\left(h_1|1\right)\textrm{Pr}\left(h_0|0\right)\Pi(h_1,h_0)\right]}{\left(\mathcal{L}_0^\SD\left[j\right]\left(\tilde{s}[j]!\right)\right)^{2}},
\end{align}
\fi
where
\ifOneCol
\begin{align}\label{chi}
\Pi(h_1,h_0)=&\;\exp\left(-2\hat{\lambda}_\N^\D[j]-\hat{\lambda}_{\s,h_1}^{\D,\Tot}[j]-\hat{\lambda}_{\s,h_0}^{\D,\Tot}[j]\right)\left(\hat{\lambda}_\N^\D[j]+\hat{\lambda}_{\s,h_0}^{\D,\Tot}[j]\right)^{\tilde{s}[j]}\nonumber\\
&\left(\hat{\lambda}_\N^\D[j]+\hat{\lambda}_{\s,h_1}^{\D,\Tot}[j]\right)^{\tilde{s}[j]}\log\left(\frac{\hat{\lambda}_\N^\D[j]+\hat{\lambda}_{\s,h_1}^{\D,\Tot}[j]}{\hat{\lambda}_\N^\D[j]+\hat{\lambda}_{\s,h_0}^{\D,\Tot}[j]}\right).
\end{align}}	
\else
\begin{align}\label{chi}
\Pi(h_1,h_0)=&\;\exp\left(-2\hat{\lambda}_\N^\D[j]-\hat{\lambda}_{\s,h_1}^{\D,\Tot}[j]-\hat{\lambda}_{\s,h_0}^{\D,\Tot}[j]\right)\nonumber\\
&\times\left(\hat{\lambda}_\N^\D[j]+\hat{\lambda}_{\s,h_0}^{\D,\Tot}[j]\right)^{\tilde{s}[j]}\left(\hat{\lambda}_\N^\D[j]+\hat{\lambda}_{\s,h_1}^{\D,\Tot}[j]\right)^{\tilde{s}[j]}\nonumber\\
&\times\log\left(\frac{\hat{\lambda}_\N^\D[j]+\hat{\lambda}_{\s,h_1}^{\D,\Tot}[j]}{\hat{\lambda}_\N^\D[j]+\hat{\lambda}_{\s,h_0}^{\D,\Tot}[j]}\right).
\end{align}}
\fi

\textcolor{black}{We observe that in \eqref{incre1}, all terms are positive except for the $\log(\cdot)$ term. Since $\log(x)>0$ when $x>1$, we separate \eqref{incre1} into two parts: $\log(\cdot)>0$ and $\log(\cdot)<0$. By doing so, we rewrite \eqref{incre1} as the sum of A and B, i.e.,
\ifOneCol	
\begin{align}\label{incre2,1}
A=\sum_{h_1=1}^{2^K}\sum_{h_0=1,\,\hat{\lambda}_{\s,h_1}^{\D,\Tot}[j]>\hat{\lambda}_{\s,h_0}^{\D,\Tot}[j]}^{2^K}\frac{\left[
\textrm{Pr}\left(h_1|1\right)\textrm{Pr}\left(h_0|0\right)\Pi(h_1,h_0)\right]}{\left(\mathcal{L}_0^\SD\left[j\right]\left(\tilde{s}[j]!\right)\right)^{2}}
\end{align}
\else
\begin{align}\label{incre2,1}
A=\sum_{h_1=1}^{2^K}\sum_{h_0=1,\,\hat{\lambda}_{\s,h_1}^{\D,\Tot}[j]>\hat{\lambda}_{\s,h_0}^{\D,\Tot}[j]}^{2^K}\hspace{-8mm}\frac{\left[
\textrm{Pr}\left(h_1|1\right)\textrm{Pr}\left(h_0|0\right)\Pi(h_1,h_0)\right]}{\left(\mathcal{L}_0^\SD\left[j\right]\left(\tilde{s}[j]!\right)\right)^{2}}
\end{align}
\fi
and
\ifOneCol	
\begin{align}\label{incre2,2}
B=&\;\sum_{h_1=1}^{2^K}\sum_{h_0=1, \,\hat{\lambda}_{\s,h_1}^{\D,\Tot}[j]<\hat{\lambda}_{\s,h_0}^{\D,\Tot}[j]}^{2^K}\frac{\left[
\textrm{Pr}\left(h_1|1\right)\right.\left.\textrm{Pr}\left(h_0|0\right)\Pi(h_1,h_0)\right]}{\left(\mathcal{L}_0^\SD\left[j\right]\left(\tilde{s}[j]!\right)\right)^{2}}.
\end{align}}
\else
\begin{align}\label{incre2,2}
B=&\sum_{h_1=1}^{2^K}\sum_{h_0=1, \,\hat{\lambda}_{\s,h_1}^{\D,\Tot}[j]<\hat{\lambda}_{\s,h_0}^{\D,\Tot}[j]}^{2^K}\hspace{-10mm}\frac{\left[
\textrm{Pr}\left(h_1|1\right)\right.\left.\textrm{Pr}\left(h_0|0\right)\Pi(h_1,h_0)\right]}{\left(\mathcal{L}_0^\SD\left[j\right]\left(\tilde{s}[j]!\right)\right)^{2}}.
\end{align}}
\fi

\textcolor{black}{We further rearrange the summation orders and exchange $h_1$ and $h_0$ in \eqref{incre2,2} to rewrite B as
\ifOneCol
\begin{align}\label{incre3}
B=&\;\sum_{h_1=1}^{2^K}\sum_{h_0=1, \,\hat{\lambda}_{\s,h_1}^{\D,\Tot}[j]>\hat{\lambda}_{\s,h_0}^{\D,\Tot}[j]}^{2^K}\frac{\left[
\textrm{Pr}\left(h_0|1\right)\right.
\left.\textrm{Pr}\left(h_1|0\right)\Pi(h_0,h_1)\right]}{\left(\mathcal{L}_0^\SD\left[j\right]\left(\tilde{s}[j]!\right)\right)^{2}}.
\end{align}}	
\else
\begin{align}\label{incre3}
B=&\;\sum_{h_1=1}^{2^K}\sum_{h_0=1, \,\hat{\lambda}_{\s,h_1}^{\D,\Tot}[j]>\hat{\lambda}_{\s,h_0}^{\D,\Tot}[j]}^{2^K}\hspace{-10.5mm}\frac{\left[
\textrm{Pr}\left(h_0|1\right)\right.
\left.\textrm{Pr}\left(h_1|0\right)\Pi(h_0,h_1)\right]}{\left(\mathcal{L}_0^\SD\left[j\right]\left(\tilde{s}[j]!\right)\right)^{2}}.
\end{align}}
\fi

\textcolor{black}{Combining \eqref{incre2,1} and \eqref{incre3} and applying $\Pi(h_1,h_0)=-\Pi(h_0,h_1)$, we have
\ifOneCol
\begin{align}\label{incre4}
\left(\frac{\mathcal{L}_1^\SD\left[j\right]}{\mathcal{L}_0^\SD\left[j\right]}\right)^{'}=&\;\sum_{h_1=1}^{2^K}\sum_{h_0=1, \,\hat{\lambda}_{\s,h_1}^{\D,\Tot}[j]>\hat{\lambda}_{\s,h_0}^{\D,\Tot}[j]}^{2^K}\left[
\vartheta(h_1,h_0)\Pi(h_1,h_0)\right]\left(\mathcal{L}_0^\SD\left[j\right]\left(\tilde{s}[j]!\right)\right)^{-2},
\end{align}	
\else\begin{align}\label{incre4}
\left(\frac{\mathcal{L}_1^\SD\left[j\right]}{\mathcal{L}_0^\SD\left[j\right]}\right)^{'}=\sum_{h_1=1}^{2^K}\sum_{h_0=1, \,\hat{\lambda}_{\s,h_1}^{\D,\Tot}[j]>\hat{\lambda}_{\s,h_0}^{\D,\Tot}[j]}^{2^K}\frac{
\left[
\vartheta(h_1,h_0)\Pi(h_1,h_0)\right]}{\left(\mathcal{L}_0^\SD\left[j\right]\left(\tilde{s}[j]!\right)\right)^{2}},
\end{align}
\fi
where $\vartheta(h_1,h_0)=\textrm{Pr}\left(h_1|1\right)\textrm{Pr}\left(h_0|0\right)-\textrm{Pr}\left(h_0|1\right)\textrm{Pr}\left(h_1|0\right)$. We find that $\eqref{incre4}>0$ holds when $\vartheta(h_1,h_0)>0$ is valid, i.e.,  where $\hat{\lambda}_{\s,h_1}^{\D,\Tot}[j]>\hat{\lambda}_{\s,h_0}^{\D,\Tot}[j]$. We note that $\hat{\lambda}_{\s,h_1}^{\D,\Tot}[j]>\hat{\lambda}_{\s,h_0}^{\D,\Tot}[j]$ leads to $\parallel\hat{\mathcal{W}}^{{\ss\RX}}_{j,h_1}\parallel_1>\parallel\hat{\mathcal{W}}^{{\ss\RX}}_{j,h_0}\parallel_1$, where $\parallel\mathbf{x}\parallel_1$ is the 1-norm of the vector $\mathbf{x}$. When $\parallel\hat{\mathcal{W}}^{{\ss\RX}}_{j,h_1}\parallel_1>\parallel\hat{\mathcal{W}}^{{\ss\RX}}_{j,h_0}\parallel_1$ holds, we have
$\textrm{Pr}\left(h_1|1\right)>\textrm{Pr}\left(h_0|1\right)$ and $\textrm{Pr}\left(h_0|0\right)>\textrm{Pr}\left(h_1|0\right)$, which leads to $\vartheta(h_1,h_0)>0$. Thus, $\vartheta(h_1,h_0)>0$ holds if $\hat{\lambda}_{\s,h_1}^{\D,\Tot}[j]>\hat{\lambda}_{\s,h_0}^{\D,\Tot}[j]$. This proves that $\left(\frac{\mathcal{L}_1^\SD\left[j\right]}{\mathcal{L}_0^\SD\left[j\right]}\right)^{'}>0$ and thus proves the decision rule for SD-ML when $\hat{\lambda}_\N^\D[j]>0$.}

\textcolor{black}{We finally prove the decision rule when $\hat{\lambda}_\N^\D[j]=0$. We recall that $\hat{\lambda}_\N^\D[j]=0$ means all previous RX symbols are ``0''. It probably occurs when all previous TX symbols are ``0'' (i.e., no ISI at $\RX_k$) if the error probability of the first phase is small. Hence, there is no likelihood that ``1'' is detected at $\RX_k$ when ``0'' is transmitted by the TX, which leads to $\textrm{Pr}\left(\hat{\mathcal{W}}^{{\ss\RX}}_{j,h}=\mathbf{0}|W_{\ss\T}[j]=0,\textbf{W}_{\ss\TX}^{j-1}\right)\approx1$ and $\textrm{Pr}\left(\hat{\mathcal{W}}^{{\ss\RX}}_{j,h}\neq\mathbf{0}|W_{\ss\T}[j]=0,\textbf{W}_{\ss\TX}^{j-1}\right)\approx0$. Using these approximations and $\hat{\lambda}_\N^\D[j]=0$, we approximate $\mathcal{L}_0^\SD\left[j\right]\approx\exp\left(0\right)\left(0\right)^{\tilde{s}[j]}$. When $\hat{\lambda}_{\N}^\A[j]=0$ and $\tilde{s}[j]=0$, $\mathcal{L}_0^\SD\left[j\right]$ is $1$, thus the decision at the FC is always $\hat{W}_{\ss\FC}[j]=0$ since $\mathcal{L}_1^\SD\left[j\right]<1$. When $\hat{\lambda}_{\N}^\A[j]=0$ and $\tilde{s}[j]>0$, $\mathcal{L}_0^\SD\left[j\right]$ is $0$, thus the decision at the FC is always $\hat{W}_{\ss\FC}[j]=1$ since $\mathcal{L}_1^\SD\left[j\right]>0$.}

\vspace{-3mm}
\section{\textcolor{black}{Proof of Lemma \ref{opti-thres}}}\label{app2}
\textcolor{black}{We take the first derivative of \eqref{Pe,noisy,SD,app} with respect to $\xi$. However, $Q^\sharp_{\ss\FC}[j]$ is a discrete function with respect to $\xi$, which makes $Q^\sharp_{\ss\FC}[j]$ not differentiable in terms of $\xi$. To tackle this challenge, we approximate the sum in \eqref{CDF,possion} with an integral with respect to $\eta$, i.e.,
\begin{align}\label{CDF,possion,2}
\Lambda\approx\!\int^{\xi}_{\eta=0}\exp\left(\!-\!\hat{\lambda}_\N^\D[j]-\!\hat{\lambda}_{\s,h}^{\D,\Tot}[j]\!\right)\frac{\left(\!\hat{\lambda}_\N^\D[j]+\!\hat{\lambda}_{\s,h}^{\D,\Tot}[j]\!\right)^{\eta}}{{\left(\eta!\right)}}d\eta.
\end{align}}

\textcolor{black}{Using the continuous approximation of $\Lambda$ in \eqref{CDF,possion,2} and ${\partial\,\int^{x}_{t=0}f(t)dt}/{\partial\,x}=f(x)$, we take the first derivative of \eqref{Pe,noisy,SD,app} with respect to $\xi$ as ${\partial\,Q^\sharp_{\ss\FC}[j]}/{\partial\,\xi}= P_1\psi_1(\xi)-(1-P_1)\psi_2(\xi)$, where $\psi_b(\xi)$, $b\in\{0,1\}$, is given by
\ifOneCol	
\begin{align}\label{Pe,noisy,SD,app,deri}
\psi_b(\xi)=&\!\sum_{h=1}^{2^K}\!\bigg[\textrm{Pr}\left(\hat{\mathcal{W}}^{{\ss\RX}}_{j,h}|W_{\ss\T}[j]=b,\textbf{W}_{\ss\TX}^{j-1}\right)\exp\left(-\hat{\lambda}_\N^\D[j]-\hat{\lambda}_{\s,h}^{\D,\Tot}[j]\right)\frac{\left(\hat{\lambda}_\N^\D[j]+ \hat{\lambda}_{\s,h}^{\D,\Tot}[j]\right)^{\xi}}{\xi!}\!\bigg]\!.
\end{align}}
\else
\begin{align}\label{Pe,noisy,SD,app,deri}
\psi_b(\xi)=&\;\sum_{h=1}^{2^K}\bigg[\textrm{Pr}\left(\hat{\mathcal{W}}^{{\ss\RX}}_{j,h} |W_{\ss\T}[j]=b,\textbf{W}_{\ss\TX}^{j-1}\right){{\left(\xi!\right)}^{-1}}\nonumber\\
&\hspace{-4mm}\times\exp\left(-\hat{\lambda}_\N^\D[j]-\hat{\lambda}_{\s,h}^{\D,\Tot}[j]\right)\left(\hat{\lambda}_\N^\D[j]+ \hat{\lambda}_{\s,h}^{\D,\Tot}[j]\right)^{\xi}\bigg].
\end{align}}
\fi

\textcolor{black}{Comparing \eqref{Pe,noisy,SD,app,deri} with \eqref{SD-ML,L,asy}, we find that $\psi_b(\tilde{s}[j])=\mathcal{L}_b^\SD\left[j\right]$. We recall that $\xi_{\ss\FC}^{\ad,\SD}[j]$ is the solution to $\mathcal{L}_1^\SD\left[j\right]=\mathcal{L}_0^\SD\left[j\right]$ in terms of $\tilde{s}[j]$. Hence, $\xi_{\ss\FC}^{\ad,\SD}[j]$ is the solution to $P_1\psi_1(\xi)-(1-P_1)\psi_2(\xi)={\partial\,Q^\sharp_{\ss\FC}[j]}/{\partial\,\xi}=0$ if $P_1=\frac{1}{2}$. Therefore, $\xi_{\ss\FC}^{\ad,\SD}[j]$ is the optimal $\xi$ which minimizes \eqref{Pe,noisy,SD,app}.}
\vspace{-3mm}
\section{\textcolor{black}{Proof of Proposition \ref{convexity}}}\label{app3}
\textcolor{black}{The problem \eqref{opti,app} has $K+1$ optimization variables and the evaluation of its Hessian requires very high computational complexity. To decrease the complexity, we first consider the simplest case with $K=2$ and investigate the Hessian of $Q^\sharp_{\ss\FC}[j]$ with respect to $S_1$ for a fixed $\xi$. To this end, we take the first derivative of $Q^\sharp_{\ss\FC}[j]$ with respect to $S_1$. In \eqref{Pe,noisy,SD,app}, $\Lambda$ is a discrete function in terms of $S_1$, which makes the derivative cumbersome. If we approximate $\Lambda$ using \eqref{CDF,possion,2}, there is no closed-form for the first derivative of \eqref{CDF,possion,2} with respect to $S_1$. To overcome this challenge, we approximate $\Lambda$ by another continuous approximation, i.e., the continuous regularized incomplete Gamma function. By doing so, we have
\begin{align}\label{Gamma}
\Lambda\approx\frac{\Gamma\left(\lceil\xi\rceil,\hat{\lambda}_\N^\D[j]+\hat{\lambda}_{\s,h}^{\D,\Tot}[j]\right)}{\Gamma\left(\lceil\xi\rceil\right)},
\end{align}
where $\Gamma\left(\gamma,\delta\right)$ is the incomplete Gamma function and the Gamma function $\Gamma\left(\gamma\right)$ is a special case of $\Gamma\left(\gamma,\delta\right)$ with $\delta=0$. Applying this approximation to \eqref{Pe,noisy,SD,app}, we obtain the continuous approximation of $Q^\sharp_{\ss\FC}[j]$. Using $\partial\Gamma\left(\gamma,\delta\right)/\partial\delta=-\exp\left(-\delta\right)\delta^{\gamma-1}$, we take the first derivative of $Q^\sharp_{\ss\FC}[j]$ as
\ifOneCol	
\begin{align}\label{1st deri}
\frac{\partial\,Q^\sharp_{\ss\FC}[j]}{\partial\,S_1}\approx&\;\frac{1}{\Gamma\left(\lceil\xi\rceil\right)}\Bigg(\sum_{a_1=0}^{1}\sum_{a_2=0}^{1}\left(\left(1-P_1\right)\alpha(a_1,a_2)-P_1\beta(a_1,a_2)\right)\exp\left(-\Xi(a_1,a_2)\right)\nonumber\\
&\times\left(\Xi(a_1,a_2)\right)^{-1+\lceil\xi\rceil}\Omega(a_1,a_2)\Bigg),
\end{align}
\else
\begin{align}\label{1st deri}
\frac{\partial\,Q^\sharp_{\ss\FC}[j]}{\partial\,S_1}\approx&\;\frac{1}{\Gamma\left(\lceil\xi\rceil\right)}\Bigg(\sum_{a_1=0}^{1}\sum_{a_2=0}^{1}\left(\left(1-P_1\right)\alpha(a_1,a_2)\right.\nonumber\\
&\left.-P_1\beta(a_1,a_2)\right)\exp\left(-\Xi(a_1,a_2)\right)\nonumber\\
&\times\left(\Xi(a_1,a_2)\right)^{-1+\lceil\xi\rceil}\Omega(a_1,a_2)\Bigg),
\end{align}
\fi
where $\Xi(a_1,a_2)=\sigma_1S_1+\sigma_2(N-S_1)+a_1\nu_1 S_1+a_2\nu_2(N-S_1)$ and $\Omega(a_1,a_2)=\sigma_1-\sigma_2+a_1\nu_1-a_2\nu_2$, where $\sigma_k$ and $\nu_k$ are given in \eqref{nu_n} and \eqref{nu}, respectively.
We then find the second derivative of $Q^\sharp_{\ss\FC}[j]$ with respect to $S_1$ as
\ifOneCol	
\begin{align}\label{2st deri}
\frac{\partial^2 Q^\sharp_{\ss\FC}[j]}{\partial {S_1}^2}=&\;-\frac{1}{\Gamma\left(\lceil\xi\rceil\right)}\Bigg(\sum_{a_1=0}^{1}\sum_{a_2=0}^{1}\exp\left(-\Xi(a_1,a_2)\right)\left(\Xi(a_1,a_2)\right)^{-2+\lceil\xi\rceil}\left(\Omega(a_1,a_2)\right)^2\varpi\Bigg),
\end{align}
\else
\begin{align}\label{2st deri}
\frac{\partial^2 Q^\sharp_{\ss\FC}[j]}{\partial {S_1}^2}=&\;-\frac{1}{\Gamma\left(\lceil\xi\rceil\right)}\Bigg(\sum_{a_1=0}^{1}\sum_{a_2=0}^{1}\exp\left(-\Xi(a_1,a_2)\right)\nonumber\\
&\times\left(\Xi(a_1,a_2)\right)^{-2+\lceil\xi\rceil}\left(\Omega(a_1,a_2)\right)^2\varpi\Bigg),
\end{align}
\fi
where
\ifOneCol
$\varpi=\left(\left(1-P_1\right)\alpha(a_1,a_2)-P_1\beta(a_1,a_2)\right)\left(1-\lceil\xi\rceil+\Xi(a_1,a_2)\right)$.
\else
\begin{align}\label{varpi}
\varpi=\!\left(\left(1\!-\!P_1\right)\alpha(a_1,a_2)\!-\!P_1\beta(a_1,a_2)\right)\left(1\!-\!\lceil\xi\rceil+\Xi(a_1,a_2)\right).
\end{align}
\fi}

\textcolor{black}{In \eqref{2st deri}, all terms are nonnegative except for $\varpi$. Thus, if $\varpi>0$ holds for each summand, \eqref{2st deri} is nonnegative. However, the condition $\varpi>0$ is not always valid for $a_1\in\{0,1\}$ and $a_2\in\{0,1\}$. Thus, for a fixed $\xi$, ${\partial^2 Q^\sharp_{\ss\FC}[j]}/{\partial {\xi}^2}$ is not always nonnegative, which means that the Hessian of $Q^\sharp_{\ss\FC}[j]$ with respect to $\mathbf{S}$ and $\xi$ is not always positive semidefinite.}

\vspace{-3mm}
\section{\textcolor{black}{Proof of Lemma \ref{molDistri}}}\label{app4}
\textcolor{black}{Using \eqref{cond2}, we simplify \eqref{1st deri} as
\ifOneCol	
\begin{align}\label{1st deri,sym}
\frac{\partial\,Q^\sharp_{\ss\FC}[j]}{\partial\,S_1}\!=\!&\frac{\exp(-\Phi_1-2\nu S_1)}{\Phi_1\Phi_2\Gamma\left(\!\lceil\xi\rceil\right)}\left(\alpha(P_1-1)+\beta P_1\!\right)\nu\left(\!\exp(2\nu S_1)\Phi_1^{\lceil\xi\rceil}\Phi_2-\exp(N\nu)\Phi_2^{\lceil\xi\rceil}\Phi_1\!\right),
\end{align}
\else
\begin{align}\label{1st deri,sym}
\frac{\partial\,Q^\sharp_{\ss\FC}[j]}{\partial\,S_1}=&\;\frac{\exp(-\Phi_1-2\nu S_1)}{\Phi_1\Phi_2\Gamma\left(\lceil\xi\rceil\right)}\left(\alpha(P_1-1)+\beta P_1\right)\nu\nonumber\\
&\times\!\left(\!\exp(2\nu S_1)\Phi_1^{\lceil\xi\rceil}\Phi_2\!-\!\exp(N\nu)\Phi_2^{\lceil\xi\rceil}\Phi_1\!\right),
\end{align}
\fi
where $\Phi_1=N(\nu+\sigma)-\nu S_1$ and $\Phi_2=N\sigma+\nu S_1$. It can be shown that $S_1=\frac{N}{2}$ is the one of the solutions of \eqref{1st deri,sym}. Hence, $Q^\sharp_{\ss\FC}[j]$ has a local minimum or maximum when $S_1=\frac{N}{2}$. We then apply \eqref{cond2} and $S_1=\frac{N}{2}$ to \eqref{2st deri} to obtain the second derivative of $Q^\sharp_{\ss\FC}[j]$ at $S_1=\frac{N}{2}$. By doing so, we have
\ifOneCol	
\begin{align}\label{2st deri,sym}
\frac{\partial^2 Q^\sharp_{\ss\FC}[j]}{\partial {S_1}^2}|_{S_1=\frac{N}{2}}=&\;\frac{4\exp(-\frac{1}{2}N(\nu+2\sigma))}{N^2(\nu+2\sigma)^2\Gamma\left(\lceil\xi\rceil\right)}\nu^2(\frac{N(\nu+2\sigma)}{2})^{\lceil\xi\rceil}\Upsilon(\xi),
\end{align}
\else
\begin{align}\label{2st deri,sym}
\frac{\partial^2 Q^\sharp_{\ss\FC}[j]}{\partial {S_1}^2}|_{S_1=\frac{N}{2}}=&\;\frac{4\exp(-\frac{1}{2}N(\nu+2\sigma))}{N^2(\nu+2\sigma)^2\Gamma\left(\lceil\xi\rceil\right)}\nonumber\\
&\times\nu^2(\frac{N(\nu+2\sigma)}{2})^{\lceil\xi\rceil}\Upsilon(\xi),
\end{align}
\fi
where all terms are nonnegative except for $\Upsilon(\xi)$. Hence if $\Upsilon(\xi)>0$, \eqref{2st deri,sym} is nonnegative and $Q^\sharp_{\ss\FC}[j]$ achieves a local minimum at $S_1=\frac{N}{2}$; otherwise, it achieves a local maximum.}
\ifOneCol
\vspace{-10mm}
\else
\vspace{-5mm}
\fi
\section{\textcolor{black}{Proof of Theorem \ref{molDistri,asy}}}\label{app5}
\textcolor{black}{Based on Lemma \ref{molDistri}, $Q^\sharp_{\ss\FC}[j]$ achieves a local minimal value at $S_1=\frac{N}{2}$ when $\Upsilon(\xi)>0$. Based on Lemma \ref{tight}, the approximation of $Q_{\ss\FC}[j]$ by $Q^\sharp_{\ss\FC}[j]$ is tight when $\xi=\xi_{\ss\FC}^{\ad,\SD}[j]$. Thus, we can prove that ${Q}_{\ss\FC}[j]$ always achieves a local minimal value at $S_1=\frac{N}{2}$ by proving that $\Upsilon(\xi)>0$ always holds when $\xi=\xi_{\ss\FC}^{\ad,\SD}[j]$. That is to say, we need to prove $\Upsilon (\xi_{\ss\FC}^{\ad,\SD}[j]) >0$. Based on the proof of Lemma \ref{opti-thres}, we also recall that $\xi_{\ss\FC}^{\ad,\SD}[j]$ is the solution to $P_1\psi_1(\xi)-(1-P_1)\psi_2(\xi)={\partial\,Q^\sharp_{\ss\FC}[j]}/{\partial\,\xi}=0$ if $P_1=\frac{1}{2}$. Thus, $\xi_{\ss\FC}^{\ad,\SD}[j]$ satisfies the condition: $\psi_1(\xi_{\ss\FC}^{\ad,\SD}[j])-\psi_2(\xi_{\ss\FC}^{\ad,\SD}[j])=0$. Applying $S_1 = N/2$ to \eqref{Pe,noisy,SD,app,deri}, we write $\psi_b(\xi_{\ss\FC}^{\ad,\SD}[j])$, where $b\in\{0,1\}$, using $\nu$ and $\sigma$ as
\ifOneCol	
\begin{align}\label{psi0}
\psi_b(\xi_{\ss\FC}^{\ad,\SD}[j])
=&\;{\left(\xi_{\ss\FC}^{\ad,\SD}[j]!\right)}^{-1}\bigg[\textrm{Pr}\left(0,0|b\right)\exp\left(-\sigma N\right)\left(\sigma N\right)^{\xi_{\ss\FC}^{\ad,\SD}[j]}+\textrm{Pr}\left(0,1|b\right)\exp\left(-\sigma N-\frac{\nu N}{2}\right)\nonumber\\
&\times\left(\sigma N+\frac{\nu N}{2}\right)^{\xi_{\ss\FC}^{\ad,\SD}[j]}+\textrm{Pr}\left(1,0|b\right)\exp\left(-\sigma N-\frac{\nu N}{2}\right)\left(\sigma N+\frac{\nu N}{2}\right)^{\xi_{\ss\FC}^{\ad,\SD}[j]}\nonumber\\
&+\textrm{Pr}\left(1,1|b\right)\exp\left(-\sigma N-\nu N\right)\left(\sigma N+\nu N\right)^{\xi_{\ss\FC}^{\ad,\SD}[j]}\bigg],
\end{align}
\else
\begin{align}\label{psi0}
&\psi_b(\xi_{\ss\FC}^{\ad,\SD}[j])\nonumber\\
=&\;{\left(\xi_{\ss\FC}^{\ad,\SD}[j]!\right)}^{-1}\bigg[\textrm{Pr}\left(0,0|b\right){}\exp\left(-\sigma N\right)\left(\sigma N\right)^{\xi_{\ss\FC}^{\ad,\SD}[j]}\nonumber\\
&+\textrm{Pr}\left(0,1|b\right)\exp\left(-\sigma N-\frac{\nu N}{2}\right)\left(\sigma N+\frac{\nu N}{2}\right)^{\xi_{\ss\FC}^{\ad,\SD}[j]}\nonumber\\
&+\textrm{Pr}\left(1,0|b\right)\exp\left(-\sigma N-\frac{\nu N}{2}\right)\left(\sigma N+\frac{\nu N}{2}\right)^{\xi_{\ss\FC}^{\ad,\SD}[j]}\nonumber\\
&+\textrm{Pr}\left(1,1|b\right)\exp\left(-\sigma N-\nu N\right)\left(\sigma N+\nu N\right)^{\xi_{\ss\FC}^{\ad,\SD}[j]}\bigg],
\end{align}
\fi
where $\textrm{Pr}\left(\hat{\mathcal{W}}^{{\ss\RX}}_{j,h} |b\right)=\textrm{Pr}\left(\hat{\mathcal{W}}^{{\ss\RX}}_{j,h} |W_{\ss\T}[j]=b,\textbf{W}_{\ss\TX}^{j-1}\right)$. In \eqref{psi0}, we have $\textrm{Pr}\left(0,1|0\right)=\textrm{Pr}\left(1,0|0\right)=\alpha$ and $\textrm{Pr}\left(0,1|1\right)=\textrm{Pr}\left(1,0|1\right)=\beta$ based on \eqref{cond2}. We then approximate $\textrm{Pr}\left(0,0|0\right)$ = $\textrm{Pr}\left(1,1|1\right)\approx1$ and $\textrm{Pr}\left(0,0|1\right)$ = $\textrm{Pr}\left(1,1|0\right)\approx0$, which is tight when the error probability of the $\TX-\RX_k$ link is small. Using these approximations, $\alpha$, and $\beta$, we rewrite $\psi_0(\xi_{\ss\FC}^{\ad,\SD}[j])$ and $\psi_1(\xi_{\ss\FC}^{\ad,\SD}[j])$ as
\ifOneCol
\begin{align}\label{psi0,re}
\psi_0(\xi_{\ss\FC}^{\ad,\SD}[j])\approx\frac{\exp\left(-\sigma N\right)\left(\sigma N\right)^{\xi_{\ss\FC}^{\ad,\SD}[j]}+2\alpha\exp\left(-\sigma N-\frac{\nu N}{2}\right)\left(\sigma N+\frac{\nu N}{2}\right)^{\xi_{\ss\FC}^{\ad,\SD}[j]}}{{\left(\xi_{\ss\FC}^{\ad,\SD}[j]!\right)}}
\end{align}	
\else
\begin{align}\label{psi0,re}
\psi_0(\xi_{\ss\FC}^{\ad,\SD}[j])\approx&\;{{\left(\xi_{\ss\FC}^{\ad,\SD}[j]!\right)}^{-1}}\exp\left(-\sigma N\right)\left(\sigma N\right)^{\xi_{\ss\FC}^{\ad,\SD}[j]}\nonumber\\
&+2\alpha{{\left(\xi_{\ss\FC}^{\ad,\SD}[j]!\right)}^{-1}}\exp\left(-\sigma N-\frac{\nu N}{2}\right)\nonumber\\
&\times\left(\sigma N+\frac{\nu N}{2}\right)^{\xi_{\ss\FC}^{\ad,\SD}[j]}
\end{align}
\fi
and
\ifOneCol
\begin{align}\label{psi1,re}
\psi_1(\xi_{\ss\FC}^{\ad,\SD}[j])\!\approx\!\frac{\exp\left(\!-\sigma N-\!\nu N\!\right)\left(\!\sigma N\!+\!\nu N\!\right)^{\xi_{\ss\FC}^{\ad,\SD}[j]}
\!+\!2\beta\exp\left(\!-\sigma N\!-\frac{\nu N}{2}\!\right)\left(\!\sigma N\!+\frac{\nu N}{2}\!\right)^{\xi_{\ss\FC}^{\ad,\SD}[j]}}{{\left(\!\xi_{\ss\FC}^{\ad,\SD}[j]\!\right)}},
\end{align}	
\else
\begin{align}\label{psi1,re}
\psi_1(\xi_{\ss\FC}^{\ad,\SD}[j])\approx&\;{{\left(\xi_{\ss\FC}^{\ad,\SD}[j]!\right)}^{-1}}\exp\left(-\sigma N-\nu N\right)\nonumber\\
&\times\left(\sigma N+\nu N\right)^{\xi_{\ss\FC}^{\ad,\SD}[j]}
+2\beta{{\left(\xi_{\ss\FC}^{\ad,\SD}[j]!\right)}^{-1}}\nonumber\\
&\hspace{-10mm}\times\exp\left(-\sigma N-\frac{\nu N}{2}\right)
\left(\sigma N+\frac{\nu N}{2}\right)^{\xi_{\ss\FC}^{\ad,\SD}[j]},
\end{align}
\fi
respectively. Using $\psi_1(\xi_{\ss\FC}^{\ad,\SD}[j])-\psi_2(\xi_{\ss\FC}^{\ad,\SD}[j])=0$ and some basic manipulations, we obtain
\ifOneCol
\begin{align}\label{beta-alpha}
\beta-\alpha = &\;\frac{1}{2}\exp\left(-\frac{\nu N}{2}\right)\left(\frac{\sigma N+\nu N}{\sigma N+\frac{\nu N}{2}}\right)^{\xi_{\ss\FC}^{\ad,\SD}[j]}\big(\exp\left(\nu N\right)\left(\frac{\sigma N}{\sigma N+\nu N}\right)^{\xi_{\ss\FC}^{\ad,\SD}[j]}-1\big).
\end{align}
\else
\begin{align}\label{beta-alpha}
\beta-\alpha = &\;\frac{1}{2}\exp\left(-\frac{\nu N}{2}\right)\left(\frac{\sigma N+\nu N}{\sigma N+\frac{\nu N}{2}}\right)^{\xi_{\ss\FC}^{\ad,\SD}[j]}\nonumber\\
&\times\left(\exp\left(\nu N\right)\left(\frac{\sigma N}{\sigma N+\nu N}\right)^{\xi_{\ss\FC}^{\ad,\SD}[j]}-1\right).
\end{align}
\fi
}

\textcolor{black}{Applying \eqref{beta-alpha} and $P_1 = \frac{1}{2}$ to \eqref{cond}, we have
\begin{align}\label{cond, ada}
\Upsilon (\xi_{\ss\FC}^{\ad,\SD}[j])=\!\frac{\theta_1\theta_2}{4}\exp\left(\!-\frac{\nu N}{2}\!\right)\left(\!\frac{\sigma N+\nu N}{\sigma N+\frac{\nu N}{2}}\!\right)^{\xi_{\ss\FC}^{\ad,\SD}[j]},
\end{align}
where $\theta_1 = \left(\exp\left(\nu N\right)\left({\sigma N}/({\sigma N+\nu N})\right)^{\xi_{\ss\FC}^{\ad,\SD}[j]}-1\right)$ and $\theta_2 = \left(2+N(\nu+2\sigma)-2\xi_{\ss\FC}^{\ad,\SD}[j]\right)$. To prove $\Upsilon (\xi_{\ss\FC}^{\ad,\SD}[j])>0$, we only need to prove $\theta_1\theta_2 >0$, since all other terms in \eqref{cond, ada} are nonnegative. If $\theta_1>0$, we have $\xi_{\ss\FC}^{\ad,\SD}[j] < {\nu N}/{\log\left((\sigma N+\nu N)/\nu N\right)}$. Applying $\log(x)\leq x-1$, where $x>0$, we further lower-bound $\xi_{\ss\FC}^{\ad,\SD}[j]$ by $\xi_{\ss\FC}^{\ad,\SD}[j] < \sigma N$, which leads to $\theta_2>0$. If $\theta_1<0$, we have $\xi_{\ss\FC}^{\ad,\SD}[j] > {\nu N}/{\log\left((\sigma N+\nu N)/\nu N\right)}$. Applying $\log(x)\geq 1-1/x$, where $x>0$, we further upper-bound $\xi_{\ss\FC}^{\ad,\SD}[j]$ by $\xi_{\ss\FC}^{\ad,\SD}[j] > (\sigma N+\nu N)$, which leads to $\theta_2<0$ if $\nu N/2>1$. Although the validity of $\nu N/2>1$ depends on the value of $\nu$ and $N$, it is generally valid. This is because $\nu N/2>1$ means that at least one signaling molecule is expected at the FC if the decision at $\RX_k$ is ``1'' and it is a reasonable condition to be satisfied. Since $\theta_1$ and $\theta_2$ are always both negative or positive, $\theta_1\theta_2 >0$ holds, which leads to $\Upsilon (\xi_{\ss\FC}^{\ad,\SD}[j])>0$. Therefore, ${Q}_{\ss\FC}[j]$ achieves a local minimum at $S_1={N}/{2}$ when $K=2$ in a symmetric topology.}

\ifOneCol
\vspace{-5mm}
\else
\vspace{-0mm}
\fi

\begin{thebibliography}{10}
\providecommand{\url}[1]{#1}
\csname url@samestyle\endcsname
\providecommand{\newblock}{\relax}
\providecommand{\bibinfo}[2]{#2}
\providecommand{\BIBentrySTDinterwordspacing}{\spaceskip=0pt\relax}
\providecommand{\BIBentryALTinterwordstretchfactor}{4}
\providecommand{\BIBentryALTinterwordspacing}{\spaceskip=\fontdimen2\font plus
\BIBentryALTinterwordstretchfactor\fontdimen3\font minus
  \fontdimen4\font\relax}
\providecommand{\BIBforeignlanguage}[2]{{%
\expandafter\ifx\csname l@#1\endcsname\relax
\typeout{** WARNING: IEEEtran.bst: No hyphenation pattern has been}%
\typeout{** loaded for the language `#1'. Using the pattern for}%
\typeout{** the default language instead.}%
\else
\language=\csname l@#1\endcsname
\fi
#2}}
\providecommand{\BIBdecl}{\relax}
\BIBdecl

\bibitem{Fang2017C}
Y.~Fang, A.~Noel, N.~Yang, A.~W. Eckford, and R.~A. Kennedy, ``Maximum
  likelihood detection for cooperative molecular communication,'' in
  \emph{Proc. IEEE ICC}, May 2018, pp. 1--7.

\bibitem{Nariman}
N.~Farsad, H.~B. Yilmaz, A.~Eckford, C.~B. Chae, and W.~Guo, ``A comprehensive
  survey of recent advancements in molecular communication,'' \emph{{IEEE}
  Commun. Surveys Tuts.}, vol.~18, no.~3, pp. 1887--1919, Aug. 2016.

\bibitem{Tadashi2010}
T.~Nakano and J.~Q. Liu, ``Design and analysis of molecular relay channels: An
  information theoretic approach,'' \emph{{IEEE} Trans. Nanobiosci.}, vol.~9,
  no.~3, pp. 213--221, Sep. 2010.

\bibitem{Atakan2008}
B.~Atakan and O.~B. Akan, ``On molecular multiple-access, broadcast, and relay
  channels in nanonetworks,'' in \emph{Proc. ICST BIONETICS}, Nov. 2008, pp.
  16:1--16:8.

\bibitem{Nakano2013}
T.~Nakano, Y.~Okaie, and A.~V. Vasilakos, ``Transmission rate control for
  molecular communication among biological nanomachines,'' \emph{{IEEE} J.
  Select. Areas Commun.}, vol.~31, no.~12, pp. 835--846, Dec. 2013.

\bibitem{Chun2013}
C.~T. Chou, ``Extended master equation models for molecular communication
  networks,'' \emph{{IEEE} Trans. Nanobiosci.}, vol.~12, no.~2, pp. 79--92,
  June 2013.

\bibitem{Koo2016}
B.~H. Koo, C.~Lee, H.~B. Yilmaz, N.~Farsad, A.~W. Eckford, and C.-B. Chae,
  ``Molecular {MIMO}: From theory to prototype,'' \emph{{IEEE} J. Select. Areas
  Commun.}, vol.~34, no.~3, pp. 600--614, Mar. 2016.

\bibitem{GC2016}
Y.~Fang, A.~Noel, N.~Yang, A.~W. Eckford, and R.~A. Kennedy, ``Distributed
  cooperative detection for multi-receiver molecular communication,'' in
  \emph{Proc. IEEE GLOBECOM}, Dec. 2016, pp. 1--7.

\bibitem{TMBMC2016}
------, ``Convex optimization of distributed cooperative detection in
  multi-receiver molecular communication,'' \emph{{IEEE} Trans. Mol. Bio.
  Multi-Scale Commun.}, vol.~3, no.~3, pp. 166--182, Sep. 2017.

\bibitem{simplified}
Y.~Fang, A.~Noel, Y.~Wang, and N.~Yang, ``Simplified cooperative detection for
  multi-receiver molecular communication,'' in \emph{Proc. IEEE ITW}, Nov.
  2017, pp. 1--5.

\bibitem{Digital}
J.~G. Proakis, \emph{Digital Communication}, 4th~ed.\hskip 1em plus 0.5em minus
  0.4em\relax New York: McGraw-Hill, 2000.

\bibitem{Deniz2016}
D.~Kilinc and O.~B. Akan, ``Receiver design for molecular communication,''
  \emph{{IEEE} J. Select. Areas Commun.}, vol.~31, no.~12, pp. 705--714, Dec.
  2013.

\bibitem{Adam2014}
A.~Noel, K.~C. Cheung, and R.~Schober, ``Optimal receiver design for diffusive
  molecular communication with flow and additive noise,'' \emph{{IEEE} Trans.
  Nanobiosci.}, vol.~13, no.~3, pp. 350--362, Sept. 2014.

\bibitem{Mahfuz2015}
M.~U. Mahfuz \emph{et~al.}, ``A comprehensive analysis of strength-based
  optimum signal detection in concentration-encoded molecular communication
  with spike transmission,'' \emph{{IEEE} Trans. Nanobiosci.}, vol.~14, no.~1,
  pp. 67--83, Jan. 2015.

\bibitem{Amit2016}
A.~Singhal, R.~K. Mallik, and B.~Lall, ``Performance analysis of amplitude
  modulation schemes for diffusion-based molecular communication,''
  \emph{{IEEE} Trans. Wireless Commun.}, vol.~14, no.~10, pp. 5681--5691, Oct.
  2015.

\bibitem{Ghavami2017}
S.~Ghavami and F.~Lahouti, ``Abnormality detection in correlated gaussian
  molecular nano-networks: Design and analysis,'' \emph{{IEEE} Trans.
  Nanobiosci.}, vol.~16, no.~3, pp. 189--202, Apr. 2017.

\bibitem{Trang2017}
T.~C. Mai, M.~Egan, T.~Q. Duong, and M.~D. Renzo, ``Event detection in
  molecular communication networks with anomalous diffusion,'' \emph{{IEEE}
  Commun. Lett.}, vol.~21, no.~6, pp. 1249--1252, June 2017.

\bibitem{Mosayebi2017}
R.~Mosayebi, V.~Jamali, N.~Ghoroghchian, R.~Schober, M.~Nasiri-Kenari, and
  M.~Mehrabi, ``Cooperative abnormality detection via diffusive molecular
  communications,'' \emph{{IEEE} Trans. Nanobiosci.}, vol.~16, no.~8, pp.
  828--842, Dec. 2017.

\bibitem{Rogers2016}
U.~Rogers and M.~S. Koh, ``Parallel molecular distributed detection with
  brownian motion,'' \emph{{IEEE} Trans. Nanobiosci.}, vol.~15, no.~8, pp.
  871--880, Dec. 2016.

\bibitem{Silva}
A.~P. de~Silva \emph{et~al.}, ``Molecular logic and computing,'' \emph{Nature
  Nanotech.}, vol.~2, no.~7, pp. 399--410, Jul. 2007.

\bibitem{Kuran2011}
M.~S. Kuran, H.~B. Yilmaz, T.~Tugcu, and I.~F. Akyildiz, ``Modulation
  techniques for communication via diffusion in nanonetworks,'' in \emph{Proc.
  IEEE ICC}, June 2011, pp. 1--5.

\bibitem{Ahmadzadeh2015C}
A.~Ahmadzadeh, A.~Noel, A.~Burkovski, and R.~Schober, ``Amplify-and-forward
  relaying in two-hop diffusion-based molecular communication networks,'' in
  \emph{Proc. IEEE GLOBECOM}, Dec. 2015, pp. 1--7.

\bibitem{Siuti}
P.~Siuti, J.~Yazbek, and T.~K~Lu, ``Synthetic circuits integrating logic and
  memory in living cells,'' \emph{Nature biotechnology}, vol.~31, pp. 448--452,
  Feb. 2013.

\bibitem{Mondrag}
O.~Mondrag{\'o}n-Palomino, T.~Danino, J.~Selimkhanov, L.~Tsimring, and
  J.~Hasty, ``Entrainment of a population of synthetic genetic oscillators,''
  \emph{Sci.}, vol. 333, no. 6047, pp. 1315--1319, Sep. 2011.

\bibitem{ShahMohammadian2013}
H.~ShahMohammadian, G.~G. Messier, and S.~Magierowski, ``Blind synchronization
  in diffusion-based molecular communication channels,'' \emph{{IEEE} Commun.
  Lett.}, vol.~17, no.~11, pp. 2156--2159, Nov. 2013.

\bibitem{Abadal2011}
S.~Abadal \emph{et~al.}, ``Bio-inspired synchronization for nanocommunication
  networks,'' in \emph{Proc. IEEE GLOBECOM}, Dec. 2011, pp. 1--5.

\bibitem{Lu2015}
Y.~Lu, M.~D. Higgins, and M.~S. Leeson, ``Comparison of channel coding schemes
  for molecular communications systems,'' \emph{{IEEE} Trans. Commun.},
  vol.~63, no.~11, pp. 3991--4001, Nov. 2015.

\bibitem{Noel2013}
A.~Noel, K.~C. Cheung, and R.~Schober, ``Using dimensional analysis to assess
  scalability and accuracy in molecular communication,'' in \emph{Proc. IEEE
  ICC}, June 2013, pp. 818--823.

\bibitem{Pierobon}
M.~Pierobon and I.~F. Akyildiz, ``Diffusion-based noise analysis for molecular
  communication in nanonetworks,'' \emph{{IEEE} Trans. Signal Processing},
  vol.~59, no.~6, pp. 2532--2547, Jun. 2011.

\bibitem{Mosayebi2014}
R.~Mosayebi \emph{et~al.}, ``Receivers for diffusion-based molecular
  communication: Exploiting memory and sampling rate,'' \emph{{IEEE} J. Select.
  Areas Commun.}, vol.~32, no.~12, pp. 2368--2380, Dec. 2014.

\bibitem{Ahmadzadeh2015}
A.~Ahmadzadeh, A.~Noel, and R.~Schober, ``Analysis and design of multi-hop
  diffusion-based molecular communication networks,'' \emph{IEEE Trans. Mol.
  Biol. Multi-Scale Commun.}, vol.~1, no.~2, pp. 144--157, June 2015.

\bibitem{Andrew2004}
S.~S. Andrews and D.~Bray, ``Stochastic simulation of chemical reactions with
  spatial resolution and single molecule detail,'' \emph{Physical Biology},
  vol.~1, no.~3, pp. 135--151, Aug. 2004.

\bibitem{Aminian}
G.~Aminian, H.~Ghourchian, A.~Gohari, M.~Mirmohseni, and M.~Nasiri-Kenari, ``On
  the capacity of signal dependent noise channels,'' in \emph{Proc. IEEE
  IWCIT}, May 2017, pp. 1--6.

\end{thebibliography}


\end{document}